\documentclass[12pt]{article} 
\pdfoutput=1
 \usepackage{amsmath}
\usepackage{amssymb}
\usepackage{graphicx}
\usepackage{tikz}
\usepackage{xcolor}
\usepackage{framed}
\usepackage{bm}
\usepackage{float}
  \usepackage{subfigure}
  \usepackage[labelfont=bf,font={small}]{caption}
 \usepackage{mciteplus}
 \usepackage{bbold}
   \definecolor{darkblue}{rgb}{0.1,0.1,.7}
 \usepackage[colorlinks, linkcolor=darkblue, citecolor=darkblue, urlcolor=darkblue, linktocpage,hyperfootnotes=false]{hyperref} 

\numberwithin{equation}{section}
 
  \usetikzlibrary{decorations.pathmorphing}
\tikzset{snake it/.style={decorate, decoration=snake}}
\usetikzlibrary{shapes}
\usepackage{empheq}

\newcommand{\tj}[6]{ \begin{pmatrix}
   #1 & #2 & #3 \\
   #4 & #5 & #6 
  \end{pmatrix}}
  \newcommand{\sj}[6]{ \begin{Bmatrix}
   #1 & #2 & #3 \\
   #4 & #5 & #6 
  \end{Bmatrix}}

			\newcommand{\U}{\scriptscriptstyle \text{U}}
	\newcommand{\M}{\scriptscriptstyle \text{M}}
	\newcommand{\PT}{\scriptscriptstyle \text{PT}}

  \newlength{\abstractwidth}
  \newcommand{\be}{\begin{equation}}
  \newcommand{\bea}{\begin{eqnarray}}
  \newcommand{\eea}{\end{eqnarray}}
  \newcommand{\beq}{\begin{equation}}
  \newcommand{\ee}{\end{equation}}
  \newcommand{\eeq}{\end{equation}}

\renewcommand{\Re}{\operatorname{Re}}
\renewcommand{\Im}{\operatorname{Im}}
\newcommand{\lb}{\langle}
\newcommand{\rb}{\rangle}

\begin{document}

\begin{titlepage}
  \bigskip

  \bigskip\bigskip

  \bigskip

\begin{center}
{\large \bf {Liouville quantum gravity -  holography, JT and matrices}}
    \bigskip
\bigskip
\end{center}

  \begin{center}

Thomas G. Mertens\footnote{\texttt{thomas.mertens@ugent.be}} and Gustavo J. Turiaci\footnote{\texttt{turiaci@ucsb.edu }} \\
  \vspace{1cm}
	  \bigskip \rm
{\small ${}^1$Department of Physics and Astronomy, Ghent University, Krijgslaan, 281-S9, 9000 Gent, Belgium}  \\
\vspace{0.1cm}
{\small ${}^2$Physics Department, University of California, Santa Barbara, CA 93106, USA}  \\
\rm

  \bigskip \rm
\bigskip
 
 \rm

\bigskip
\bigskip

  \end{center}

\vspace{1.5cm}
  \begin{abstract}

We study two-dimensional Liouville gravity and minimal string theory on spaces with fixed length boundaries. We find explicit formulas describing the gravitational dressing of bulk and boundary correlators in the disk. Their structure has a striking resemblance with observables in 2d BF (plus a boundary term), associated to a quantum deformation of $SL(2,\mathbb{R})$, a connection we develop in some detail. For the case of the $(2,p)$ minimal string theory, we compare and match the results from the continuum approach with a matrix model calculation, and verify that in the large $p$ limit the correlators match with Jackiw-Teitelboim gravity. We consider multi-boundary amplitudes that we write in terms of gluing bulk one-point functions using a quantum deformation of the Weil-Petersson volumes and gluing measures. Generating functions for genus zero Weil-Petersson volumes are derived, taking the large $p$ limit. Finally, we present preliminary evidence that the bulk theory can be interpreted as a 2d dilaton gravity model with a $\sinh \Phi$ dilaton potential.

 \medskip
  \noindent
  \end{abstract}
\bigskip \bigskip \bigskip

  \end{titlepage}

   \tableofcontents


\newpage
\section{Introduction and summary}

One of the most exciting developments the past few years, is the discovery of exactly solvable models of quantum gravity, starting with Kitaev's SYK models \cite{Kitaevtalks,*Sachdev:1992fk,*Polchinski:2016xgd,*Maldacena:2016hyu}, going through bulk Jackiw-Teitelboim (JT) gravity \cite{Jackiw:1984je,*Teitelboim:1983ux,Almheiri:2014cka,*Jensen:2016pah,*Maldacena:2016upp,*Engelsoy:2016xyb} and its correlation functions \cite{Bagrets:2016cdf,Stanford:2017thb, Mertens:2017mtv,Lam:2018pvp,Mertens:2018fds,Blommaert:2018oro,Kitaev:2018wpr, *Yang:2018gdb,Iliesiu:2019xuh}, and leading to the inclusion of higher genus and random matrix descriptions \cite{Saad:2019lba}, making contact with the black hole information paradox in its various incarnations \cite{Saad:2019pqd, Almheiri:2019qdq, *Penington:2019kki, Marolf:2020xie}.
It goes without saying that finding other models that are solvable to the same extent would be highly valuable, in particular to test the robustness of the ideas. For example, it is important to have a similar non-perturbative definition of theories of gravity as in \cite{Saad:2019lba} that are also coupled to matter.  

In the same work \cite{Saad:2019lba}, it was proposed that JT gravity can be viewed as a parametric limit of the older minimal string model.
The latter can be viewed as a double-scaled matrix integral \cite{Brezin:1990rb, *Douglas:1989ve, *Gross:1989vs} that in the continuum description becomes a non-critical string theory described by Liouville CFT, coupled to a minimal model and the $bc$ ghost sector. We will call this combination \emph{Liouville gravity} in what follows. Since there is a substantial amount of evidence in favor of a random matrix description of these models, finding JT gravity within a limiting situation illustrates that it is in hindsight not a surprise at all that JT gravity is a matrix integral.
\\~\\
In this work, we will develop these UV ancestors of JT gravity in more detail. We will enlarge our scope slightly: instead of restricting to only minimal models to complete the Liouville CFT, we will consider a generic matter CFT for the first few sections. In that case, we do not have a (known) matrix description to guide us. At times, we will restrict to the minimal string and find perfect agreement between continuum and matrix descriptions. 
A particular emphasis is placed on correlation functions within these theories and how precisely they approach the JT correlation functions in a certain limit.
We also highlight how the Riemann surface description of JT gravity at higher topology also generalizes (in fact, quantum deforms) to these models leading to generalizations of the Weil-Petersson (WP) volumes to glue surfaces together.
\\~\\
Let us sketch the set-up in more detail.
Consider a disk-shaped worldsheet with coordinates $(z,\bar{z})$ and boundary coordinate $x$. Within Liouville gravity, we are allowed to insert closed string tachyon vertex operators $\mathcal{T}_{i}$ and open string tachyon vertex operators $\mathcal{B}_{i}$. Denoting these operator insertions collectively by $\mathcal{O}$, we will define the disk amplitudes $\mathcal{A}_\mathcal{O}(\ell_1,\ldots, \ell_n)$ with fixed length boundaries $\ell_1 \hdots \ell_n$ (see discussion around \eqref{introbdycond} for more details on the boundary conditions) as
\begin{figure}[h]
\centering
\raisebox{18mm}{$\mathcal{A}_\mathcal{O}(\ell_1,\ldots, \ell_n) \quad = \quad$}
\begin{tikzpicture}[scale=0.8]
\draw[fill=blue!40!white,opacity=0.7] (0,0) ellipse (1.5 and 1.5);
\draw[fill] (-0.5,0) circle (0.06);
\node at (0,-1.985) {};
\node at (-0.1,0) {\small $\mathcal{T}_1$};
\draw[fill] (0.5,0) circle (0.06);
\node at (0.9,0) {\small $\mathcal{T}_2$};
\node at (0.75,-0.5) {\small $...$};
\draw[fill] (-1.3,-0.75) circle (0.06); 
\node at (-1.65,-0.8) {\small $\mathcal{B}_1$};
\node at (-1.85,0) {\small $\ell_1$};
\node at (-1,1.6) {\small $\ell_2$};
\node at (1,1.6) {\small $\ell_3$};
\draw[fill] (-1.3,0.75) circle (0.06); 
\node at (-1.65,0.8) {\small $\mathcal{B}_2$};
\draw[fill] (0,1.5) circle (0.06); 
\node at (0,1.8) {\small $...$};
\end{tikzpicture}
\end{figure}

\noindent Since the string worldsheet theory is treated as 2d gravity (by imposing the Virasoro constraints), the operator insertions of interest $\mathcal{B}_i$ and $\mathcal{T}_i$ have to be worldsheet coordinate-invariant. 
The familiar strategy from string theory is to restrict these to conformal weight one (in both holomorphic and anti-holomorphic sectors), and then integrate them over the entire worldsheet:
\begin{equation}\label{defopintro}
\mathcal{B}= \oint_{\partial \Sigma} dx \, \Phi_{\rm M}(x)e^{\beta \phi(x)}   ,\qquad  \mathcal{T}= \int_\Sigma d^2z \,\mathcal{O}_{\rm M}(z,\bar{z}) e^{2 \alpha \phi(z,\bar{z})}.
\end{equation}
Here $\Phi_{\rm M}$ and $\mathcal{O}_{\rm M}$ denote boundary and bulk matter operators, $\phi$ is the Liouville field (scale factor in physical metric) and the parameters $\beta$ and $\alpha$ are tuned to the matter operator to make the integrand marginal in both cases. These operators will be labeled by the Liouville parameters corresponding to the matter operators $\alpha_M$ and $\beta_M$ (see \eqref{bulklioupar2} and \eqref{bdylioupar} for the definition). The conventional interpretation of these formulas is that the bare matter operators $\Phi_{\rm M}$ and $\mathcal{O}_{\rm M}$ (as objects in only the matter CFT), are gravitationally dressed by the Liouville vertex operators $e^{\beta \phi(x)}$ and $e^{2 \alpha \phi(z,\bar{z})}$ to produce observable worldsheet diff-invariant operators. From this perspective, the matter fields are the more fundamental objects and we will indeed reach this conclusion throughout our work as well.
As well-known in string theory, we can use the SL$(2,\mathbb{R})$ isometries of the disk to gauge-fix the worldsheet location of three degrees of freedom (where a bulk operator counts as two, and a boundary operator as one). If one has more operator insertions, there are non-trivial integrations left over the moduli space of the punctured disk. Throughout this work, we only focus on the case without moduli integration. This leaves only four disk configurations which we explicitly investigate.
In the final section of this work, we investigate higher topology, and in particular the annulus diagram which has a single worldsheet modulus.
\\~\\
It should be emphasized that the worldsheet boundary coordinates $x_i$ (and their moduli) and the physical distances $\ell_i$ are distinct. They are only related by the non-local (and not so restrictive) constraints:
\begin{equation}\label{introbdycond}
\ell_i = \int_{x_i}^{x_{i+1}}dx\hspace{0.1cm} e^{b\phi(x)}
\end{equation}
in terms of the Liouville field $\phi$ appearing in the Liouville gravity models we will consider.
For all disk cases we study, the worldsheet coordinate $x$-dependence drops out due to gauge-fixing, but the final result depends explicitly on the physical distances $\ell$. In this sense, even though boundary operators are integrated over the worldsheet as in \eqref{defopintro}, they behave as local insertions in the physical space and their gravitational dressing has the effect of fixing geodesic distances between them. Moreover, even though the worldsheet theory is a CFT, the boundary amplitudes as a function of physical lengths do not respect conformal symmetry (see for example \eqref{twoa} below).
For the annulus amplitude, there is a single worldsheet modulus $\tau$ that needs to be integrated over. Doing so leads in the end to an amplitude that depends on the physical lengths of both boundaries of the annulus.
\\~\\
Next we present a summary of the main results regarding fixed length amplitudes, some known some new, that are computed in this paper. We introduce the quantities:
\beq
\mu_B(s) = \kappa \cosh 2\pi b s,~~~\kappa \equiv \frac{\sqrt{\mu}}{\sqrt{\sin \pi b^2}},
\eeq
where $\mu$ is the bulk cosmological constant, $\mu_B(s)$ is the boundary cosmological constant for FZZT boundaries labeled by $s$, and $b$ is defined through the central charge of the Liouville field $c_{\rm L}=1+6Q^2$, with $Q=b+1/b$. 

\paragraph{Partition Function:} We compute the marked partition function 
\bea
Z(\ell) = N \mu^{\frac{Q}{2b}} \int_{0}^{\infty} ds ~e^{-\ell \mu_B(s)}\rho(s) ,
\eea
where we define the spectral weight 
\beq
\rho(s)\equiv \sinh 2\pi b s \sinh\frac{2\pi s}{b},
\eeq
which coincides with the Virasoro modular S-matrix $S_0{}^s=\rho(s)$, and $N$ is a length independent normalization. After performing the integral, the partition function can be put in the more familiar form $Z(\ell) \sim \frac{1}{\ell} \mu^{\frac{1}{2b^2}}K_{1/b^2}(\kappa \ell)$. This quantity was previously obtained by \cite{Fateev:2000ik} (and from the dual matrix integral by \cite{Moore:1991ir}). We present a more systematic derivation which we found to be more useful in order to generalize this to correlation functions.  \\
Following \cite{Saad:2019lba} we interpret $\mu_B(s)$ as the energy of the boundary theory dual to the bulk gravity, $\rho(s)$ as a density of states, and $\ell$ as an inverse temperature.

\paragraph{Bulk one-point function:} We compute the fixed length partition function with a bulk insertion $\mathcal{T}_{\alpha_M}$, and $P$ is the Liouville momentum associated to $\alpha_M$. This can be depicted as
\begin{equation}  
\left\langle \mathcal{T}_{\alpha_M}\right\rangle_\ell =~ \begin{tikzpicture}[baseline={([yshift=-.5ex]current bounding box.center)}, scale=0.7]
\draw[fill=blue!40!white,opacity=0.7] (0,0) ellipse (1.5 and 1.5);
\draw[fill] (0,0) circle (0.06);
\node at (-1.8,0.2) {\small $\ell$};
\node at (0.5,0) {\small $\mathcal{T}$};
\end{tikzpicture}
\end{equation}
Repeating the previous procedure we obtain 
\begin{equation}
\label{eq:b1}
\left\langle \mathcal{T}_{\alpha_M}\right\rangle_\ell = \frac{2}{b} \int_{0}^{\infty} ds\hspace{0.1cm} e^{-\ell \mu_B(s)} \cos 4 \pi P s.
\end{equation}
The integrand coincides with the Virasoro modular S-matrix $S_P{}^s=\cos 4 \pi P s$. We interpret the bulk operator as creating a defect (for $P$ imaginary) or a hole (for $P$ real) on the physical space. This interpretation is consistent with classical solutions of the Liouville equation, and also becomes clear in the JT gravity limit \cite{Mertens:2019tcm}.

\paragraph{Boundary two-point function:} The two point function between boundary operators, labeled by $\beta_M$, inserted between segments of fixed physical length is defined from the following diagram 
\begin{equation}  
\hspace{-0.2cm}\mathcal{A}_{\beta_M}(\ell_1,\ell_2) =~ \begin{tikzpicture}[baseline={([yshift=-.5ex]current bounding box.center)}, scale=0.7]
\draw[fill=blue!40!white,opacity=0.7] (0,0) ellipse (1.5 and 1.5);
\draw[fill] (-1.5,0) circle (0.06);
\draw[fill] (1.5,0) circle (0.06); 
\node at (-1.9,0) {\small $\mathcal{B}$};
\node at (1.9,0) {\small $\mathcal{B}$};
\node at (0,-1.8) {\small $\ell_1$};
\node at (0,1.8) {\small $\ell_2$};
\end{tikzpicture}
\end{equation}
We obtain
\beq
\label{twoa}
\mathcal{A}_{\beta_M}(\ell_1,\ell_2)= N_{\beta_M} \int ds_1 ds_2 \rho(s_1) \rho(s_2)\hspace{0.05cm} e^{-\mu_B(s_1)\ell_1} e^{-\mu_B(s_2)\ell_2}\hspace{0.05cm}\mathcal{M}_{\beta_M}(s_1,s_2)^2,
\eeq
where $N_{\beta_M}$ is a length independent constant and we define the amplitude
\beq
\label{twoam}
\mathcal{M}_{\beta_M}(s_1,s_2) \equiv \frac{\prod_{\pm\pm}S_b\left(\beta_M \pm i s_1 \pm i s_2\right)^{1/2}}{S_b(2\beta_M)^{1/2}},
\eeq
where $S_b(x)$ is the double sine function. Its definition and properties that will be relevant in this paper can be found in Appendix B.1 of \cite{Mertens:2017mtv}. The appearance of this structure was derived somewhat cavalier in \cite{Mertens:2019tcm}, and we substantiate it here. 

Following \cite{Saad:2019lba}, the amplitude $\mathcal{M}_{\beta_M}(s_1,s_2)$ can be interpreted as a matrix element of operators in the dual boundary theory between energy eigenstates. We interpret this result as an exact expression for the gravitational dressing by Liouville gravity of boundary correlators (notice that the boundary lengths are not necessarily large and therefore this corresponds to gravity in a finite spacetime region). 

Another motivation for studying these correlators is the resemblance with exact results in double-scaled SYK derived in \cite{Berkooz:2018jqr,*Berkooz:2018qkz,*Berkooz:2020xne}, which we hope to come back to in future work.

\paragraph{Boundary three-point function:} The fixed length boundary three-point function is defined as 
\begin{equation}  
\hspace{-0.2cm}\mathcal{A}_{123}(\ell_1,\ell_2,\ell_3) =~ \begin{tikzpicture}[baseline={([yshift=-.5ex]current bounding box.center)}, scale=0.7]
\draw[fill=blue!40!white,opacity=0.7] (0,0) ellipse (1.5 and 1.5);
\draw[fill] (0,1.5) circle (0.06);
\draw[fill] (-1.3,-0.75) circle (0.06); 
\draw[fill] (1.3,-0.75) circle (0.06); 
\node at (-1.7,-0.8) {\small $\mathcal{B}_1$};
\node at (1.75,-0.8) {\small $\mathcal{B}_3$};
\node at (0,-1.85) {\small $\ell_1$};
\node at (-1.6,1) {\small $\ell_2$};
\node at (1.6,1) {\small $\ell_3$};
\node at (0,1.85) {\small $\mathcal{B}_2$};
\end{tikzpicture}
\vspace{-0.4cm}
\end{equation}
and we get
\bea
\label{threea}
\mathcal{A}_{123}(\ell_1,\ell_2,\ell_3) &=& N_{\beta_1\beta_2\beta_3} \int ds_1 ds_2 ds_3 \rho(s_1) \rho(s_2)\rho(s_3) e^{- \mu_B(s_1)\ell_1}e^{- \mu_B(s_2)\ell_2}e^{- \mu_B(s_3)\ell_3} \nonumber\\
&&\times \mathcal{M}_{\beta_{M2}}(s_2,s_3)\mathcal{M}_{\beta_{M1}}(s_1,s_2)\mathcal{M}_{\beta_{M3}}(s_1,s_3)  \sj{\beta_{M1}}{\beta_{M2}}{\beta_{M3}}{s_3}{s_1}{s_2},
\eea
where $N_{\beta_1\beta_2\beta_3}$ is a length independent constant. The quantity appearing in the second line is the quantum deformed $6j$ symbols computed by Teschner and Vartanov \cite{Teschner:2012em, *Vartanov:2013ima} (this quantity is proportional to a Virasoro fusion kernel). This expression gives the universal Liouville gravitational dressing of boundary three-point functions. 

\paragraph{Bulk-boundary correlator:}
The fixed length bulk-boundary two-point function is defined by 
\begin{equation}  
\hspace{0cm}\mathcal{A}_{\alpha_M, \beta_M}(\ell) =\hspace{0.5cm} \begin{tikzpicture}[baseline={([yshift=-.5ex]current bounding box.center)}, scale=0.7]
\draw[fill=blue!40!white,opacity=0.7] (0,0) ellipse (1.5 and 1.5);
\draw[fill] (1.5,0) circle (0.06); 
\node at (0,-1.985) {};
\node at (0,1.8) {\small $\ell$};
\draw[fill] (0,0) circle (0.06);
\node at (-0.5,0) {\small $\mathcal{T}$};
\node at (1.9,0) {\small $\mathcal{B}$};
\end{tikzpicture}
\vspace{-0.4cm}
\end{equation}
where $\alpha_M$ (with momentum $P$) and $\beta_M$ label the bulk and boundary insertions. We obtain
\begin{align}
\label{twoabb}
\mathcal{A}_{\alpha_M, \beta_M}(\ell) = N_{\beta_M,P}\int_{0}^{+\infty} ds_1 ds_2 \rho(s_1) \rho(s_2) e^{-\mu_B(s_1) \ell} \, \frac{S_P{}^{s_2}}{S_0{}^{s_2}} \, \mathcal{M}_{\beta_M/2}(s_1,s_2)^2,
\end{align}
in terms of the Virasoro modular S-matrices defined above. 
\\~\\
We will also define the JT classical limits of these equations, where we will reproduce known expressions found in \cite{Mertens:2019tcm,Mertens:2017mtv,Iliesiu:2019xuh}. 
\\~\\
If we take the specific case of the minimal string (where the matter sector is a minimal model), we have the power of the matrix model at our disposal to aid our investigation. In particular, the set of minimal string boundary primaries correspond to setting $\beta_M = -bj$, for $j\in \mathbb{N}/2$. The two-point amplitude \eqref{twoam} becomes degenerate (due to a singularity in the denominator) and using the matrix description we will derive the answer:
\begin{equation}
\mathcal{M}_{\beta_M}(s_1,s_2)^2 = (2j)! \sum_{n=-j}^{j}\frac{\delta(s_1-s_2-in b)}{\prod_{\stackrel{m=-j}{m\neq n}}^{j} (\cosh 2\pi b (s+i nb) - \cosh 2\pi b (s+imb))}.
\end{equation}
These delta-functions have to be interpreted as causing a contour shift within the double integral \eqref{twoa}. One can also take the degenerate limit directly in \eqref{twoam} using quantum group methods, and we will find agreement. 
Taking the JT classical limit for these correlators, we find the degenerate Schwarzian bilocal correlators, for which the first case $j=1/2$ was studied in appendix D of \cite{Mertens:2019tcm}, and the generic case is studied in \cite{Mertens:2020pfe}.

\begin{figure}[h]
\centering
\includegraphics[width=0.3\textwidth]{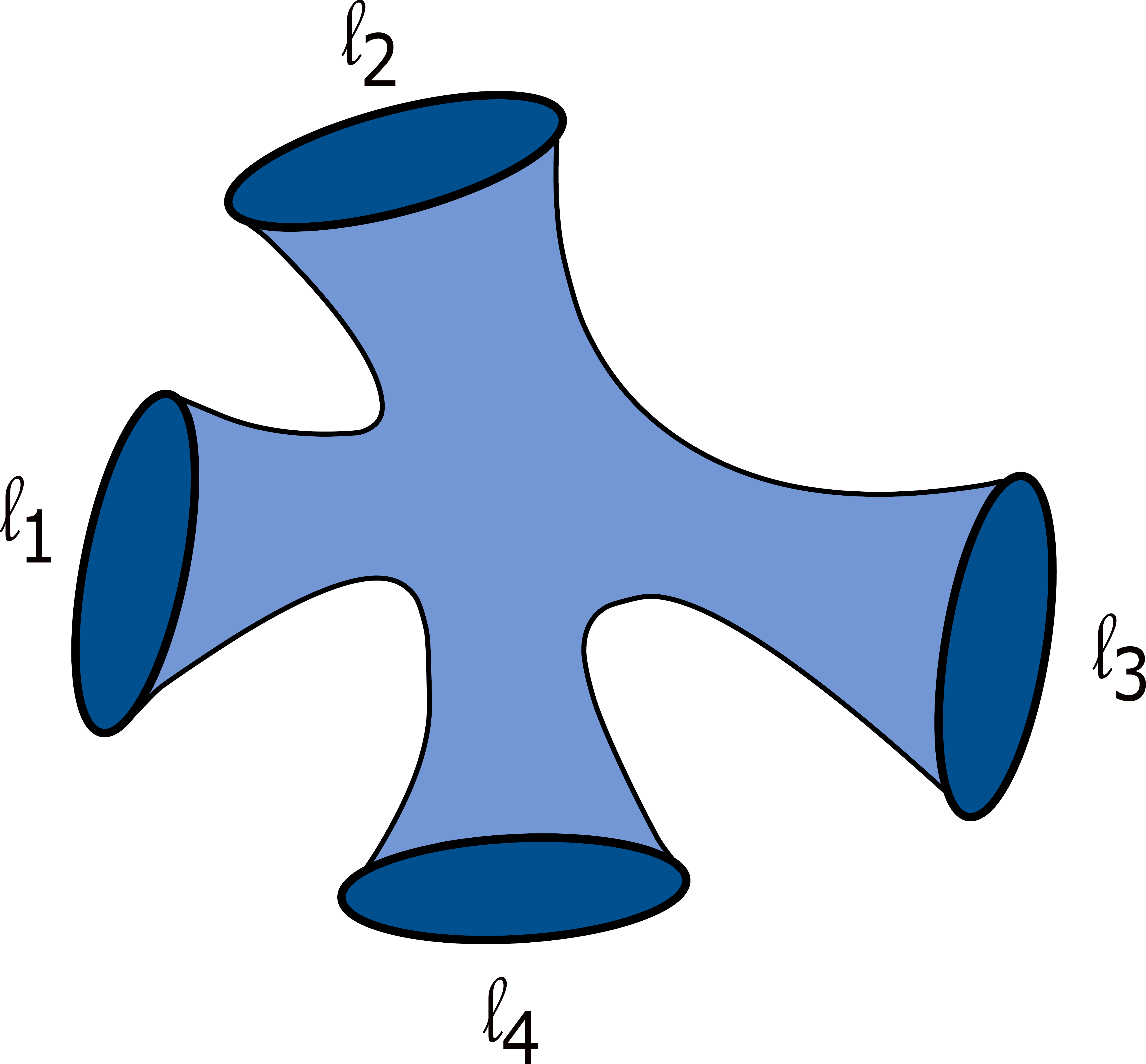}
\caption{Genus zero $n$-boundary loop amplitude (here $n=4$).}
\label{multiboundaryQ}
\end{figure}

Next to these amplitudes, we also analyze multi-boundary amplitudes for the minimal string. A four-boundary example is drawn in Figure \ref{multiboundaryQ}.
For $n$ circular boundaries, we find the genus $g$ amplitude is of the form:
 \bea\label{eq:nloopcorr}
 \Big\lb \prod_{i=1}^{n} Z(\ell_i) \Big\rb_{g, \, {\rm conn.}}  \sim  \prod_{i=1}^{n}\int_0^\infty \lambda_i d\lambda_i  \tanh \pi \lambda_i \,V_{g,n}(\bm{\lambda}) \,  \left\langle \mathcal{T}_{\alpha_{Mi}}\right\rangle_{\ell_i},
 \eea
where $\left\langle \mathcal{T}_{\alpha_{Mi}}\right\rangle$ is the bulk one-point function \eqref{eq:b1} with $P_i = b\lambda_i /2$ (which we interpret as a Liouville gravity trumpet partition function), the quantity $V_{g,n}(\bm{\lambda})$ is a symmetric polynomial of order $n+3g-3$ in the $\lambda_i^2$ and a quantum deformation of the WP volumes. The measure factor $\lambda_i d\lambda_i  \tanh \pi \lambda_i$ generalizes the classical gluing formula for Riemann surfaces $b_i db_i$, where $b_i$ is the circumference of the gluing geodesic. Indeed, for large values of $\lambda_i$ (the classical JT limit), these formulas reduce to these classical WP gluing formulas.  \\
In particular, we focus on the genus zero contributions, for which we give a general formula for the deformed volumes (and therefore by taking the appropriate limit, an explicit formula for the classical WP volumes). For higher genus, we argue they also take the form \eqref{eq:nloopcorr}. It would be interesting to develop a more geometrical interpretation of this quantum deformation of the WP volumes. Such derivation would confirm the choice of normalization of the one-point function and the integration measure in \eqref{eq:nloopcorr} \footnote{The ambiguity arises since, for example, the final answer (except for the special case of two boundaries and no handles) is unchanged under $d\mu(\lambda) \to f(\lambda) d\mu(\lambda)$ and $\lb \mathcal{T} \rb \to f(\lambda)^{-1} \lb \mathcal{T} \rb$, for an arbitrary $f(\lambda)$ that goes to one in the JT gravity limit. We argue below the choice in \eqref{eq:nloopcorr} is the most natural one.}. 
\\~\\
The organization of the paper and summary of some more results is as follows. 
In \textbf{section \ref{sec:review}} we give a quick review on the non-critical string, Liouville gravity and the minimal string. The knowledgeable reader can skip this section, although we do fix conventions and write down previous results that will be essential later on. 
In \textbf{section \ref{sec:diskZ}} we describe a systematic way to compute fixed length amplitudes and illustrate it by reproducing known formulas for the fixed length partition function. 
In \textbf{section \ref{sec:diskcorr}} we compute explicitly fixed length boundary correlation functions with and without bulk insertions. We also define and take the JT gravity limit of these observables. 
\textbf{Section \ref{s:qg}} explains the structure of these equations as coming from a constrained version of the $\mathcal{U}_q(\mathfrak{sl}(2,\mathbb{R}))$ quantum group. In particular, the vertex function is reproduced from a 3j-symbol computation with Whittaker function insertions. In \textbf{section \ref{sec:MM}} we show for the case of the minimal string how to produce the correlators directly from the matrix model. We check that the quantum group formulas from the previous section lead to the same structure. 
Finally in \textbf{section \ref{sec:othertopo}} we study other topologies. We give a streamlined derivation of the cylinder amplitude. We also review the exact result presented in \cite{Ambjorn:1990ji, Moore:1991ir} for the $n$ boundary-loop correlator at genus zero for the minimal string theory and discuss its decomposition in terms of gluing measures, bulk one-point functions and quantum deformed WP volume factors. By taking the JT gravity limit we give a very simple generating function of WP volumes for $n$ geodesic boundaries at genus zero. 
In \textbf{section \ref{sec:conclusions}} we end with a discussion and open problems for future work. In particular, we argue that the bulk gravity can be rewritten in terms of a 2d dilaton gravity model with a sinh dilaton potential. In the appendices, we include some related topics that would otherwise distract from the story. In particular, we discuss the role of poles in the complex $\mu_B$ plane as one transforms to fixed length amplitudes, we discuss degenerate bulk one-point functions, and degenerate (ZZ) branes as boundary segments. For the multi-boundary story for unoriented surfaces, we compute the crosscap spacetime contribution, which we show matches with a GOE/GSE matrix model calculation.

\section{Non-critical strings and 2d gravity} \label{sec:review}
This section contains review material on Liouville gravity and minimal string theory. We first discuss the bulk stories in \ref{s:qlg} and \ref{s:mst}, and then the boundary versions in \ref{s:bdy}.

\subsection{Quantum Liouville gravity}
\label{s:qlg}
We study two dimensional theories on Riemann surfaces $\Sigma$ with dynamical gravity, by summing over all metrics $g_{\mu\nu}(x)$ (in Euclidean signature) modulo diffeomorphisms. We also add a matter theory with fields $\chi(x)$ living on the Riemann surfaces with action $S_M[\chi;g]$. The starting point is the path integral
\beq\label{eq:deftheory}
Z=\sum_{\rm topologies} \int \frac{\mathcal{D}g \mathcal{D}\chi}{{\rm Vol}({\rm Diff})} e^{ - S_M[\chi;g] - \mu_0 \int_\Sigma d^2x \sqrt{g}},
\eeq
where $\mu_0$ is the bare cosmological constant. We will focus only on the case where the matter sector is a CFT with central charge $c_M$. We will also consider minimal models as matter CFT which might not have a path integral representation. 

Following \cite{Polyakov:1981rd, *Distler:1988jt, *David:1988hj} we can gauge fix conformal gauge $g_{\mu\nu}=e^{2 b\phi(x)} \hat{g}_{\mu\nu}(x)$ with $\phi$ a dynamical scale factor, $b$ a normalization to be fixed later, and $\hat{g}$ a fiducial metric. This has the effect of adding the usual $bc$-ghosts with central charge $c_{\rm gh} = -26$ and a Liouville mode coming in part from the conformal anomaly in the path integral measure and also from the bare cosmological constant. One ends up with an action consisting of the matter on the fixed fiducial metric $S_M[\chi; \hat{g}]$, the ghost action, and a Liouville field theory with action \cite{Polyakov:1981rd}
\beq
S_L[\phi] = \frac{1}{4\pi} \int_{\Sigma} \left[ (\hat{\nabla} \phi)^2 + Q \hat{R} \phi + 4 \pi \mu e^{2 b \phi} \right] . 
\eeq 
This can be interpreted as CFTs living on the fiducial metric. It is important the matter sector is a CFT so that no explicit interactions appear between matter and the Liouville field. The renormalized bulk cosmological constant is $\mu$ and scale invariance fixes the background charge $Q = b + b^{-1}$.  The central charge of the Liouville mode is $c_L = 1+6Q^2$. The three sectors are coupled through the conformal anomaly cancellation 
\beq
c_M + c_L + c_{\rm gh} =0.
\eeq

The results in this paper are mostly independent of the details of the matter CFT but we will refer to two cases for concreteness. We will analyze timelike Liouville CFT as matter, with action 
\beq\label{eq:timeLioaction}
S_M[\chi] = \frac{1}{4\pi} \int_{\Sigma} \left[ -(\hat{\nabla} \chi)^2 - q \hat{R} \chi + 4 \pi \mu_M e^{2 b \chi} \right].
\eeq 
For simplicity we can also set its cosmological constant term $\mu_M$ to zero, in which case the theory becomes the usual Coulomb gas. The central charge for this theory is $c_M = 1- 6 q^2$. The matter and Liouville background charges are related from the anomaly cancellation 
\beq
c_M + c_L =26,~~~\Rightarrow~~~q = 1/b - b, 
\eeq
which for $\mu_M\neq 0$ is consistent with the choice of the exponential interaction in \eqref{eq:timeLioaction}. This theory is equivalent to a Liouville CFT with $\tilde{b} = i b$, $\tilde{Q} = i q$ and $\tilde{\mu}=\mu_M$. The case with non-vanishing matter cosmological constant was analyzed in detail in \cite{Zamolodchikov:2005fy}. In the next section we will also consider the case of a $(p,q)$ minimal model. 

Now we will go through the construction of physical operators in these theories. First, generic bulk operators of the Liouville CFT and matter CFT, seen as two independent field theories, can be written as 
\bea\label{bulklioupar1}
{\rm Liouville:}&&~~~\hspace{0.1cm}~V_\alpha = \exp{(2 \alpha \phi)}~~~~~~~~\hspace{0.1cm}\Delta_\alpha = \alpha(Q-\alpha),\\
\label{bulklioupar2} \hspace{-0.3cm}{\rm Matter:}&&~~\mathcal{O}_{\alpha_M} = \exp{(2 \alpha_M \chi)}~~~~\Delta_{\alpha_M} = \alpha_M(q+\alpha_M),
\eea
where we also wrote their scaling dimension under worldsheet conformal transformations. When we consider other matter CFT we will still label their operators by the parameter $\alpha_M$. It is customary to also introduce the Liouville momentum and energy $\alpha= Q/2 + i P$ and $ \alpha_M= -q/2 + i E$. These can be interpreted as target space energy and momentum $(E,P)$ in a Minkowski 2D target space $(X^0,X^1)=(\chi, \phi)$ with a linear dilaton background. 

If gravity was not dynamical, the only operators of the theory would be the matter $\mathcal{O}_{\alpha_M}$. When gravity is turned on diffeomorphism invariant observables are made out of physical operators that are marginal. The gravitational dressing necessary for this is achieved by combining matter and Liouville operators into the bulk vertex operator 
\beq\label{tachyondef}
\mathcal{T}_{\alpha_M} \sim \int_{\Sigma}  \hspace{0.1cm}\mathcal{O}_{\alpha_M}(x) \hspace{0.01cm}V_\alpha(x),
\eeq
with a normalization that will be fixed later. After gauge fixing, we can replace the integral by a local insertion with the ghosts $\mathcal{T}_{\alpha_M} \sim c \bar{c} \hspace{0.1cm}\mathcal{O}_{\alpha_M} \hspace{0.01cm}V_\alpha$. In the context of non-critical string theory, these insertions create bulk tachyons which will be labeled by its matter content. The parameter $\alpha$ controlling the gravitational dressing is fixed through the relation \cite{Knizhnik:1988ak}
\beq
\Delta_{\alpha_M} + \Delta_\alpha = 1 ,~~~\Rightarrow~~~\alpha_+=b-\alpha_M,~~\alpha_-=\frac{1}{b}+\alpha_M.
\eeq
For fixed $\mathcal{O}_{\alpha_M}$ these two choices are related through $\alpha_+ = Q-\alpha_-$, which up to reflection coefficients creates the same operator. For a given $\Delta_{\alpha_M}$ there are also two possible choices of $\alpha_M$ (related by $\alpha_M\to -q -\alpha_M$) giving four choices of pairs $(\alpha_M,\alpha)$ all related through Liouville reflection relations. In terms of momenta the dressing condition can be nicely summarized as $P^2 =E^2$ which is the on-shell condition of a massless field moving in the target space with 2-momentum $(E,P)$. Up to this identification between $\alpha_M$ and $\alpha$, when computing correlators of $\mathcal{T}_{\alpha_M}$ the answer factorizes into a matter, Liouville and ghost contributions, before the integration over the moduli. 

A simple operator that we will use later is the area operator which can be defined as $\hat{A} = \int_\Sigma V_b$.  This can also be written after gauge fixing in the form of a tachyon vertex operator as above, which corresponds to picking the identity in the matter sector $\mathcal{T}_{\rm id} \sim c\bar{c}\hspace{0.1cm} V_b$. This operator measures the total area of the surface in terms of the physical metric.

Before we moving on, we will enumerate some special set of operators in both the matter and Liouville sectors that will be useful to distinguish later on:
\paragraph{Degenerate Liouville operators:} These operators, labeled by two positive integers $m\geq1$ and $n\geq1$, are defined through the parameter 
\beq\label{eq:liouvdeg}
\alpha_{(m,n)}= - \frac{(n-1)b}{2} - \frac{(m-1)b^{-1}}{2},~~~{\rm and}~~\alpha_{(m,n)} \to Q-\alpha_{(m,n)}.
\eeq

\paragraph{Degenerate matter operators:} We can analogously define operators which are degenerate in the matter sector also labeled by positive integers $m\geq1$ and $n\geq1$ 
\beq\label{eq:mattdeg}
\alpha_{M(m,n)} =- \frac{(n-1) b}{2} + \frac{(m-1)b^{-1}}{2},~~~{\rm and}~~\alpha_{M(m,n)} \to - q-\alpha_{M(m,n)}.
\eeq

Its important to notice that these operators never appear together in a tachyon vertex operator. We can easily see from the expressions above that if the matter content corresponds to a degenerate operator, then the Liouville dressing will be generic. One the other hand, if the Liouville dressing is degenerate, the matter operator will be generic instead. We can easily see this in the semiclassical (also related to JT gravity) limit:

\paragraph{Semiclassical limit:} Following \cite{Saad:2019lba} we will be interested in the limit $b\to0$ for which $c_M \to -\infty$ and $c_L \to \infty$. In this limit we will parametrize light matter operators as $\alpha_M = b h$, where $h$ is a continuous parameter which is fixed in the $b\to0$ limit. They are dressed by Liouville operators with $\alpha = b(1-h)$. In this limit, $h$ corresponds to the dimension of the matter operator $\Delta_{\alpha_M} \to h$, while the Liouville field has $\Delta_\alpha \to 1-h$. Degenerate matter operators have $h_{Mn} =\frac{1-n}{2}=0,-\frac{1}{2},-1,-\frac{3}{2},\ldots$, while Liouville degenerate operators have $h_{Ln} = \frac{1+n}{2}=1,\frac{3}{2}, 2, \ldots$. These carry a single index since the other set from \eqref{eq:liouvdeg} or \eqref{eq:mattdeg} become infinitely heavy.

\subsection{Minimal string theory} 
\label{s:mst}
In this section we review the definition of the minimal string theory. This corresponds to the same theory of 2D gravity as before, but the matter CFT now consists on the $M_{p,p'}$ minimal model, for any $p'>p\geq 2$ coprime. The Liouville-like parametrization of the physical quantities that characterize this theory will be very useful later. For example, the central charge can still be written as $c_M = 1 - 6q^2$, where $q=1/b-b$ and $b=\sqrt{p/p'}$, which also matches the parameter $b$ of the gravitational Liouville mode, canceling the conformal anomaly. 

The matter CFT for the $(p,p')$ minimal string has a discrete and finite set of operators $\mathcal{O}_{n,m}$. These can still be parametrized through the Liouville-like parameter $\alpha_M$. The spectrum of the minimal model consists of the matter degenerate states with label $\alpha_{M(n,m)}$ and dimension $\Delta_{n,m}$ given by
\beq
\mathcal{O}_{n,m}:~~~\alpha_{M(n,m)} =- \frac{(n-1)b}{2} + \frac{(m-1)b^{-1}}{2},~~~\Delta_{n,m} = \frac{(nb^{-1} -m b)^2-(b^{-1}-b)^2}{4}.
\eeq 
where $n=1,\ldots, p'-1$ and $m=1,\ldots p-1$. Due to the reflection property the operators $\mathcal{O}_{n,m} \equiv \mathcal{O}_{p'-n,p-m}$ are identified this gives a total of $(p'-1)(p-1)/2$ operators. For some purposes, it is useful to define a fundamental domain $(n,m)\in E_{p'p}$ defined by $1\leq n \leq p'-1$ and $1\leq m \leq p-1$ and $p' m < p n$. We can construct physical string theory vertex operators $\mathcal{T}_{n,m}$ for each primary $\mathcal{O}_{n,m}$ by adding the gravitational dressing and integrating over the worldsheet as in equation \eqref{tachyondef}. 

Since we will need them later, we will quote results for the torus characters for these degenerate representations 
\beq\label{degcharacters}
\chi_{n,m}(q) = \frac{1}{\eta(q)} \sum_{k\in\mathbb{Z}} (q^{a_{n,m}(k)}-q^{a_{n,-m}(k)}),~~~~a_{n,m}(k) =\frac{( 2 p' p k + p n - p' m)^2}{4 p'p},
\eeq
where $q=e^{2\pi i \tau}$ and $\tau$ is the torus moduli. We will also need the modular S-matrix describing their transformation under $\tau \to - 1/\tau$, which is given by 
\beq
S_{n,m}^{n',m'} = 2 \sqrt{\frac{2}{p'p}}(-1)^{1+mn'+n m'} \sin\Big( \pi \frac{p}{p'} n' n \Big)\sin\Big( \pi \frac{p'}{p}m'm \Big).
\eeq
More results regarding these representations such as their fusion coefficients $\mathcal{N}_{n_1,m_1;n_2,m_2}^{n_3,m_3}$ can be found in \cite{DiFrancesco:1997nk}.

We will be mostly interested in the $(2,2\mathfrak{m}-1)$ minimal string which is known to be dual to a single-matrix model \cite{Moore:1991ir}.  This theory has $\mathfrak{m}-1$ bulk tachyons labeled by a single integer  
\beq
\mathcal{T}_{n} \equiv \mathcal{T}_{n,1}\sim \int_{\Sigma} \hspace{0.1cm} \mathcal{O}_{n,1} \hspace{0.1cm} e^{2(b-\alpha_{M(n,1)})\phi},
\eeq
where $n=1,\ldots, \mathfrak{m}-1$. The matter sector for these operators has parameter $\alpha_{M(n,1)} = \frac{1-n}{2}b$ and its Liouville dressing insertion has $\alpha_{n,1} = (1+n)b/2$. We have chosen these parameters in order to have a smooth semiclassical limit. 

We will be interested in the $\mathfrak{m}\to \infty$ limit of the $(2,2\mathfrak{m}-1)$ minimal string, which is equivalent to JT gravity \cite{Saad:2019lba}. This limit, since $b=\sqrt{2/(2\mathfrak{m}-1)}$, corresponds to $c_M \to -\infty$ and $c_L \to \infty$. We will focus on `light' operators $\mathcal{T}_n$ with fixed $n$ in the $k\to\infty$ limit. These are the semiclassical operators defined in the previous section with parameter $h=n/2$.  Another interesting limit is given by heavy operators with $n/\mathfrak{m}$ fixed in the large $\mathfrak{m}$ limit.

\subsection{2D gravity on the disk}
\label{s:bdy}
We will be mostly interested in observables on the disk. We quickly review here results for Liouville theory with boundaries, focusing mostly on the gravitational part. The simplest boundary condition for the Liouville mode corresponds to the FZZT brane \cite{Fateev:2000ik}. This is labeled by a single parameter $\mu_B$ referred to as the boundary cosmological constant. A path integral representation is given by the Liouville Lagrangian plus the following boundary term 
\beq
S_L^{\rm bdy} [\phi] = \frac{1}{2\pi} \oint_{\partial \Sigma} \left[ Q \hat{K} \phi + 2\pi \mu_B e^{b \phi} \right].
\eeq
It is convenient to parametrize the boundary cosmological constant in terms of the FZZT parameter $s$ as 
\beq
\mu_B = \kappa \cosh 2\pi b s,~~~~\kappa\equiv\frac{\sqrt{\mu}}{\sqrt{\sin \pi b^2}}.
\eeq
It will also be useful to keep the parameter $\kappa=\mu_B(s=0)$, with an implicit dependence on the bulk cosmological constant $\mu$ and $b$. In the case of timelike Liouville matter we can introduce analogous branes labeled by another continuous parameter we will call $\tilde{s}$. 

This boundary condition can be understood from the point of view of the boundary conformal bootstrap \cite{Fateev:2000ik}. Each boundary condition is related to a Liouville primary field with parameter $\alpha =\frac{Q}{2} + i s(\mu_B)$, analogously to the rational case \cite{Cardy:1989ir}. A different set of boundary conditions is given by the ZZ brane, which are labeled by degenerate representations \cite{Zamolodchikov:2001ah}. The FZZT boundary conditions can be represented through Cardy boundary states \cite{Cardy:1989ir}
\bea
|{\rm FZZT}(s) \rb &=& \int_0^{\infty} dP \hspace{0.1cm}\Psi_s(P)   |P\rb\hspace{-0.1cm}\rb,\\
\Psi_s(P) &=&(\pi \mu \gamma(b^2))^{-iP/b} \frac{\Gamma(1+2i Pb)\Gamma(1+2iP/b)}{2^{1/4}(-2 i \pi P)} \cos 4 \pi s P
\eea
where $|P\rb\hspace{-0.1cm}\rb$ denotes the Ishibashi state \cite{Ishibashi:1988kg} corresponding to the primary $P$ and the wavefunction $\Psi_s(P)$ was found in \cite{Fateev:2000ik}.

A similar set of branes can be defined for the matter sector when written as a time-like Liouville theory. In the case of the minimal string we can also write boundary conditions in terms of boundary states. Their form for the minimal model sector is
\beq
|n,m \rb = \sum_{n',m'}  \frac{S_{n,m}^{n',m'}}{(S_{1,1}^{n',m'})^{1/2}}   |n',m' \rb\hspace{-0.1cm}\rb,
\eeq
written in terms of the modular S-matrix. They are also labeled by primary operators \cite{Cardy:1989ir}.

We will be interested in the case of bulk and boundary correlators on the disk (following for example \cite{Kostov:2003uh}). The Liouville parametrization of boundary changing operators is 
\bea
{\rm Liouville:}&&~~~\hspace{0.1cm}~B_\beta^{s_1s_2} = \exp{( \beta \phi)}~~~~~~~~\hspace{0.1cm}\Delta_\beta = \beta(Q-\beta),\\
\label{bdylioupar}\hspace{-0.3cm}{\rm Matter:}&&~~~~~\Phi_{\beta_M}^{\tilde{s}_1\tilde{s}_2} = \exp{(\beta_M \chi)}~~~~~\hspace{-0.06cm}\Delta_{\beta_M} = \beta_M(q+\beta_M),
\eea
where we indicated explicitly the boundary conditions $s_i$/$\tilde{s}_i$ between which these operators interpolate. With this normalization, degenerate operators for both theories can be written in terms of the same expression as bulk operators so $\beta_{(n,m)}$ and $\beta_{M(n,m)}$ are equivalent to \eqref{eq:liouvdeg} and \eqref{eq:mattdeg}. Since it will be important later, we quote here the parameter for matter degenerates 
\beq
\beta_{{\rm M}(m,n)} =- \frac{(n-1) b}{2} + \frac{(m-1)b^{-1}}{2},
\eeq
with $(n,m)$ a pair of positive integers. Similar operators can be written for the minimal string $\Phi_{(n,m)}^{n_1,m_1;n_2,m_2}$ which now generate a finite discrete set of dimension $\Delta_{(n,m)}$ interpolating between $(n_1,m_1)$ and $(n_2,m_2)$ branes.  

We construct physical open tachyon vertex operators by gravitational dressing
\beq
\mathcal{B}_{\beta_M} \sim \oint_{\partial\Sigma} \hspace{0.1cm}\Phi_{\beta_M} \hspace{0.1cm}B_\beta,
\eeq
where from now on we omit the boundary conditions labels on each side of the insertion. After gauge fixing this is $\mathcal{B}_{\beta_M} \sim c \hspace{0.1cm}\Phi_{\beta_M}\hspace{0.1cm}B_\beta$. The relation between $\beta_M$ and the dressing parameter $\beta$ is the same as for the bulk operators, and we will pick $\beta_M=b-\beta$. Physical correlators factorize into the ghost, matter and Liouville contribution up to a possible integral over moduli. For the minimal string we have a discrete set $\mathcal{B}_{n,m}$ and for the $(2,2\mathfrak{m}-1)$ case we have $\mathcal{B}_{n}\equiv \mathcal{B}_{n,1}$. 

A special operator that we will make use of analogous to the area operator in the bulk is $\mathcal{B}_{\rm id} \sim c B_b^{s_1,s_2} = c e^{b\phi}$, which we will refer to as the boundary \emph{marking operator}. It is the gravitationally dressed version of the matter identity operator $\mathbf{1}_M$. Before gauge fixing, this operator can also be written as $\hat{\ell} = \oint B_b$ which measures the physical length of the boundary. 

Finally, we will need the boundary correlators of Liouville CFT for an FZZT boundary \cite{Fateev:2000ik, Ponsot:2001ng}. This is simplified if we choose the fiducial metric space to be the upper half plane $(z,\bar{z})$ with ${\rm Im}(z)\geq0$ and a boundary labeled by $z=\bar{z} = x$. The bulk one point function is 
\beq
\lb V_\alpha(z,\bar{z}) \rb_s = \frac{U_s(\alpha)}{|z-\bar{z}|^{2\Delta_\alpha}}, 
\eeq
with 
\beq
\label{Lonep}
U_s(\alpha) = \frac{2}{b}(\pi \mu \gamma(b^2))^{(Q-2\alpha)/2b} \Gamma(2b \alpha - b^2) \Gamma\Big(\frac{2\alpha}{b}-\frac{1}{b^2}-1\Big)\cosh 2\pi (2\alpha-Q)s,
\eeq
The boundary two point function is
\beq
\lb B_{\beta_1}^{s_1s_2}(x)B_{\beta_2}^{s_2s_1}(0)\rb = \frac{\delta(\beta_2 + \beta_1-Q)+ d(\beta|s_1,s_2)\delta(\beta_2-\beta_1)}{|x|^{2\Delta_{\beta_1}}}.
\eeq
where we define the quantity\footnote{There is an implicit product over all four sign combinations of the $S_b$ in this and in subsequent similar equations.}
\beq \label{liouvillebdy2pt}
d(\beta|s_1,s_2) = (\pi \mu \gamma(b^2)b^{2-2b^2})^{\frac{Q-2\beta}{2b}} \frac{\Gamma_b(2\beta-Q)\Gamma_b^{-1}(Q-2\beta)}{S_b(\beta \pm i s_1 \pm i s_2)}.
\eeq
The bulk-boundary two point function is of the form 
\beq
\lb V_\alpha(z,\bar{z})B_\beta^{ss}(x)\rb_s = \frac{R_s(\alpha,\beta)}{|z-\bar{z}|^{2\Delta_\alpha-\Delta_\beta}|z-x|^{2\Delta_\beta}}
\eeq
with 
\bea\label{bbdy}
R_s(\alpha,\beta)&=&2\pi (\pi \mu \gamma(b^2)b^{2-2b^2})^{\frac{Q-2\alpha-\beta}{2b}} \frac{\Gamma_b^3(Q-\beta)}{\Gamma_b(Q)\Gamma_b(Q-2\beta)\Gamma_b(\beta)} \frac{\Gamma_b(2\alpha-\beta)\Gamma_b(2Q-2\alpha-\beta)}{\Gamma_b(2\alpha) \Gamma_b(Q-2\alpha)}\nonumber\\
&&\times \int_{\mathbb{R}} dt \hspace{0.1cm} e^{4\pi i t s} \frac{S_b(\frac{1}{2}(2\alpha+\beta-Q)+i t)S_b(\frac{1}{2}(2\alpha+\beta-Q)-it)}{S_b(\frac{1}{2}(2\alpha-\beta+Q)+it) S_b(\frac{1}{2}(2\alpha-\beta+Q)-it)}
\eea 

We will look at the boundary two-point function with $\beta_1=\beta_2$. A naive application of the formula given above would predict a divergent factor of $\delta(\beta_2-\beta_1)\to\delta(0)$. This zero-mode divergence is canceled when one divides by the full group of diffeomorphisms (an analogous thing was observed recently in \cite{Erbin:2019uiz} for the case of the bosonic critical string). The correct answer is given by 
\beq
\lb  \mathcal{B}_{\beta_M} \mathcal{B}_{\beta_M} \rb = 2(Q-2\beta) d(\beta| s_1,s_2) \times ({\rm matter}),
\eeq
as explained for example in \cite{Aharony:2003vk, *Alexandrov:2005gf}. This result can be obtained by taking a derivative of the two point function with respect to the cosmological constant, producing a three point function with all symmetries fixed, which can then be integrated obtaining the relation above.
The on-shell condition relating $\beta$ with $\beta_M$ produces a cancellation of the worldsheet coordinate dependence $x$, after including the ghost two-point function. The last factor in the equation above comes from the matter normalization.

\section{Disk partition function}\label{sec:diskZ}
In this section we will analyze the disk partition function for the minimal string and Liouville gravity for fixed length boundary conditions. 

\subsection{Fixed length boundary conditions} \label{s:bosdisk}

We will start by defining the fixed length boundary condition in the disk. We will mostly focus on the Liouville sector and therefore the answer will be valid for both the time-like Liouville string and the minimal string. 

The starting point is the disk with FZZT brane boundary conditions, specified by the boundary cosmological constant $\mu_B$. It will be useful to distinguish two different notions of partition function of the disk. The first is the unmarked partition function $Z(\mu_B)^{\U}$. We will refer to the second type as the mark partition function $Z(\mu_B)^{\M}$ defined by 
\beq
Z(\mu_B)^{\M} \equiv \partial_{\mu_B}Z(\mu_B)^{\U}= \left\langle c \hspace{0.1cm}e^{b\phi}\right\rangle_{\mu_B}.
\eeq
This is equivalent to the partition function on a marked disk, where a boundary base point has been chosen, and we do not consider translations of the boundary coordinate as a gauge symmetry \cite{Moore:1991ir}. We will refer to insertions of $e^{b\phi}$ as marking operators. This corresponds to inserting a factor of $\ell$ in the length basis. The fixed length partition function is then defined by the inverse Laplace transform 
\beq\label{eq:deffixlength}
Z(\ell) \equiv -i \int_{-i\infty}^{i \infty} d\mu_B e^{\mu_B \ell} Z(\mu_B)^{\M}.
\eeq
This is explained, for example, by Kostov in \cite{Kostov:2002uq}. One can check from the path integral definition of Liouville theory that this integral when combined with the boundary term produces a fixed length delta function, justifying this formula. 

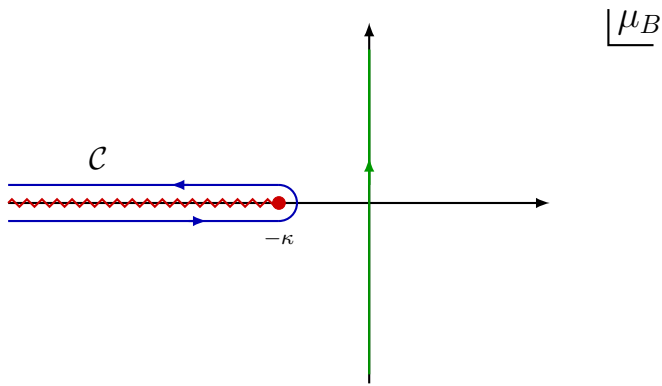
\begin{figure}[t!]
\centering
\begin{tikzpicture}[scale=1.2]
\draw[->,thick, -latex] (0,-2) -- (0,2); 
\draw[->,thick, -latex] (-4,0) -- (2,0); 
\draw[red!80!black, thick,decoration = {zigzag,segment length = 2mm, amplitude = 0.5mm},decorate] (-4,0) -- (-1,0);
\filldraw[red!80!black] (-1,0) circle (0.07);
\draw[blue!70!black,thick] (-4,0.2) -- (-2,0.2);  
\draw[blue!70!black,thick,latex-] (-2.2,0.2) -- (-1,0.2);  
\draw[blue!70!black,thick] (-1,0.2) arc (90:-90:0.2);
\draw[blue!70!black,thick] (-1,-0.2) -- (-2,-0.2);  
\draw[blue!70!black,thick, latex-] (-1.8,-0.2) -- (-4,-0.2);
\draw[green!60!black,thick, -latex] (0,-1.9) -- (0,0.5);
\draw[green!60!black,thick] (0,0.2) -- (0,1.7);
\draw[green!60!black,thick] (0,0.2) -- (0,1.7);
\node at (-3,0.5) {$\mathcal{C}$};
\node at (-1,-0.4) {\scriptsize $-\kappa$};
\node at (3,2) {\large $\mu_B$};
\draw[thick] (2.65,1.75+0.4) -- (2.65,1.75) -- (2.65+0.5,1.75);
\end{tikzpicture}
\caption{Contour deformation from the original one (in green) to a deformed one that wraps the negative real axis (blue line). The segment $(-\kappa,0)$ has no branch cut and the contour can be further deformed to the semi-infinite interval $(-\infty,-\kappa)$.}
\label{contourDeformfirst}
\end{figure}

The first step is then to compute the FZZT partition function $Z(\mu_B)^{\U}$. Following the calculation of Seiberg and Shih done in \cite{Seiberg:2003nm}, its useful to differentiate with respect to the bulk cosmological constant in order to fix all the symmetries in the problem
\bea\label{eq:unmarkedZ}
\partial_\mu Z(\mu_B)^{\U} &=& \lb c \bar{c}\hspace{0.1cm} e^{2b\phi} \rb_{\mu_B}  \\
&=& \frac{2}{b} (\pi \mu \gamma(b^2))^{\frac{1}{2b^2}-\frac{1}{2}} \Gamma(b^2) \Gamma(1-b^{-2}) \cosh 2 \Big( b-\frac{1}{b} \Big) \pi s 
\eea
where in the second line we pick a normalization such that the result is precisely the bulk cosmological constant one-point function derived in \cite{Fateev:2000ik} (Seiberg and Shih make a different normalization choice). Integrating this with respect to the cosmological constant $\mu$ we obtain the unmarked disk partition function
\beq
\label{ses}
Z(\mu_B)^{\U} = (\pi \mu \gamma(b^2))^{\frac{1-b^2}{2b^2}}  \frac{4 \Gamma(b^2) \Gamma(1-b^{-2}) \mu b^2}{b(1+b^2)} \left(b^2 \cosh{2\pi b s} \cosh \frac{2\pi s}{b} - \sinh{2\pi b s} \sinh \frac{2\pi s}{b}\right),
\eeq
where the FZZT parameter should be understood as implicitly depending on $\mu_B$ and $\mu$. We compute now the marked partition function differentiating with respect to $\mu_B$ which simplifies the $\mu_B$ dependence considerably
\beq
\label{Zmarked}
Z(\mu_B)^{\M} \sim \mu^{\frac{1}{2b^2}} \cosh \frac{2 \pi s}{b},
\eeq
where we omit the $s$ independent prefactor that we will put back later. The next step is to compute the integral defined in \eqref{eq:deffixlength}. This can be done by deforming the contour around the negative real axis, as shown in figure \ref{contourDeformfirst}.
This allows us to write the integral as
\begin{equation}
\label{toplug}
Z(\ell) =-i\int_{-\kappa}^{-\infty}d\mu_B e^{\mu_B \ell}~ \text{Disc}\left[Z(\mu_B)^{\M}\right]
\end{equation}
in terms of the discontinuity $\text{Disc}\left[Z(\mu_B)^{\M}\right]$ along the negative real axis of the marked partition function.

A first observation, as shown in figure \ref{contourDeformfirst}, is that the branch cut along the negative real axis starts at $\mu_B =-\kappa$, where $\kappa \equiv \sqrt{\mu/\sin \pi b^2} = \mu_B(s=0)$. The value of $ s \sim \cosh^{-1} (\mu_B/\kappa)$ on the negative real axis for $\mu_B \in \left(-\kappa,\kappa\right)$ is purely imaginary and conjugate above and below the real axis. Since any \emph{even} function of $s$ has no discontinuity, $\text{Disc}\left[Z(|\mu_B|<\kappa)^{\M}\right]=0$.

In what follows we will be mostly interested in the $\ell$ dependence of the final answer. On the interval $(-\infty,-\kappa)$, we can use the fact that $\text{arccosh}(\frac{\mu_B}{\kappa} + i\varepsilon) = \text{arccosh}\frac{\left|\mu_B\right|}{\kappa} \pm i\pi$. Then the discontinuity of an arbitrary function $F(s)$ on this interval is given by ${\rm Disc}[F(s)]=F(s+i/2b) -F(s-i/2b)$. Using this fact we can compute explicitly the discontinuity as
\begin{equation}
{\rm Disc} \left[\cosh \Big( \frac{1}{b^2} \text{arccosh} \frac{\mu_B}{\kappa} \Big) \right]= 2i \sin \frac{\pi}{b^2} \sinh \Big( \frac{1}{b^2}\text{arccosh}\frac{ \left|\mu_B\right|}{\kappa} \Big).
\end{equation}
We can use this to compute $\text{Disc}\left[Z(\mu_B)^{\M} \right]$ and inserting the answer into \eqref{toplug} we find the fixed-length marked disk amplitude 
\beq
\label{fixedldisk}
Z(\ell) = N \mu^{\frac{1}{2b^2}} \int_{\kappa}^{\infty} d\mu_B e^{-\ell \mu_B} \sinh \Big( \frac{1}{b^2} \text{arccosh} \frac{\mu_B}{\kappa} \Big).
\eeq
This answer is consistent with the result of \cite{Fateev:2000ik}. Keeping track of the prefactor appearing in \eqref{eq:unmarkedZ}, the normalization is given by $N = (\pi \gamma(b^2))^{\frac{1}{2b^2}} \frac{8 \pi (1-b^2)}{b \Gamma(b^{-2})}$. Written in terms of the FZZT $s$ variable the partition function is 
\bea\label{partfuncs}
Z(\ell) \sim \mu^{\frac{1}{2b^2}+\frac{1}{2}} \int_{0}^{\infty} ds ~e^{-\ell \kappa \cosh(2\pi b s)}\rho(s),~~~~\rho(s)\equiv \sinh 2\pi b s \sinh\frac{2\pi s}{b}.
\eea
In the language of \cite{Saad:2019lba} where the boundary is identified as Euclidean time of a dual theory, we see $\ell$ can be interpreted as an inverse temperature $\beta \to \ell $, while $\mu_B$ is identified with the eigenvalue of the boundary Hamiltonian $E \to \mu_B = \kappa \cosh 2\pi b s$ \footnote{Interestingly, the density of states is equal to the Plancherel measure on the principal series irreps of the quantum group $\mathcal{U}_q(\mathfrak{sl}(2,\mathbb{R}))$ \cite{Ponsot:1999uf} as a function of representations labeled by $s$. It is also equal to the vacuum modular S-matrix $S_0{}^s$. We expand on this in section \ref{s:qg}.}.  In terms of the energy $E$, we write:
\bea\label{partfuncs2}
Z(\beta) \sim \mu^{\frac{1}{2b^2}} \int_{\kappa}^{\infty} dE ~e^{-\beta E}\rho_0(E),~~~~\rho_0(E) = \sinh\Big(\frac{1}{b^2} \text{arccosh}\frac{E}{\kappa}\Big).
\eea
We will review some interesting properties of this expression in section \ref{sec:proprho}. The integral can be done explicitly using the identity 
\beq
\int_{0}^{+\infty}ds e^{-\ell \kappa \cosh 2 \pi  b s} \sinh 2 \pi b s \sinh \frac{2\pi s}{b} = \frac{1}{2\pi b^3} \frac{1}{\kappa \ell} K_{\frac{1}{b^2}}(\kappa \ell),
\eeq
where the right hand side involves a modified Bessel function of the second kind. 

More generally, if we consider the $M$-marked fixed length partition function, then we would write:
\begin{equation}
\label{genmark}
Z(\ell) \sim \frac{1}{\ell^{2-M}} K_{i\lambda}(\kappa \ell), \qquad i\lambda= 1/b^2.
\end{equation}
This formula holds since taking $\mu_B$-derivatives to bring down $\oint e^{b\phi}$ corresponds in the fixed length basis to just including factors of $\ell$. In our case we set $M=1$. The unmarked Seiberg-Shih partition function \eqref{ses}, when transformed to the fixed length basis, corresponds to setting $M=0$ in \eqref{genmark}.

\subsection{Marking operators}\label{s:bosmark}
\begin{figure}[t!]
\centering
\begin{tikzpicture}[scale=0.7]
\draw[fill=blue!50!white,opacity=0.7] (0,0) ellipse (2 and 1);
\draw[fill] (-2,0) circle (0.08); 
\node at (-2.6,0) {$e^{b\phi}$};
\node at (-1.4,-1.1) {$\mu_{B3}$};
\draw[fill] (1,0.86) circle (0.08);
\node at (1.4,1.3) {$e^{b\phi}$};
\draw[fill] (1,-0.86) circle (0.08);
\node at (1.3,-1.2) {$e^{b\phi}$};
\node at (-1,1.2) {$\mu_{B1}$};
\node at (2.6,0) {$\mu_{B2}$};
\draw[-latex] (3.5,0) -- (4.5,0);
\draw[fill=blue!50!white,opacity=0.7] (7.5,0) ellipse (2 and 1);
\node at (10.5,1) {\small $\ell_1 + \ell_2 + \ell_3$};
\end{tikzpicture}
\caption{FZZT brane segments between $n$ marking operators leads upon transforming to the fixed length basis with length $\ell \equiv \sum_j \ell_j$. In the figure we show an example with $n=3$.}
\label{markingthree}
\end{figure}
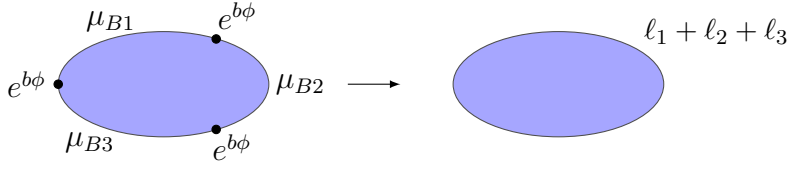

In this section, we demonstrate that inserting more marking operators $c e^{b\phi}$ between generic FZZT brane segments does not affect the final answer for the fixed length partition function. More precisely, the boundary $n$-point function of $n$ marking operators, in the fixed length basis, is precisely given by the fixed-length disk partition function itself \eqref{fixedldisk}, see figure \ref{markingthree}. Notice that this is different than marking by differentiating with respect to $\mu_B$ as in \eqref{genmark}. As explained before, these operators are physical by themselves and correspond to the dressed identity operator in the matter sector $\mathbf{1}_M$. The resulting equality we mention here is then indeed expected.

We illustrate this fact first with the simplest case of two operator insertions, after gauge fixing $\lb [c e^{b\phi}] [c e^{b\phi}] \rb$. The Liouville CFT boundary two-point function is given in \eqref{liouvillebdy2pt} specialized to $\beta=b$, and its contribution to the full 2D quantum gravity two-point function is given by $2(Q-2b)d(b|s_1,s_2)$. We can simplify this expression considerably using
\beq
\frac{1}{S_b(b \pm i s_1 \pm i s_2)} =\frac{\sinh \frac{\pi}{b} (s_1-s_2) \sinh \frac{\pi}{b} (s_1+s_2)}{\sinh \pi b (s_1-s_2) \sinh \pi b(s_1+s_2)}= \kappa \frac{\cosh \frac{2\pi}{b} s_1 - \cosh \frac{2\pi}{b} s_2}{\mu_{B}(s_1)-\mu_{B}(s_2)},
\eeq
giving
\begin{equation}\label{eq:lft2ptmbo}
d(b|s_1,s_2) = \left[(\pi \gamma(b^2))^{\frac{1}{2b^2}-\frac{1}{2}} \Gamma(b^2)\Gamma(1-b^{-2}) \frac{\sqrt{\sin(\pi b^2)}}{\pi} \right] \mu^{\frac{1}{2b^2}} \frac{\cosh \frac{2\pi}{b} s_1 - \cosh \frac{2\pi}{b} s_2}{\mu_{B1}-\mu_{B2}}.
\end{equation}

The definition of the fixed length amplitude for two marking operator insertions between two intervals of length $\ell_1$ and $\ell_2$ is given by 
\begin{equation}
\label{twolength}
\mathcal{A}_b(\ell_1,\ell_2) = i^{-2} \prod_{i=1,2}\int_{-i\infty}^{+i \infty}d\mu_{Bi} e^{\mu_{Bi} \ell_i} 2(Q-2b)d(b|s_1,s_2)
\end{equation}
Repeating the procedure outlined in the previous section and taking the double discontinuity, we find
\begin{align}
{\rm Disc}\left[\frac{\cosh \frac{2\pi}{b} s_1 - \cosh \frac{2\pi}{b} s_2}{\mu_{B1}-\mu_{B2}} \right] = - 2i \sin \frac{\pi}{b^2} \sinh \Big( \frac{1}{b^2}\text{arccosh} \frac{|\mu_B|}{\kappa} \Big)2\pi i \delta(\mu_{B1}-\mu_{B2}),
\end{align}
which is non vanishing only for $\mu_{B1} = \mu_{B2} < - \kappa$. Plugging this into the expression \eqref{twolength} after deforming the contour and using the delta function to do one of the integrals, we get the fixed-length amplitude with two marking operator insertions:
\bea
\label{markedsame}
\mathcal{A}_b(\ell_1,\ell_2) &=& N \mu^{\frac{1}{2b^2}} \int_{\kappa}^{\infty} d\mu_B e^{-(\ell_1+\ell_2) \mu_B} \sinh \Big( \frac{1}{b^2} \text{arccosh} \frac{\mu_B}{\kappa} \Big)  \nonumber\\
&=&Z(\ell_1 + \ell_2),
\eea
where we also checked that the final $b$ dependent prefactor in the equation above, derived from \eqref{eq:lft2ptmbo}, coincides with the one in the partition function derived from \eqref{Zmarked}.

This result can be generalized to an arbitrary number of marking operators. Hosomichi wrote down a generalization to an arbitrary $n$-point correlator of such $\beta=b$ insertions \cite{Hosomichi:2008th} interpolating between FZZT boundaries of parameter $\mu_{Bi}=\mu_i$,
\begin{eqnarray}
 \left\langle {}^{\mu_1}[e^{b\phi_1}]{}^{\mu_2} \hdots {}^{\mu_n}[e^{b\phi_n}]{}^{\mu_1}\right\rangle = \frac{(-)^{\frac{n(n-1)}{2}}}{\Delta(\mu_i)}\text{det}\left(\begin{array}{ccccc}
1 & \mu_1 & \hdots & \mu_1^{n-2} & Z^{\M}(s_1) \\
\vdots & \vdots & \vdots &\vdots & \vdots \\
1 & \mu_n & \hdots & \mu_n^{n-2} & Z^{\M}(s_n)
\end{array}\right),
\end{eqnarray}
where we indicated by the indices the parameters that each operator interpolates between. The transformation to fixed length generalize immediately and yields the same outcome \eqref{markedsame}, which means that all of them are equal to the (singly-marked) partition function. The main result of this section is the check that 
\beq\label{eq:markedtrivial}
\mathcal{A}_b(\ell_1,\ldots, \ell_n) = Z(\ell_1 + \ldots + \ell_n).
\eeq
This result has a simple explanation from the perspective of the matrix integral when applied to the minimal string that we mention in section \ref{sec:MMpf}.

\subsection{Properties of the density of states}\label{sec:proprho}

In this section we will present some properties regarding the density of states. We will first work out the JT gravity limit of these expressions, as defined by Saad Shenker and Stanford \cite{Saad:2019lba}. To begin, we will rescale the energy and boundary length in the following way 
\beq\label{JTparam}
E = \kappa (1+ 2\pi^2b^4 E_{\rm JT}),~~~~\ell =\frac{\ell_{\rm JT}}{2 \pi^2 \kappa b^4}.
\eeq
In terms of these variables the partition function can be written as 
\beq
\label{defrho0}
Z(\beta) \sim e^{- \ell_{JT} E_0} \int_0^\infty dE_{\rm JT}\hspace{0.1cm} e^{-\ell_{\rm JT} E_{\rm JT}}  \sinh\Big(\frac{1}{b^2} \text{arccosh}\big(1 + 2 \pi^2 b^4 E_{\rm JT} \big)\Big), 
\eeq
where the edge of the energy spectrum normalized to be conjugate to the rescaled length $\ell_{\rm JT}$ is given by $E_0 = 1/2\pi^2 b^4$. So far this is an exact rewriting. Now we can take the JT limit defined by $b\to0$ with $\ell_{\rm JT}$ fixed, which implies the integral is dominated by $E_{\rm JT}$ fixed in the limit. The density of states is approximately 
\beq\label{jtdosap}
\rho_0 (E) \approx \sinh 2 \pi \sqrt{E_{\rm JT}}, 
\eeq
which precisely coincides with the JT gravity answer, as first pointed out in \cite{Saad:2019lba}. We will take this as a definition of JT gravity limit in the case of more general observables, where all boundary length go to infinity as $b$ goes to zero, following equation \eqref{JTparam}. 

We can easily reproduce this result from the partition function written in terms of the parameter $s$ as in equation \eqref{partfuncs}. In this case the density of states is $\rho(s) = \sinh 2 \pi b s \sinh \frac{2\pi s}{b}$ and the energy $ E(s) = \kappa \cosh(2 \pi b s)$. When we pick the boundary length such that $\ell_{\rm JT}$ is fixed, the integral is dominated by $s = b k$, where we keep $k$ fixed as $b\to0$. In this limit we get $\rho(s) \sim k \sinh(2 \pi k)$ and $\ell (E(s)-\kappa)\sim  \ell_{\rm JT} k^2$, reproducing the previous result after the $E_{\rm JT} = k^2$ identification. This representation will be more useful when applied to more general observables. 

This derivation was done for a general Liouville gravity in the small $b$ limit. When applied beyond the minimal string theory its interpretation is not clear since the theory is not dual to a single matrix integral anymore. The minimal string corresponds to $b^2=2/(2\mathfrak{m}-1)$. In this case the density of states is a polynomial in $\sqrt{E}$ of order $2\mathfrak{m}-1$, since it can be rewritten as 
\beq
\rho_\mathfrak{m}(E) = \frac{1}{\sqrt{2E}}(T_\mathfrak{m}(1+E)-T_{\mathfrak{m}-1}(1+E)), 
\eeq
where $T_p(\cos \theta) = \cos( p\theta)$ is the Chebyshev polynomial of the first kind. In the JT gravity limit $\mathfrak{m}$ is large and the series becomes approximately infinite reproducing \eqref{jtdosap}. 

\begin{figure}[h]
\centering
\includegraphics[width=0.45\textwidth]{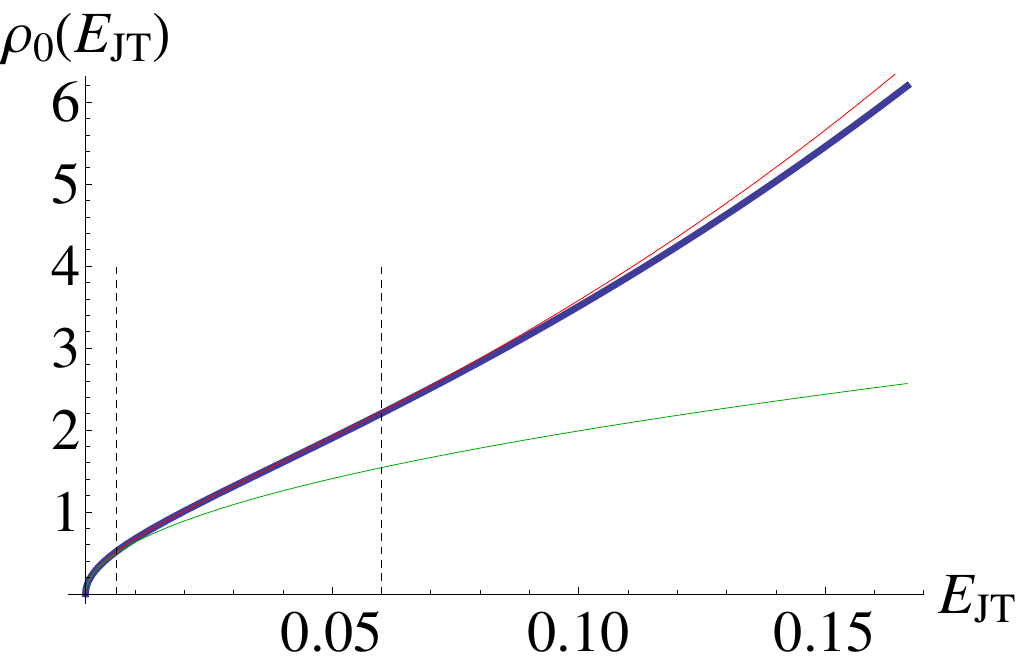}
\caption{(Blue) Energy density of states $\rho_0(E_{JT})$ defined in \eqref{defrho0} with $b=1/2$. (Red) JT limit which focusses on the middle region. (Green) spectral edge limit.}
\label{rho}
\end{figure}

Having presented the JT limit we will now give a more global picture of the density of states for general $b$. The energy density of states is sketched in Figure \ref{rho}.
This quantity has three regimes, the small $E$ regime close to the spectral edge where  $\rho_0 \sim 2\pi \sqrt{E_{JT}}$, the intermediate JT range where effectively $E_{JT} \ll 1/b^2$, and the UV regime where a different power-law behavior is present $\rho_0(E) \sim E^{1/b^2}$ (this is evident for the minimal string but still true for arbitrary $b$).

An interesting feature is that the UV rise of the spectral density in this theory is slower than that of JT gravity, which has Cardy scaling $\sim e^{2\pi \sqrt{E}}$ at high energies. Since by the UV/IR connection in holography, the high energy states probe the asymptotic region, we propose that the bulk asymptotic region becomes strongly coupled and the geometry deviates from AdS. We will discuss further how this happens in the conclusion.

The saddle of the above Laplace integral \eqref{partfuncs2} gives the energy-temperature relation:
\begin{equation}
\label{firstlaw}
\sqrt{E^2-\kappa^2} = \frac{1}{b^2 \beta},
\end{equation}
where $\beta = \ell_{\rm JT}$. As above, this law changes qualitatively from $\sqrt{E_{JT}} \sim \beta^{-1}$, the AdS$_2$ JT black hole first law, into $E_{JT} \sim \beta^{-1}$ at high energies. This suggests the possibility that the UV region close to the boundary of the space is strongly coupled, even in the JT gravity limit. It is important to explain this entire thermodynamic relation as a black hole first law of the bulk gravity system. We comment on how this works in the conclusion. 

\section{Disk correlators}\label{sec:diskcorr}
In this section, we extend the discussion to a larger class of correlators. We discuss the fixed length amplitudes of the bulk one-point function in \ref{sec:bulk1pt}, the boundary two-point function in \ref{s:bostwo}, the boundary three-point function in \ref{s:three} and the bulk-boundary two-point function \ref{s:bbtwo}. \\
Since the fixed length amplitudes are found by Fourier transforming the FZZT branes, one can also wonder whether the degenerate ZZ-branes have any relation to the fixed length branes directly. This question is only tangentially related to our main story, and we defer some of the details to appendix \ref{s:degbrane}.

\subsection{Bulk one-point function} \label{sec:bulk1pt}
In this section we will compute the fixed length partition function in the presence of a bulk tachyon insertion $\mathcal{T}_{\alpha_M}$ with dimension $\Delta_{\alpha_M}$. In general now we will get a contribution from the matter sector given by the matter one-point function. 

First we will compute the bulk Liouville one-point function for an FZZT boundary. 
We will normalize the tachyon vertex, after gauge fixing, in the following way 
\beq
\label{defbulk}
\mathcal{T}_{\alpha_M} = N_{\alpha_M} c\bar{c}\hspace{0.1cm}\mathcal{O}_{\alpha_M=-\frac{q}{2}+iP}\hspace{0.1cm}V_{\alpha=\frac{Q}{2}+iP},
\eeq 
where
\beq\label{onepointins}
N_{\alpha_M} = \frac{(\pi \mu \gamma(b^2))^{\frac{i P}{b}}}{4\pi^2 b}\frac{\Gamma(-2iP/b)}{\Gamma(1+2iP b)} \frac{1}{\text{(matter)}}.
\eeq
We divided out by the factor from the matter one-point function. In the case of the minimal string calculation of $\left\langle \mathcal{T}_{(n,m)}\right\rangle_\ell$ the matter contribution is given by the Cardy wavefunction $S_{\text{\tiny$(n',\hspace{-0.04cm}m')$}}{}^{\text{\tiny$(n,\hspace{-0.04cm}m)$}}/(S_{\text{\tiny$(1,\hspace{-0.04cm}1)$}}{}^{\text{\tiny$(n,\hspace{-0.04cm}m)$}})^{1/2}$ where the matter boundary state is a Cardy state associated to the primary $(n',m')$.
The fixed length amplitude with the bulk insertion is given by the same inverse Laplace transform as the partition function with respect to the boundary cosmological constant
\bea
\left\langle \mathcal{T}_{\alpha_M}\right\rangle_\ell &=&-i \int_{-i\infty}^{+i \infty}d\mu_B e^{\mu_B \ell} \partial_{\mu_B}\left[\left\langle \mathcal{T}_{\alpha_M}\right\rangle_{\mu_B}\right].
\eea
Inserting the Liouville contribution \eqref{Lonep}, the marked partition function with the bulk insertion is proportional to
\begin{equation}
\label{fion}
\partial_{\mu_B} \left[\cos 4\pi P s\right] = - \frac{2P}{b\kappa} \frac{\sin 4\pi P s}{ \sinh 2\pi b s}.
\end{equation}
Notice that this amplitude is actually marked twice now; we will explicitly see it in the final formula below.
We can again deform the contour as we did for the partition function. The integrand is meromorphic (and actually analytic) in the complex $\mu_B$ plane except for a branch cut at negative values. The discontinuity is given by 
\begin{equation}\label{eq:disccos}
\text{Disc }\partial_{\mu_B} \left[\cos 4\pi P s\right]= \frac{2P}{b\kappa} 2i \sinh \frac{2\pi P}{b} \frac{\cos 4\pi P s }{ \sinh 2\pi b s},
\end{equation}
valid for $\mu_B<-\kappa$. For $\mu_B \in (-\kappa,0)$, the function \eqref{fion} has no discontinuity as is readily checked, and seen immediately since \eqref{fion} is even in $s$. Finally the bulk one-point function at fixed length is given by 
\begin{equation}
\label{bulkone}
\left\langle \mathcal{T}_{\alpha_M}\right\rangle_\ell = \frac{2}{b} \int_{0}^{\infty} ds\hspace{0.1cm} e^{-\ell \kappa \cosh(2\pi b s)} \cos 4 \pi P s.
\end{equation}
 This integral can be done explicitly:\footnote{Using the identity
\beq
\label{idthree}
\int_{0}^{+\infty}ds e^{-\ell \kappa\cosh 2 \pi  b s} \cos 2\pi b \lambda s = \frac{1}{2\pi b} K_{i\lambda}(\kappa \ell).
\eeq
}
\beq
\left\langle \mathcal{T}_{\alpha_M}\right\rangle_\ell   = \frac{1}{\pi b^2} K_{\frac{2i P}{b}}(\kappa \ell).
\eeq
Notice that no prefactors of $1/\ell$ appear, comparing to \eqref{genmark}, making this amplitude interpretable as a twice-marked amplitude. 
Intuitively, one marking is just as the partition function, the second marking happens because of the non-trivial bulk insertion that creates a branch cut in the chiral sector of the geometry that has to intersect the boundary somewhere, marking it a second time. We develop this intuition in appendix \ref{app:mark}.

It was mentioned below equation \eqref{partfuncs} that the integrand of the disk partition function in terms of $s$ is the vacuum modular S-matrix. Here, in the presence of a bulk state of momentum $P$, we find a similar structure with the non-vacuum modular S-matrix $S_P{}^{s}$ appearing. 

One can parametrize microscopic bulk operators by setting $P = i \frac{\theta}{2b}$, in terms of a new parameter $\theta$. For the particular case of $\theta \in \mathbb{N}$, the Liouville one-point amplitude $U_s(\alpha)$ is divergent. 
We argue in Appendix \ref{app:unif} that one should not additionally mark the boundary in this case. We do this by arguing that this case is embedded in the degenerate Virasoro Liouville insertions. We complement this argument by a bulk Liouville geometry discussion. The analogous expressions are written in \eqref{degvir} and \eqref{bulkexc}.

\subsection{Boundary two-point function} 
\label{s:bostwo}
In this section we will compute the boundary two-point function between generic operators, for a fixed length boundary.  We will consider a general matter operator labeled by the parameter $\beta_M$ and include its gravitational dressing Liouville operator with parameter $\beta$
\begin{equation}
\label{tocu}
\mathcal{A}_{\beta_M} ( \ell_1, \ell_2) = \left\langle \mathcal{B}^{+}_{\beta_M} \hspace{0.1cm}\mathcal{B}^{-}_{\beta_M}\right\rangle_{\ell_1,\ell_2},
\end{equation}
where we defined the boundary tachyon operators 
\bea
\mathcal{B}^{+}_{\beta_M} &=&(\pi \mu \gamma(b^2))^{\frac{2\beta-Q}{4b}}  \frac{\Gamma(b(Q-2\beta))}{\pi} c \hspace{0.1cm}e^{\beta\phi} \hspace{0.1cm}e^{\beta_M\chi}, \\
\mathcal{B}^{-}_{\beta_M} &=&(\pi \mu \gamma(b^2))^{\frac{2\beta-Q}{4b}}  \frac{\Gamma(b^{-1}(Q-2\beta))}{\pi} c \hspace{0.1cm}e^{\beta\phi} \hspace{0.1cm}e^{(-q-\beta_M)\chi},
\eea
where we included the leg-pole factor in the definition of the insertion. Since we will eventually consider light matter operators we will pick the Liouville dressing with $\beta=b-\beta_M$. We will omit the labels $+/-$ on the operators when its clear by context.

\begin{figure}[t!]
\centering
\begin{tikzpicture}[scale=0.9]
\draw[fill=blue!50!white,opacity=0.7] (0,0) ellipse (2 and 1);
\draw[fill] (-2,0) circle (0.08); 
\node at (-2.7,0) {$e^{\beta_M \chi}$};
\draw[fill] (1,0.86) circle (0.08);
\node at (1.5,1.3) {$e^{(-q-\beta_M)\chi}$};
\node[red!70!black] at (1.3,-1.2) {$1$};
\node[red!70!black] at (-1,1.2) {$\beta_M$};
\draw[-latex] (3.5,0) -- (4.5,0);
\draw[thick, black] (6,-0.5)--(9,-0.5);
\draw[thick, black] (7.5,-0.5)--(7.5,-0.5+1.5);
\node[red!70!black] at (6.3,-0.18) {\small $1$};
\node[red!70!black] at (8.7,-0.18) {\small $\beta_M$};
\node[red!70!black] at (7.5+0.5,0.8) {\small $\beta_M$};
\node at (7.5,1.4) {$e^{\beta_M \chi}$};
\end{tikzpicture}
\caption{Matter Coulomb gas two-point function with a vacuum brane $\mathbf{1}$ injected with charge $\beta_M$ to form the state $\beta_M$-brane and then back.}
\label{mattertwo}
\end{figure}
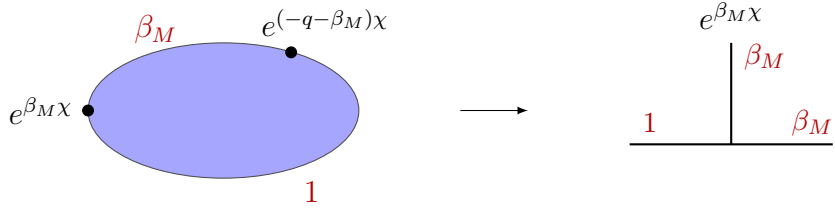

It is easy to account for the matter contribution since its independent of the boundary and bulk cosmological constant. In fact we can choose the matter operator to be normalized such that the boundary two point function has unit prefactor
\begin{equation}
\left\langle e^{\beta_M \chi} e^{ (-q - \beta_M) \chi}\right\rangle_M = \frac{1}{x^{\Delta_{\beta_M}}}.
\end{equation}
This correlator corresponds to the vacuum brane changing to the state $\beta_M$ brane and then back according to the fusion $\mathbf{1} \times \beta_M \to \beta_M$ and $\beta_M \times (-q-\beta_M) \to \mathbf{1}$ (see figure \ref{mattertwo}).
Likewise for the ghost sector. This leaves again only the Liouville sector as the source of non-trivial dependence on the boundary lengths. 

For these reasons we will focus again only on the Liouville sector. Starting with the boundary two-point function
\begin{equation}
\label{lou2}
d(\beta|s_1,s_2) = (\pi \mu \gamma(b^2)b^{2-2b^2})^{\frac{Q-2\beta}{2b}} \frac{\Gamma_b(2\beta-Q)\Gamma_b^{-1}(Q-2\beta)}{S_b(\beta \pm i s_1 \pm i s_2)},
\end{equation}
and denoting
\begin{equation}
 D_{s_1,s_2} \equiv \frac{1}{S_b(\beta \pm i s_1 \pm i s_2)} = S_b(Q - \beta \pm i s_1 \pm i s_2),
\end{equation}
we can compute the fixed length amplitude with boundary segments $\ell_1$ and $\ell_2$ by computing the Fourier transform:
\bea
\mathcal{A}_{\beta_M} ( \ell_1, \ell_2) &=&(\pi \mu \gamma(b^2))^{\frac{2\beta-Q}{2b}}2(Q-2\beta)\frac{\Gamma(b(Q-2\beta))}{\pi}\frac{\Gamma(b^{-1}(Q-2\beta))}{\pi} \nonumber\\
&&\times i^{-2}\prod_{i=1,2}\int_{-i\infty}^{+i \infty}d\mu_{Bi} e^{\mu_{Bi} \ell_i} d(\beta|s_1,s_2),
\eea
where we included all the prefactors coming from the Liouville mode. We can again deform the contour to wrap the negative real axis. The main quantity to compute (up to prefactors) is the following discontinuity of the product of double sine functions
\begin{equation}
\prod_{i=1,2}\int_{0}^{+\infty}d\mu_B e^{-\mu_{Bi} \ell_i} \text{Disc } D_{s_1,s_2}. 
\end{equation}
The discontinuity of the object $D_{s_1,s_2}$ can be found by subtracting the terms with $s_i \pm \frac{i}{2b}$,namely
\begin{equation}
\text{Disc } D_{s_1,s_2} \equiv D_{s_1  + \frac{i}{2b}, s_2 + \frac{i}{2b}} - D_{s_1  +\frac{i}{2b}, s_2 -\frac{i}{2b}} - D_{s_1  - \frac{i}{2b}, s_2 + \frac{i}{2b}} + D_{s_1  - \frac{i}{2b}, s_2 - \frac{i}{2b}}.
\end{equation}
Using the shift formulas that this double sine function satisfies
\begin{align}
\label{Sshift}
S_b(b+x) = 2\sin \pi b x S_b(x), \qquad S_b\Big(\frac{1}{b}+x\Big) = 2 \sin \frac{\pi x}{b} S_b\left(x \right),
\end{align}
the discontinuity can be tremendously simplified into\footnote{This kind of relation is actually much more general. For example replace $b\to1/b$ from equations (3.18), (3.20) and (3.24) of \cite{Kostov:2003uh}.}
\beq\label{discmain}
\text{Disc} \Big[D_{s_1,s_2}\Big] =\Big[16 \sin \frac{2 \pi \beta}{b} \sin \frac{\pi}{b}(b^{-1}-2\beta)\Big] \sinh \frac{2\pi s_1}{b} \sinh \frac{2\pi s_2}{b}~S_b\left(b -\beta\pm i s_1 \pm i s_2\right),
\eeq
where the factors in brackets depend only on $\beta$ and $b$ and the rest include all the $\mu_B$ dependent terms that will affect the length dependence of the final answer. Note the first term in the argument of the double sine functions was shifted from $Q-\beta \to b-\beta=\beta_M$ which is precisely the Liouville parameter associated to the matter operator. This will be important when taking the JT gravity limit. 

It is straightforward to check that in the range $\mu_{Bi} \in \left(-\kappa,0\right)$, one has instead a pure imaginary value of $s_i $ and its conjugate below the real axis. Since $D_{s_1,s_2}=D_{-s_1,s_2}$, $D_{s_1,s_2}=D_{s_1,-s_2}$, etc, there is again no discontinuity along this interval. Even though there are no more branch cuts, in this case there are now poles coming from the double sine functions. We can define the original $\mu_B$ contour in a way that does not pick them and the matrix model calculation of section \ref{sec:MM} supports this definition. Alternatively we will also show in Appendix \ref{app:poles} that they are negligible in the JT gravity limit. 

The final answer for the two-point amplitude is 
\beq\label{eq:2pt}
\boxed{\mathcal{A}_{\beta_M}(\ell_1,\ell_2)= N_{\beta_M}\kappa^2 \int ds_1 ds_2 \rho(s_1) \rho(s_2)\hspace{0.05cm} e^{-\mu_B(s_1)\ell_1} e^{-\mu_B(s_2)\ell_2}\hspace{0.05cm}\frac{S_b\left(\beta_M \pm i s_1 \pm i s_2\right)}{S_b(2\beta_M)}}.
\eeq
The prefactor in the right hand side can be obtained keeping track of it at each step of the calculation. Surprisingly all terms conspire to simplify drastically into the $\beta$ independent prefactor $N_{\beta_M}=16 \pi b^2$. In the case of the minimal string this factor should be multiplied by the matter contribution to the two point function. 

When viewed as a holographic theory, the result \eqref{eq:2pt} can be interpreted (read from left to right) as a sum over two intermediate channels with their respective densities of states, their propagators over lengths $\ell_i$ weighted by energies $\mu_B(s_i)$, and a matrix element squared of the matter operator between energy eigenstates given by the product of double sine functions.

Finally we can analyze the UV behavior. We will pick $\ell_1<\ell_2$ and call $\tau\equiv \ell_1$ and $\beta=\ell_1+\ell_2$. The UV behavior without gravity is given by $G_0(\tau) \sim 1/\tau^{2h}$ for very small $\tau\to 0$. This arise from a combination of the fact that even though the density of states grows exponentially $\rho(E) \sim e^{\sqrt{E}}$ the matrix elements decay too, up to a power law $\rho(E)|\lb E | \mathcal{O} | E \rb|^2 \sim  E^{2h-1} $ at high energies. The situation when quantum gravity is turned on is surprisingly not too different. Now the density of states grows as a power law at large energies $\rho(E) \sim E^{p/2}$. We can use the asymptotics of the double sine function $S_b(x) = e^{i \delta(b)} e^{\mp i x (x-Q)}$ when ${\rm Im}(x)\to \pm \infty$, where $\delta(b)$ is a phase that is independent of $x$. We find that the amplitude goes as $S_b(\ldots)\sim E^{-p/2} E^{\frac{2}{b}\beta_M-1}$. The slower growth of the density of states is exactly compensated by a slower decay of matrix elements. This gives an asymptotics that is very similar to the case without gravity $G(\tau) \sim 1/ \tau^{2 h_{\rm eff}}$, with an effective gravitational dress scaling dimension $h_{\rm eff} \equiv \beta_M/b$. This is given, as a function of the bare scaling dimension $\Delta = \beta_M(q+\beta_M)$ as 
\beq
h_{\rm eff} =\frac{\sqrt{2b^2(2\Delta-1)+b^4+1}-1+b^2}{2b^2},
\eeq
where we picked the root that has a smooth $b\to0$ limit. When gravity is weakly coupled $b\to0$ and $h_{\rm eff} (b\to0) \sim \Delta$. On the other hand when gravity is strong we get $h_{\rm eff}(b\sim 1) \sim \sqrt{\Delta}$ but the qualitative behavior in the UV is the same. In any case, including quantum gravity does not seem to smooth out the UV divergence.

\subsection{Boundary three-point function}
\label{s:three}
In this subsection we will compute the boundary three point function between three operators with matter parameters $\beta_{M1}$, $\beta_{M2}$ and $\beta_{M3}$, which we will denote as
\beq
\mathcal{A}_{123}(\ell_1,\ell_2,\ell_3) \equiv \lb \mathcal{B}_{\beta_{M1}} \mathcal{B}_{\beta_{M2}} \mathcal{B}_{\beta_{M3}}  \rb,
\eeq
 and can be obtained as an inverse Laplace transform of FZZT boundary conditions as before. 
 
 The expressions required in this calculation are very involved so we will focus only on the length dependence to simplify the presentation. The first object we need is the Liouville three-point function between operators of parameter $\beta_1$, $\beta_2$ and $\beta_3$ which should be thought of as a function of the matter parameter $\beta_i = b - \beta_{Mi}$. The Ponsot-Teschner \cite{Ponsot:2001ng} boundary three-point function is 
\beq\label{bdy3pt}
C_{\beta_3\beta_2\beta_1}^{s_3s_2s_1} = \frac{g_{Q-\beta_3}^{s_3s_1}}{g_{\beta_2}^{s_3s_2}g_{\beta_1}^{s_2s_1}} F_{s_2 \beta_3}\left[{}^{\beta_2}_{s_3} \hspace{0.1cm}{}^{\beta_1}_{s_1}\right],
\eeq
where following \cite{Ponsot:2001ng} we define 
\beq
g_{\beta}^{s_2s_1}\equiv (\pi \mu \gamma(b^2)b^{2-2b^2})^{\beta/2b} \frac{\Gamma_b(Q) \Gamma_b(Q-2\beta) \Gamma_b(Q+ 2i s_1) \Gamma_b(Q-2is_2)}{\Gamma_b(Q-\beta\pm i s_1 \pm i s_2)}.
\eeq
The fusion matrix appearing in the right hand side of \eqref{bdy3pt} was also computed by Ponsot and Teschner previously in \cite{Ponsot:1999uf}. We can rewrite this boundary three point function in the following suggestive way  
\bea
C_{\beta_3\beta_2\beta_1}^{s_3s_2s_1} &=&\frac{S_b(2\beta_1)^{\frac{1}{2}}S_b(2\beta_2)^{\frac{1}{2}}S_b(2\beta_3)^{\frac{1}{2}}}{\sqrt{2\pi}} C_{\beta_1,\beta_2,\beta_3}{}^{\frac{1}{2}}\nonumber\\
&&\times \Big[S_b(\bar{\beta}_2\pm is_2\pm is_3)S_b(\bar{\beta}_1\pm is_1\pm is_2)S_b(\bar{\beta}_3\pm is_1\pm is_3) \Big]^{\frac{1}{2}} \sj{\bar{\beta}_1}{\bar{\beta}_2}{\bar{\beta}_3}{s_3}{s_1}{s_2},\nonumber
\eea
where we defined $\bar{\beta}=Q-\beta$ and used that $\Gamma_b(Q)^2= 2 \pi/\Upsilon'(0)$. The factor appearing in the first line is the DOZZ structure constant 
\beq
C_{\beta_1,\beta_2,\beta_3} = \frac{(\pi \mu \gamma(b^2) b^{2-2b^2})^{(Q-\beta_{123})/b}\Upsilon'(0)\Upsilon(2\beta_1)\Upsilon(2\beta_2)\Upsilon(2\beta_3)}{\Upsilon(\beta_{1+2-3})\Upsilon(\beta_{3+2-1}) \Upsilon(\beta_{3+1-2})\Upsilon(\beta_{123}-Q)}.
\eeq
The final term is the b-deformed $6j$-symbol of SL$(2,\mathbb{R})$ computed by Teschner and Vartanov \cite{Teschner:2012em}.

Now we can compute the discontinuity of the boundary OPE along the negative $\mu_B$ axis. We can do this by applying three times equation (3.24) of \cite{Kostov:2003uh} and the result, up to a $s$ independent prefactor, is 
\beq
{\rm Disc}[C_{\beta_3\beta_2\beta_1}^{s_3s_2s_1}] \sim \sinh \frac{2\pi s_1}{b} \sinh \frac{2\pi s_2}{b} \sinh \frac{2\pi s_3}{b} C_{\beta_3+\frac{1}{b}\beta_2+\frac{1}{b}\beta_1+\frac{1}{b}}^{s_3s_2s_1}.
\eeq
Putting everything together and using the relation $\beta=b-\beta_M$ we can write a final answer for the boundary three point function 
\bea\label{eq:3ptbdyfinal}
\mathcal{A}_{123}(\ell_1,\ell_2,\ell_3) &=& N_{\beta_1\beta_2\beta_3} \int \prod_{i=1}^3 \Big[ds_i\rho(s_i) e^{- \mu_B(s_i)\ell_i}\Big] \nonumber\\
&&\hspace{-4cm}\times \big[S_b(\beta_{M2}\pm is_2\pm is_3)S_b(\beta_{M1}\pm is_1\pm is_2)S_b(\beta_{M3}\pm is_1\pm is_3) \big]^{\frac{1}{2}} \sj{\beta_{M1}}{\beta_{M2}}{\beta_{M3}}{s_3}{s_1}{s_2} .
\eea
where the prefactor $N_{\beta_1\beta_2\beta_3}$ includes contributions from both the Liouville and the matter sectors. Interestingly it is proportional to the square root of the DOZZ structure constant. This prefactor is important since it quantifies the bulk coupling between the three particles created by the boundary operators, but to estimate its size it's important to include properly the matter contribution, which depends on the theory. 

\subsection{Bulk boundary two-point function}
\label{s:bbtwo}
The bulk-boundary two-point function we will consider is of the form
\begin{equation}
\left\langle \mathcal{T}_{\alpha_M} \, \mathcal{B}_{\beta_M}^{+}\right\rangle_{\ell}.
\end{equation}
We will take the bulk operator with $\alpha = Q/2 + i P$, $\beta_1 = Q/2 + is$ as the FZZT boundary label and $\beta = b- \beta_M$ for the boundary operator. \\

The Liouville amplitude was listed in \eqref{bbdy}. We transform this to fixed length by evaluating the discontinuity across the branch cut on the negative real $\mu_B$-axis. To do this, the following functional discontinuity relation can be used:\footnote{The analogous relation for a shift in $b$ was written in eq (3.20) in \cite{Kostov:2003uh}, in turn extracted from the Teschner trick computation of \cite{Hosomichi:2001xc}. We corrected a typo in that equation in the middle Gamma-function in the denominator.}
\begin{align}
R_{s+\frac{i}{2b}} - R_{s-\frac{i}{2b}} &= \sinh \frac{2\pi}{b}s  \,\, R_{s}(\alpha,\beta+1/b) \\
&\times 2\pi \left( \frac{\mu}{\pi \gamma(-b^2)}\right)^{1/2} \frac{\Gamma(1-\frac{2}{b}\beta)\Gamma(1-\frac{1}{b^2} - \frac{2}{b}\beta)}{\Gamma^2(1-\frac{1}{b}\beta)\Gamma(1 - \frac{1}{b}\beta - \frac{2}{b}\alpha + \frac{1}{b}Q) \Gamma(1- \frac{1}{b}\beta  + \frac{2}{b}\alpha - \frac{1}{b}Q)}. \nonumber
\end{align} 
The resulting bulk-boundary two-point function has the following complicated form:
\begin{align}
\label{oneone}
&\int_{0}^{+\infty} ds \rho(s) e^{-\mu_B(s) \ell} \, \Gamma(b(Q-2\beta)) \frac{1}{4\pi^2 b} \frac{\Gamma(-2iP/b)}{\Gamma(1+2iPb)}\\
&\times\frac{\Gamma(1-2b^{-1}\beta)\Gamma(1-b^{-2}-2b^{-1}\beta)}{\Gamma^2(1-b^{-1}\beta)\Gamma(1-b^{-1}\beta-2b^{-1}\alpha+b^{-1}Q)\Gamma(1-b^{-1}\beta+2b^{-1}\alpha-b^{-1}Q)} \nonumber \\
&\times  \frac{\Gamma_b^3(\beta_M)}{\Gamma_b(Q) \Gamma_b(-Q+2\beta_M)\Gamma_b(Q-\beta_M)} \frac{\Gamma_b(2\alpha-Q + \beta_M)\Gamma_b(Q-2\alpha+\beta_M)}{\Gamma_b(2\alpha) \Gamma_b(Q-2\alpha)} \,\, I_{\beta_1\alpha}(\beta + 1/b). \nonumber
\end{align}
The first line contains the legpole factors of the boundary operator, and the normalization of the bulk operator \eqref{onepointins}. The second line contains the prefactors coming from deforming the contour. The final line is the Liouville bulk-boundary two-point function in terms of the modular $S$-matrix, defined by Teschner and Vartanov as \cite{Teschner:2012em}:
\begin{align}
S^{\PT}_{\beta_1 \beta_2}&(\alpha_0) \equiv  \frac{S_0{}^{\beta_2} e^{\frac{\pi i}{2} \Delta \alpha_0}}{S_b(\alpha_0)} I_{\beta_1\beta_2}(\alpha_0) \\
&= \frac{S_0{}^{\beta_2} e^{\frac{\pi i}{2} \Delta \alpha_0}}{S_b(\alpha_0)} \int_{\mathbb{R}}dt e^{2\pi t (2\beta_1-Q)}\frac{S_b(\frac{1}{2}(2\beta_2+\alpha_0-Q)+it)}{S_b(\frac{1}{2}(2\beta_2-\alpha_0+Q)+it)}\frac{S_b(\frac{1}{2}(2\beta_2+\alpha_0-Q)-it)}{S_b(\frac{1}{2}(2\beta_2-\alpha_0+Q)-it)}.
\end{align}
The integral $I_{\beta_1\alpha}(\beta + 1/b) $ can be evaluated as:
\begin{align}
I_{\beta_1\alpha}(\beta + 1/b) &= \int_{\mathbb{R}}dt e^{2\pi t (2\beta_1-Q)}\frac{S_b(\frac{1}{2}(2\alpha+(\beta+1/b)-Q)+it)}{S_b(\frac{1}{2}(2\alpha-(\beta+1/b)+Q)+it)}\frac{S_b(\frac{1}{2}(2\alpha+(\beta+1/b)-Q)-it)}{S_b(\frac{1}{2}(2\alpha-(\beta+1/b)+Q)-it)} \nonumber \\
&= \frac{1}{S_b(\beta_M)^2}\int_{\mathbb{R}}dt e^{4\pi t i P} S_b( \beta_M/2 \pm i s \pm it),
\end{align}
where we used the property $I_{\beta_1\beta_2}(\alpha_0) = S_b(\alpha_0)^2 I_{\beta_2\beta_1}(Q-\alpha_0)$ to swap the roles of $\alpha$ and $\beta_1$. This allows for a well-defined JT limit below. 
Using the shift identities and the gamma-function reflection identity, the integrand of \eqref{oneone} can be simplified into:
\begin{align}
N_{\beta_M,P} \int_{\mathbb{R}}dt\hspace{0.1cm} e^{4\pi t i P} \frac{S_b( \beta_M/2 \pm i s \pm it)}{S_b(\beta_M)},
\end{align}
which contains (in order) the prefactor $2/b$ for the bulk operator, the boundary operator, and a coupling between these in the third prefactor, given by 
\beq
\label{prebb}
N_{\beta_M, P}= \frac{2}{b} \frac{\Gamma_b(1/b +\beta_M)}{\Gamma_b(1/b + 2 \beta_M)}\frac{\Gamma_b(\frac{1}{b} + \beta_M \pm 2iP)}{\Gamma_b(\frac{1}{b} \pm 2iP)}.
\eeq

Upon using $t\to -t$ to write the $t$-integral over $\mathbb{R}^+$, we can write:
\begin{equation}
e^{4\pi t i P} \,\, \to \,\,\cos 4\pi P t = S_P{}^t = \frac{S_P{}^t}{S_0{}^t} S_0{}^t,
\end{equation}
in terms of the Virasoro modular $S$-matrix, where $S_0{}^t = \rho(t) = \sinh 2 \pi b s \sinh \frac{2\pi s}{b}$. We then obtain for the full result \eqref{oneone}:
\begin{equation}\label{eq:bulkbdyfinal}
\boxed{
\left\langle \mathcal{T}_{\alpha_M} \, \mathcal{B}_{\beta_M}^{+}\right\rangle = N_{\beta_M, P} \int_{0}^{+\infty} ds dt \rho(s) \rho(t) e^{-\mu_B(s) \ell} \, \frac{S_P{}^t}{S_0{}^t} \,   \frac{S_b(\beta_M/2 \pm i s \pm it)}{S_b(\beta_M)}}.
\end{equation}

As a check on this formula, taking $\beta_M \to 0$, we can use the identity
\begin{equation}
\lim_{\beta_M \to 0}\frac{S_b(\beta_M/2 \pm i s \pm it)}{S_b(\beta_M)} = \frac{\delta(s-t)}{S_0{}^{t}},
\end{equation}
to obtain
\begin{align}
\frac{2}{b} \int_{0}^{+\infty} ds e^{-\mu_B(s) \ell} S_P{}^t,
\end{align}
which is indeed the bulk one-point function we derived in section \ref{sec:bulk1pt}.

\subsection{JT gravity limit}
In this section we take the semiclassical limit of the formulas derived above, for which the central charge of the Liouville mode becomes large. We will see in each case a match with the analogous calculation done previously in JT gravity. 

\begin{center}
\textbf{Bulk one-point function}
\end{center}

We will begin with the bulk one-point function 
\begin{equation}
\left\langle \mathcal{T}\right\rangle_\ell = \frac{2}{b} \int_{0}^{+\infty} ds e^{-\ell \kappa \cosh(2\pi b s)} \cos 4 \pi P s, 
\end{equation}
We take the $b\to0$ limit and write it in terms of $\ell_{\rm JT}$ (see section \ref{sec:proprho} for its definition in terms of $\ell$). In order to have a non-trivial limit, we consider heavy matter operators such that the Liouville momenta scales as $P=\lambda/2b$, with finite $\lambda$. Then the one-point function becomes 
\begin{equation}
\left\langle \mathcal{T}\right\rangle_\ell = 2 \int_{0}^{+\infty} dk e^{-\ell_{\rm JT} k^2 } \cos 2 \pi \lambda k. 
\end{equation}
This expression coincides with the JT gravity partition function on a single trumpet of geodesic length $2\pi \lambda$. Therefore in this limit the bulk operator has the effect of creating a macroscopic hole of a given length. 
These single defect partition functions in JT gravity are known to be related to functional integrals within the different Virasoro coadjoint orbits \cite{Mertens:2019tcm},\footnote{See also \cite{Nayak:2019evx}.} where the choice of defect selects a particular orbit. For $\lambda \in \mathbb{R}$,  these can be identified with the hyperbolic orbits of the Virasoro group.

On the other hand, for imaginary $\lambda \equiv i\theta$ this partition function is equivalent to the JT gravity calculation with a conical defect inside the disk, with angular identification $\varphi \sim \varphi + 2 \pi \theta$. These are identified with functional integrals along the elliptic coadjoint orbits of the Virasoro group. For $\theta\in \mathbb{N}$, these become replicated geometries. Taking the JT limit of \eqref{bulkexc}, we get:
\begin{equation}
\left\langle \mathcal{T^{\U}}\right\rangle_\ell = 4 \int_{0}^{+\infty} dk e^{-\ell_{\rm JT} k^2 } k \sinh 2 \pi n k, 
\end{equation}
matching the JT exceptional elliptic defect amplitudes discussed in \cite{Mertens:2019tcm}.

Starting instead with \eqref{degvir}, and setting $n=\lambda/b^2$ with $\lambda$ a new continuous quantity, one gets the limit:
\begin{equation}
\left\langle \mathcal{T^{\text{deg}}}\right\rangle_\ell = 4 \int_{0}^{+\infty} dk e^{-\ell_{\rm JT} k^2 } \sinh 2 \pi \lambda k \sinh 2 \pi n k, 
\end{equation}
which we proposed in \cite{Mertens:2019tcm} to be related to the exceptional hyperbolic Virasoro coadjoint orbits. 

In conclusion, the insertion of a bulk operator has the effect of creating a hole (for real $P$) or a localized conical defect (for imaginary $P$). We checked this in the semiclassical JT limit but this is consistent with the classical solution of the Liouville equation, see for example the discussion in \cite{Moore:1991ir}.

\begin{center}
\textbf{Boundary two-point function}
\end{center}

Now we will take the JT gravity limit of the two point function computed in \eqref{eq:2pt}. We will take the matter operator with parameter $\beta_M = b h$ and keep $h$ fixed in the $b\to0$ limit. We will also take the boundary length to be large with $\ell_{{\rm JT}1}$ and $\ell_{{\rm JT}2}$ fixed. Then up to only $b$ dependent terms, we can write the two point function as 
\beq
\label{JTtwo}
\mathcal{A}_{\beta_M}(\ell_1,\ell_2) \sim \mu^{\frac{1}{2b^2}} \int dk_1 dk_2 \rho_{\rm JT}(k_1)\rho_{\rm JT}(k_2)e^{-k_1^2 \ell_{{\rm JT}1}}e^{-k_2^2 \ell_{{\rm JT}2}}\frac{\Gamma(h \pm i k_1 \pm i k_2)}{\Gamma(2h)},
\eeq
where $\rho_{\rm JT}(k) = k \sinh 2\pi k$ and we used the small $b$ asymptotic of the double sine function $S_b(bx) \propto \Gamma(x)$. This expression coincides with the JT gravity two point function computed in \cite{Mertens:2017mtv,Lam:2018pvp}. 

In the limit of large $\ell_{{\rm JT}1}$ and $\ell_{{\rm JT}2}$ this formula simplifies further since the Schwarzian mode becomes weakly coupled. Renaming $\tau = {\rm min} ( \ell_{{\rm JT}1}, \ell_{{\rm JT}2})$ and $\beta=\ell_{{\rm JT}1}+\ell_{{\rm JT}2}$, for large $\beta,\tau$ we get $\mathcal{A} \sim ( \sin \frac{\pi}{\beta}\tau)^{-2h}$. This is precisely the boundary correlator one would get if the gravitational mode would be turned off. In order to obtain this limit we need $b$ to be small. Therefore in general theories there is no regime where the gravitational dressing becomes weakly coupled \footnote{Similar drastic effects of gravitational dressings can happen also in higher dimensions \cite{Lewkowycz:2016ukf}.}. 

\begin{center}
\textbf{Boundary three-point function}
\end{center}
Following the previous calculation we take the limit of the three-point function \eqref{eq:3ptbdyfinal} when the three boundary length to be large with fixed $\ell_{{\rm JT}i}$ and $\beta_{{\rm M}i}= b h_i$ for $i=1,2,3$. The integrals are then dominated by $s_i = b k_i$. Ignoring length independent prefactors, using the asymptotics of the double sine functions we can get 
\bea
\lb \mathcal{B}_1 \mathcal{B}_2 \mathcal{B}_3 \rb&\sim& \int \prod_{i=1}^3 \Big[dk_i\rho_{\rm JT}(k_i) e^{- \ell_{{\rm JT}i}k_i^2}\Big] \nonumber\\
&&\hspace{-4cm}\times \big[\Gamma(h_2\pm ik_2\pm ik_3)\Gamma(h_1\pm ik_1\pm ik_2)\Gamma(h_3\pm ik_1\pm ik_3) \big]^{\frac{1}{2}} \sj{h_1}{h_2}{h_3}{k_3}{k_1}{k_2}_{\text{SL}(2,\mathbb{R})}, 
\eea
where the expression involves now the $6j$-symbol of the classical group SL$(2,\mathbb{R})$ between three principal series representations labeled by $k_i$ and three discrete representations labeled by $h_i$. This is precisely the same structure as the JT gravity three-point function computed in equation (4.35) of \cite{Iliesiu:2019xuh}. 

\begin{center}
\textbf{Bulk-boundary two-point function}
\end{center}

Finally we will take the JT limit of the bulk boundary correlator given in equation \eqref{eq:bulkbdyfinal}. We set $\beta_M = bh$, and $P = \lambda/2b$. It is instructive to work this out for $h \in \mathbb{N}$. In this particular case, the last factor of the prefactor \eqref{prebb} simplifies to:
\begin{equation}
\frac{\Gamma_b(\frac{1}{b} + \beta_M \pm i \frac{\lambda}{b})}{\Gamma_b(\frac{1}{b} \pm i \frac{\lambda}{b})} \to 2\pi b \left(\frac{\sinh \pi \lambda}{\pi \lambda} \right)^{h},
\end{equation}
for a hyperbolic (macroscopic) defect with geodesic circumference $2\pi \lambda$. 
\\~\\
For an elliptic (microscopic) insertion, we set $\lambda = i \theta$, and obtain instead $2\pi b \left(\frac{\sin \pi \theta}{\pi \theta} \right)^{h}$. Notice that this factor vanishes for $\theta \in \mathbb{N}_0$, which are precisely the values of the Virasoro exceptional elliptic coadjoint orbits.
The other prefactors scale in uninteresting ways and can be absorbed in normalization of the bulk and boundary operators separately. 

To find a finite result, we rescale $t\to bt$ and use the small $b$-asymptotics of the $S_b$-function to get:
\begin{align}
\label{11lim}
\left\langle \mathcal{T}_{\alpha_M} \, \mathcal{B}_{\beta_M}^{+} \right\rangle &= 2\pi b \left(\frac{\sinh \pi \lambda}{\pi \lambda} \right)^{h} \int_{0}^{+\infty} dk dt \rho_{\rm JT}(k) \rho_{\rm JT}(t) e^{- \ell k^2} \chi_t(\lambda)  \frac{\Gamma(h/2 \pm i k \pm it)}{\Gamma(h)},
\end{align}
in terms of the character insertion $\chi_t(\lambda)$ for $\lambda$ a hyperbolic conjugacy class element \cite{Mertens:2019tcm}:
\begin{equation}
\chi_t(\lambda) = \frac{\cos 2 \pi \lambda t}{t \sinh 2\pi t}.
\end{equation}
The $t$-momentum variable has no exponential factor, and hence no boundary segment.

\begin{figure}[h!]
\centering
\begin{tikzpicture}[scale=1]
\draw[fill=blue!40!white,opacity=0.7] (0,0) ellipse (1.5 and 1.5);
\draw[red] (0,0.5) -- (1.5,0) -- (0,-0.5);
\draw[red] (0,-0.5) arc (180+94:180-94:0.5);
\draw[fill] (1.5,0) circle (0.06); 
\node at (0,-1.985) {};
\node at (0,1.8) {\small $\ell$};
\node at (0.6,0) {\small $t$};
\node at (-0.8,0.8) {\small $k$};
\draw[fill,green!40!black] (0,0) circle (0.06);
\node[green!40!black] at (-0.23,0) {\small $\lambda$};
\node at (1.9,0) {\small $\mathcal{B}$};
\end{tikzpicture}
\caption{Schwarzian limit of the modular S-matrix, and hence the bulk-boundary propagator. The answer is given by the expectation value of a boundary-anchored bilocal line (red line) encircling the defect (green dot). This line separates two regions with energy parameters $k$ (region without defect) and $t$ (region with defect).}
\label{modularS}
\end{figure}
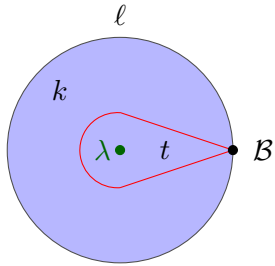

The Schwarzian diagram is sketched in Figure \ref{modularS} with a bilocal line lasso-ing around the defect.
Notice that the bilocal line has \emph{half} the value of $h$ of the boundary operator. This can be appreciated by viewing this single boundary operator as the renormalized point-split version of two boundary operators with half the value of $h$ as:
\begin{equation}
:\lim_{x_2\to x_1} e^{\frac{\beta_M}{2}\chi_1}e^{\frac{\beta_M}{2}\chi_2}:\,\, \equiv \,\, e^{\beta_M \chi}.
\end{equation}
leading indeed to the vertex functions present in \eqref{11lim}. We also remark that this renormalization removes the coincident UV divergence of the two constituent boundary operators which would correspond in the JT limit to a contractible bilocal line (i.e. \emph{not} encircling the defect).
 
\section{A quantum group perspective}
\label{s:qg}
We have seen that the propagation factors in the amplitudes $e^{-\mu_B(s) \ell}$ (as in e.g. \eqref{twoa}) contain in the exponent the factor $\cosh 2 \pi b s$, and the measure is $\rho(s) = \sinh 2\pi b s \sinh \frac{2 \pi s}{b}$. In this section we highlight the quantum group structure that underlies these expressions. \\
The quantity $C_s \equiv \cosh 2 \pi b s$ is identified with the Casimir $C_s$ of the (continuous) self-dual irreps $\mathcal{P}_s$ labeled by $s$ of $\mathcal{U}_q(\mathfrak{sl}(2,\mathbb{R}))$ with $q=e^{\pi i b^2}$. The associated Plancherel measure on this set of representations is
\begin{equation}
d\mu(s) = ds \sinh 2\pi b s \sinh \frac{2 \pi s}{b}.
\end{equation}
This class of representations is characterized by the following \cite{Ponsot:1999uf,Ponsot:2000mt,*Bytsko:2002br,*Bytsko:2006ut,*Ip}:
\begin{itemize}
\item
It is a \emph{positive} representation, in the sense that all generators are represented by positive self-adjoint operators.
\item 
They are closed under tensor product in the sense:
\begin{equation}
\mathcal{P}_{s_1} \otimes \mathcal{P}_{s_2} \simeq \int^{\oplus} d\mu(s) \mathcal{P}_s.
\end{equation}
\item 
They are simultaneously representations of the dual quantum group $\mathcal{U}_{\tilde{q}}(\mathfrak{sl}(2,\mathbb{R}))$ where $\tilde{q} = e^{\pi i b^{-2}}$. Hence they can be viewed naturally as representations of the modular double $\mathcal{U}_{q}(\mathfrak{sl}(2,\mathbb{R})) \otimes \mathcal{U}_{\tilde{q}}(\mathfrak{sl}(2,\mathbb{R}))$.
\end{itemize}
This means the expressions \eqref{eq:b1}, \eqref{twoa}, \eqref{threea} and \eqref{twoabb} have the same group theoretic structure as those of 2d Yang-Mills or 2d BF theory, but based on the modular double of $\mathcal{U}_{q}(\mathfrak{sl}(2,\mathbb{R}))$ as underlying quantum group structure. Notice that the restriction to only these self-dual representations is a strong constraint on the group-theoretic structure. But it is one that is necessary to make contact with geometric notions, as can be seen through the link with Teichm\"uller theory \cite{Nidaiev:2013bda}. Roughly speaking, the positivity constraint ensures one only has eigenstates of positive geodesic distance.
\\~\\
JT gravity can be realized in a similar group theoretical language, based on the subsemigroup SL${}^+(2,\mathbb{R})$ structure \cite{Blommaert:2018iqz}, where the defining representation of the subsemigroup consists of all positive $2\times 2$ matrices. This positivity is directly related to having only hyperbolic monodromies and hence only smooth (i.e. not punctured) geometries.
Additionally, one has to impose gravitational boundary conditions at all holographic boundaries. These boundary conditions enforce a coset structure of the underlying group and reduce the complete set of intermediate states from the full space of irrep matrix elements $R_{ab}(g)$ (by the Peter-Weyl theorem), to the double coset matrix elements $R_{00}(x)$ where both indices are fixed by the gravitational constraints. \\
From a SL${}^+(2,\mathbb{R})$ perspective, the generators $J^+$ and $J^-$ are constrained as $J^+=1$, $J^-=1$ for resp. the ket and the bra of the matrix element. This corresponds to imposing constraints on the parabolic generators, and we call the resulting matrix element a mixed parabolic matrix element. In the mathematics literature, such matrix elements are called Whittaker functions.
\\~\\
The vertex function in JT gravity $\frac{\Gamma(h \pm i k_1 \pm i k_2)^{1/2}}{\Gamma(2h)^{1/2}}$ is known to correspond to the integral definition of a 3j-symbol. For a compact group, one writes the expression as:
\begin{equation}
\int d g R_{1,m_1n_1}(g) R_{2,m_2n_2}(g) R_{3,m_3n_3}(g) = \tj{R_1}{R_2}{R_3}{m_1}{m_2}{m_3}\tj{R_1}{R_2}{R_3}{n_1}{n_2}{n_3}.\label{3R}
\end{equation}
In the JT gravity case, we have insertions of two principal series representation mixed parabolic matrix elements, and one insertion of a discrete representation (corresponding to the operator insertion):
\begin{equation}
\label{JT3j}
\int d x R_{k_1,00}(x) R_{h,00}(x) R_{k_2,00}(x) = \int_{-\infty}^{+\infty}dx\, K_{2ik_1}(e^{x}) e^{2 h x} K_{2ik_2}(e^{x}) = 
2^{2h-3}\frac{\Gamma(h \pm i k_1 \pm i k_2)}{\Gamma(2h)} .
\end{equation}
\\~\\
\noindent We here illustrate that this structure persists in the $q$-deformed case and in particular to the vertex functions we wrote down \eqref{twoam} in this work. The Whittaker function of the principal series representation of $\mathcal{U}_q(\mathfrak{sl}(2,\mathbb{R}))$ was derived in \cite{Kharchev:2001rs}:
\begin{equation}
\label{qmelb}
\psi^{\epsilon}_s(x) = e^{\pi i 2 s x} \int_{-\infty}^{+\infty} \frac{d\zeta}{(2\pi b)^{-2i\zeta/b-2is/b}} S_b(-i\zeta) S_b(-i 2 s -i \zeta ) e^{-\pi i \epsilon (\zeta^2 + 2s \zeta)} e^{2\pi i \zeta x} ,
\end{equation}
where $\epsilon = \pm 1$. In the notation of \cite{Kharchev:2001rs}, this corresponds to choosing $g = (2\pi b)^{1/b}$. It satisfies the following finite difference equation:
\begin{align}
\label{wdw}
\left(1+(2\pi b)^2 e^{2\pi b x - i \pi b^2} \right) \psi_s^{-}(x-ib) + \psi_s^{-}(x+ib) &= 2\cosh 2\pi b s \psi^{-}_s(x), \\
\psi_s^{+}(x-ib) + \left(1+(2\pi b)^2 e^{2\pi b x + i \pi b^2} \right) \psi_s^{+}(x+ib) &= 2\cosh 2\pi b s \psi^{+}_s(x),
\end{align}
which boils down from the Casimir equation on $\mathcal{U}_q(\mathfrak{sl}(2,\mathbb{R}))$ by constraining a parabolic generator in both the left- and right-regular representation. The rhs contains the Casimir eigenvalue in the irrep $s$. In the classical $b\to 0$ limit, this structure is precisely the same as how one constrains the $\mathfrak{sl}(2,\mathbb{R})$ Casimir equation to produce the 1d Liouville equation. Indeed, the classical $b\to 0$ limit transforms the finite difference equations both into the 1d Liouville differential equation. The options $\epsilon =\pm 1$ can be viewed as different discretizations (quantum versions) of the same classical problem. At the level of the eigenfunctions, one has the limiting behavior:
\begin{equation}
\lim_{b\to 0}\psi^{\epsilon}_s \left(\frac{x}{\pi b}\right) = \frac{1}{\pi b} K_{2 i s/b}\left(\frac{2}{b}e^{x}\right).
\end{equation}
Setting $s=bk$ and shifting $x$, the function $K_{2 i k}\left(e^{x}\right)$ is known as the Whittaker function of SL${}^+(2,\mathbb{R})$ and was inserted in \eqref{JT3j}. It is equally the 1d Liouville Schr\"odinger eigenfunction.\footnote{Crucially, in the same notation, the Whittaker function of SL$(2,\mathbb{R})$ is $\cosh \pi k \, K_{2i k}(e^{x})$ and this difference in prefactor in the end produces the SL$(2,\mathbb{R})$ Plancherel measure $d\mu(k) = dk \frac{k \sinh 2\pi k}{\cosh^2 \pi k} = 2 dk k \tanh \pi k$, in stark contrast to the SL${}^+(2,\mathbb{R})$ Plancherel measure $d\mu(k) = dk k \sinh 2\pi k$, relevant for gravity. One may encounter this Whittaker function with an additional factor of $e^x$ present: this compensates for the Haar measure on the group (coset) manifold, and one can choose to remove it and simultaneously take a flat measure in the $x$-integral as we have done.}
The modified Bessel function has a Mellin-Barnes integral representation as:
\begin{equation}
\label{mbk}
K_\nu(z) = \frac{1}{4\pi i }\left(\frac{z}{2}\right)^{\nu} \int_{-i\infty}^{+i\infty} dt \Gamma(t) \Gamma(t-\nu) \left(\frac{z}{2}\right)^{-2t},
\end{equation}
and the above formula \eqref{qmelb} is its $q$-deformed version. We need to scale $s \to b k$ in order to obtain a finite classical limit. 

By analogy with the lhs of \eqref{JT3j}, we hence compute the integral of two Whittaker functions, and one discrete insertion, of the type ($\beta_M = b h$):
\begin{equation}
\int_{-\infty}^{+\infty} dx\, \psi^{\epsilon}_{s_1} (x)  \psi^{\epsilon * }_{s_2} (x) e^{2 \beta_M \pi x}.
\end{equation}
Inserting the explicit expression \eqref{qmelb}, one can evaluate the $x$-integral as:
\begin{equation}
\int_{-\infty}^{+\infty} dx e^{\pi i (2s_1 - 2s_2 + 2 \zeta_1 - 2 \zeta_2) + 2 \beta_M \pi x} = \delta(\zeta_1-\zeta_2 + s_1 - s_2 - i \beta_M).
\end{equation}
We get:
\begin{align}
&\int_{-\infty}^{+\infty} dx \psi^{\epsilon}_{s_1} (x) \psi^{\epsilon * }_{s_2} (x) e^{2 \beta_M \pi x} = e^{-\pi i \epsilon(\beta_M^2-s_1^2+s_2^2+2i s_1 \beta_M)} \\
&\times \int_{-\infty}^{+\infty} \frac{d\zeta_1}{(2\pi b)^{2\beta_{M}/b}} e^{\pi 2 \epsilon \beta_M \zeta_1} S_b(-i\zeta_1) S_b(-i \zeta_1-2is_1) S_b(i\zeta_1+is_1 -is_2 + \beta_M)S_b(i\zeta_1+is_1 + is_2 + \beta_M). \nonumber
\end{align}
The $q$-deformed first Barnes lemma is:
\begin{align}
\label{qbarne}
\int d\tau e^{\pi \tau (\alpha+\beta+\gamma+\delta)}&S_b(\alpha+i\tau)S_b(\beta+i\tau)S_b(\gamma-i\tau)S_b(\delta-i\tau) \\
&= e^{\pi i (\alpha\beta-\gamma\delta)} \frac{S_b(\alpha+\gamma)S_b(\alpha+\delta)S_b(\beta+\gamma)S_b(\beta+\delta)}{S_b(\alpha+\beta+\gamma+\delta)}. \nonumber
\end{align}
Using \eqref{qbarne}, we can do the remaining integral and obtain finally:\footnote{The prefactor is immaterial and can be absorbed into the normalization of the boundary operator. Reinstating the parameter $g$ of \cite{Kharchev:2001rs}, the prefactor would be $\frac{1}{g^{2\beta_M}}$ instead.}
\begin{equation}\label{idwhitsb}
\boxed{
\int_{-\infty}^{+\infty} dx \hspace{0.1cm} \psi^{\epsilon}_{s_1} (x) \psi^{\epsilon * }_{s_2} (x) e^{2 \beta_M \pi x} = \frac{1}{(2\pi b)^{2\beta_{M}/b}} \frac{S_b(\beta_M \pm is_1 \pm is_2)}{S_b(2\beta_M) }}\, .
\end{equation}
Following the structure of \eqref{JT3j}, we interpret this as the square of the 3j-symbol with two mixed parabolic entries, and one discrete parabolic entry, of the quantum group $\mathcal{U}_q(\mathfrak{sl}(2,\mathbb{R}))$. As a check, taking the $b\to 0$ limit of both sides, we get the equality:
\begin{equation}
\left(\frac{b}{2}\right)^{2h}\frac{1}{(\pi b)^3} \int_{-\infty}^{+\infty} dx K_{2ik_1}(e^{x}) K_{2ik_2}(e^{x}) e^{2hx} = b^{2h} \frac{1}{(2\pi b)^3} \frac{\Gamma(h\pm i k_1 \pm ik_2)}{\Gamma(2h)},
\end{equation}
matching back onto \eqref{JT3j}
 
 \subsection{Wheeler-DeWitt wavefunction} 
\begin{figure}[h!]
\begin{center}
  \begin{tikzpicture}[scale=0.75]
\pgftext{\includegraphics[scale=0.3]{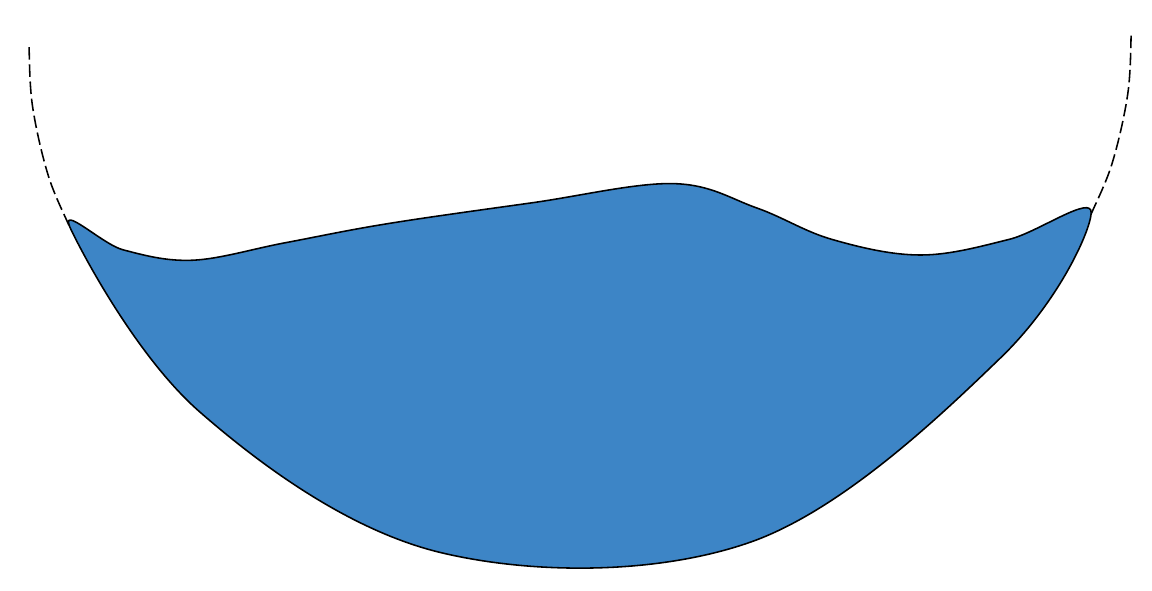}} at (0,0);
\draw (2,-1) node  {\small $\beta/2$};
\draw (0,1) node  {\small $L$};
  \end{tikzpicture}
 \caption{\label{fig:wdw} Depiction of the geometry creating the Hartle-Hawking state $\Psi_\beta(L)$. The state is labeled by a parameter $\beta$ that gives the proper length of the boundary segment preparing the state. The constant time slice is labeled by $L$, which is related to the geodesic distance along the slice.}
\end{center}
\end{figure}

 We have computed above the partition function on a hyperbolic Euclidean disk with a fixed length boundary. We can cut this disk along a bulk geodesic with length function $L$, that joins two boundary points separated by a distance $\beta/2$. This can be interpreted as a Euclidean preparation of the Wheeler-DeWitt (WdW) wavefunction $\Psi_\beta(L)$ corresponding to the two-sided black hole, see figure \ref{fig:wdw}. This wavefunction has been studied in the context of JT gravity in \cite{Harlow:2018tqv}.\footnote{This is different than the radial-quantization WdW wavefunction studied for example in \cite{Maldacena:2019cbz,Iliesiu:2020zld}.}  Based on the properties of the Whittaker function $\psi_s(x)$ above, we propose the following identification 
 \beq
 \Psi_\beta (L) = \int ds\hspace{0.1cm} e^{- \frac{1}{2}\beta \mu_B(s)} \rho(s) \psi^+_s(L), 
 \eeq
where we take $\epsilon=+1$ for concreteness. When we take the JT gravity limit, the density of states becomes the Schwarzian density of states, while the Whittaker function becomes a Bessel function derived directly from JT gravity in \cite{Harlow:2018tqv}. We have identified the group (coset) parameter $x$ of the Whittaker function with the argument of the wavefunction $L$. In the classical $b\to 0$ limit, this quantity is related to the boundary-to-boundary geodesic length $d$ as $x \to e^{d/2}$. The wavefunction can also be interpreted as the Euclidean partition function in the disk with an end-of-the-world brane. 

 To verify this identification we can rewrite the exact two point function \eqref{twoa} in the following form
 \beq
 \lb \mathcal{B}\hspace{0.05cm} \mathcal{B} \rb = \int dL  \hspace{0.1cm} e^{2 \beta_M \pi L} \hspace{0.1cm}\Psi_{\ell_1}(L)^\dagger \hspace{0.1cm} \Psi_{\ell_2}(L), 
 \eeq
 where we used the relation \eqref{idwhitsb}. This expression can be interpreted as gluing two portions of the disk along their bulk geodesic with the inclusion of the matter propagator $e^{2 \beta_M \pi L}$. This is structurally identical to the JT gravity expressions, and it would be interesting to give a more rigorous derivation from Liouville gravity.
 
 Finally, the wavefunction $\Psi_\beta(L)$ proposed here satisfies an interesting equation. We can rewrite the wavefunction for the same Hartle-Hawking state in an energy basis, which becomes the Whittaker function $\Psi_{E=\mu_B(s)}(L) = \psi^+_s(L)$ and satisfies the difference equation \eqref{wdw}. In terms of the fixed length basis this equation is 
 \beq
 \Psi_\beta(L-ib) + \big(1+(2\pi b)^2e^{2\pi b L + i \pi b^2} \big) \Psi_\beta(L+ib) = 4 \frac{\partial}{\partial \beta} \Psi_\beta(L),
 \eeq
 which can be viewed as a discretized (due to the $q$-deformation) ancestor of the Wheeler-DeWitt equation. This suggests that Liouville quantum gravity effectively discretizes the spacetime in a way we do not understand sufficiently, and this discreteness might be related to the quantum group structure present in the theory.  

\subsection{Degenerate fusion algebra}
Modified Bessel functions satisfy the following identity:
\begin{equation}
\label{besselprop}
K_{\alpha+1}(x) - K_{\alpha-1}(x) = \frac{2\alpha}{x}K_{\alpha}(x),
\end{equation}
which can be proved directly from the Mellin-Barnes representation \eqref{mbk}. This identity is important since they act as the degenerate fusion rules that directly lead to the degenerate $h\in -\mathbb{N}/2$ vertex functions for JT gravity \cite{Mertens:2020pfe}, where the vertex function in e.g. \eqref{JTtwo} is singular. Following a similar strategy with \eqref{qmelb}, one can prove the following fusion property for $\epsilon = \pm 1$:
\begin{equation}
\label{qfus}
\psi^{\epsilon}_{s+ib/2}(x) - \psi^{\epsilon}_{s-ib/2}(x) = \frac{\sinh 2 \pi b s}{ \pi i b\,  e^{\pi b x}} \psi^{\epsilon}_s(x).
\end{equation}
This relation is the basis to derive the minimal string correlators where $\beta_M \in - b\, \mathbb{N}/2$ from the continuum approach directly. The trick is to successively apply it to compute ($j \in \mathbb{N}/2$):
\begin{equation}
\int_{-\infty}^{+\infty} dx \psi^{\epsilon}_{s_1}(x) \psi^{\epsilon * }_{s_2}(x) e^{- 2 \pi b j x},
\end{equation}
until we reach
\begin{equation}
\int_{-\infty}^{+\infty} dx \psi^{\epsilon}_{s_1}(x) \psi^{\epsilon * }_{s_2}(x) = \frac{\delta(s_1-s_2)}{S_0^{s_1}}.
\end{equation}
After providing a matrix model computation of these minimal string correlators, we will come back to this approach using \eqref{qfus} and check explicitly that they match indeed.

\section{Dual matrix models}\label{sec:MM}
In this section we will give a matrix model interpretation of some of the results in the previous sections for the case of the $(2,p)$ minimal string. This case is special since the dual is a single matrix model. The discrete calculation of disk boundary correlators was proposed in \cite{Kostov:2002uq} (see also \cite{Hosomichi:2008th, Ishiki:2010wb, Bourgine:2010ja}). Besides the explicit checks, the new ingredient is to interpret the dual matrix as a boundary Hamiltonian in the sense of holography, as suggested by \cite{Saad:2019lba}. Then we will see boundary correlators of the bulk theory are equal to boundary correlators of random operators. 

\subsection{Partition function} \label{sec:MMpf}
Motivated by \cite{Saad:2019lba} we will denote the random matrix as $H$ since we will interpret it as a boundary random Hamiltonian. The matrix model dual of a marked disk partition function is 
\beq
Z(\mu_B) = \left\langle \text{Tr}\hspace{0.1cm}\frac{1}{H-\mu_B}\right\rangle.
\eeq
After inverse Laplace transforming the fixed length partition function is instead 
\beq
Z(\ell) =  \left\langle \text{Tr}\hspace{0.1cm}e^{-\ell H}\right\rangle.
\eeq
By choosing an appropriate potential for the matrix model ensemble we can make this match with the continuum answer  in the double scaling limit. 

Before moving on, we want to show that the result \eqref{eq:markedtrivial} can actually be easily deduced using the matrix model language. According to this formulation of the theory, the $n$ marking operator correlator is given by the expectation value of the following product of matrices
\begin{equation}\label{eq:markcorrmamo}
\left\langle {}^{\mu_1}e^{b\phi_1}{}^{\mu_2} \hdots {}^{\mu_n}e^{b\phi_n}{}^{\mu_1}\right\rangle = \left\langle \text{Tr}\hspace{0.1cm}\frac{1}{(H-\mu_1)\hdots (H-\mu_n)}\right\rangle.
\end{equation}
Instead of finding the expectation value first, we can inverse Laplace transform directly the matrix model observable
\begin{equation}
\left\langle \text{Tr}\hspace{0.1cm}e^{-(\ell_1+\hdots + \ell_n)H}\right\rangle,
\end{equation}
which makes manifest that depends only on the total boundary length and is consistent with \eqref{eq:markedtrivial}, since the operator $\text{Tr}\hspace{0.1cm}e^{-\ell H}$ is dual to inserting a fixed length $\ell$ boundary.

\subsection{Amplitudes}
The matrix model dual to the minimal string with boundary insertions can be written by introducing vector degrees of freedom 
\beq\label{eq:genfuncmm}
e^Z = \int DH D\bar{v}Dv \hspace{0.1cm} e^{- L {\rm Tr} V(H) - \bar{v}_a C^{ab}(H) v_b},
\eeq
where $v_a$ are $N$ dimensional vectors and $a=1,\ldots, N_f$. For example, the FZZT unmarked boundary partition function can be obtained by taking a single vector $N_f=1$ and an interaction $C(H)=\mu_B-H$. Similarly the boundary correlator of $n$ marking operators in the previous section can be obtained still by a single vector and a higher order polynomial interaction $C(H) = (\mu_{B1}-H)(\mu_{B2}-H)\ldots (\mu_{Bn}-H)$ which should be compared to \eqref{eq:markcorrmamo}. We will follow the presentation in \cite{Ishiki:2010wb}.

For the insertion of the two point function corresponding to $\mathcal{B}_{2,1}$ we need two vectors and the following interaction 
\beq\label{matrixvectorinter}
C(H) = \begin{pmatrix}
\mu_B(s_1)-H&c^{12}\\
c^{21} & F_2(H)\\
\end{pmatrix},~~~~F_2(H)=\prod_{\pm} (\mu_B(s_2 \pm i b)-H).
\eeq
For this choice \eqref{eq:genfuncmm} is a generating function of  $\mathcal{B}_{2,1}$ correlators for which $c^{12}$ and $c^{21}$ are sources and boundary conditions shift from $\mu_B(s_1) \to \mu_B(s_2)$. For the minimal string matrix model this produces the same answer as the star polymer operators in the context of the loop gas formalism \cite{Kostov:2002uq}. For example, the two point function is 
\beq
\lb\mathcal{B}_{2,1}\mathcal{B}_{2,1} \rb = \left\langle \text{Tr}\hspace{0.1cm}\frac{1}{(H-\mu_B(s_1))}\frac{1}{(H-\mu_B(s_2-ib))}\frac{1}{(H-\mu_B(s_2+ib))}\right\rangle.
\eeq
This can be compared directly in the fixed cosmological constant basis to the results from the continuum Liouville approach. Instead we will transform the observable directly into fixed length basis. For this we need to perform the inverse Laplace transform of the previous formula for the operator inside the trace
\begin{align}
\int_{-i\infty}^{+i\infty}dy \frac{e^{-y \ell_1}}{(y-H)}\int_{-i\infty}^{+i\infty}dx \frac{1}{(\cosh(2\pi b (s_2+ib/2)) - H) (\cosh( 2\pi b (s_2-ib/2) - H)} e^{-x\ell_2},
\end{align}
where for simplicity we set $\kappa=1$  and define $x=\cosh 2\pi b s_2$, $y=\cosh 2\pi b s_1$. The $y$-integral directly gives the marked length $\ell_1$ operator $e^{-\ell_1 H}$. The denominator can be written as $x^2 - 2H \cos \pi b^2 x + H^2 - \sin^2 \pi b^2$ and the integral can then be directly evaluated by residues, picking up two pole contributions, yielding
\begin{equation}
\lb\mathcal{B}_{2,1}\mathcal{B}_{2,1} \rb = \left\langle \text{Tr}\hspace{0.1cm} e^{-\ell_1 H} e^{-\ell_2 H \cos \pi b^2} \frac{\sin \left( \ell_2 \sin \pi b^2 \sqrt{H^2-1}\right)}{\sin \pi b^2 \sqrt{H^2-1}} \right\rangle.
\end{equation}
This is for the matrix $H$ underlying the minimal string matrix integral. If we now identify
\begin{equation}
H \leftrightarrow \cosh  2\pi  b s = \mu_B, \qquad \sqrt{H^2-1} \leftrightarrow \sinh 2\pi b s,
\end{equation}
we get for the full result at leading order in the genus expansion, using the leading density of states
\begin{align}
\lb\mathcal{B}_{2,1}\mathcal{B}_{2,1} \rb &=  \int_0^\infty ds \rho(s) \hspace{0.1cm} e^{-\ell_1 \cosh  2\pi  b s } e^{-\ell_2 \cosh 2 \pi  b s \cos \pi b^2} \frac{\sin \left( \ell_2 \sin \pi b^2 \sinh 2\pi b s \right)}{\sin \pi b^2 \sinh 2\pi b s} \nonumber \\
\label{tocmp}
&= \int_0^\infty ds \rho(s) \hspace{0.1cm} e^{-\ell_1 \cosh  2\pi  b s }\left[  \frac{e^{-\ell_2 \cosh  2\pi  b (s+ib/2)}}{\sin \pi b^2 \sinh 2\pi b s} - \frac{e^{-\ell_2 \cosh  2\pi  b (s-ib/2)}}{\sin \pi b^2 \sinh 2\pi b s}\right]. 
\end{align}
Following the interpretation of \cite{Saad:2019lba} of the random matrix as a random Hamiltonian we can interpret the boundary correlator as inserting an operator. Since they match for fixed FZZT boundaries this correlator matches with the fixed length two-point function when $\beta_M = - b/2 $ corresponding to $\mathcal{B}_{2,1}$. 

This correlator has a very simple JT gravity limit. Following the previous discussion, we set $s= bk$, with fixed $k$ as $b\to 0$ and define the renormalized length $\ell_{{\rm JT}i} \equiv 2\pi^2 b^4\ell_i$. This gives the simple answer 
\bea
\lb\mathcal{B}_{2,1}\mathcal{B}_{2,1} \rb_{(2,p\to\infty)} &=& \int_0^\infty kdk \hspace{0.1cm} \sinh 2 \pi k e^{-(\ell_{{\rm JT}1}+\ell_{{\rm JT}2}) k^2}\, e^{ \frac{1}{4} \ell_{{\rm JT}2}} \frac{\sin \left( \ell_{{\rm JT}2} k \right)}{k}\\
&\sim& \Big( \frac{\beta}{\pi} \sin \frac{\pi \tau}{\beta} \Big) e^{\frac{\tau(\beta-\tau)}{4\beta}},
\eea
where in the second line we defined $\tau =\ell_{{\rm JT}2}$ and $\beta= \ell_{{\rm JT}1} + \ell_{{\rm JT}2}$. This is precisely equal to the exact Schwarzian two point function for operators of dimension $\Delta=-1/2$. This is equivalent to equation (D.7) of \cite{Mertens:2019tcm}, for $C_{\rm there}=1/2$. As explained there, only operators with negative half integer dimension have such a simpler form, and these correspond to the minimal model CFT dimensions. 

This discussion can be extended to higher degenerate insertions $\mathcal{B}_{2j+1,1}$ where $\beta_M = -bj$ and $j\in \mathbb{N}/2$. This can be achieved still with two vectors interacting through the same two by two matrix in \eqref{matrixvectorinter}, but with $F_j(H) =\prod_{n=-j}^{j}(\cosh(2\pi b (s+inb)) - H)$. The two-point function of $\mathcal{B}_{2j+1,1}$ corresponds then to the matrix integral insertion
\begin{align}
\lb \mathcal{B}_{2j+1,1}\mathcal{B}_{2j+1,1}\rb = \left\lb \text{Tr}\hspace{0.1cm}\frac{1}{\cosh 2\pi b s_1-H}(2j)! \prod_{n=-j}^{j} \frac{1}{\cosh(2\pi b (s_2+inb)) - H}\right\rb .
\end{align}
 Transferring to the fixed length basis, one has to perform the integral
\begin{align}
(2j)!\int_{-i\infty}^{+i\infty}dy \frac{e^{-y \ell_1}}{(y-H)}\int_{-i\infty}^{+i\infty}dx \frac{1}{\prod_{n=-j}^{j}(\cosh(2\pi b (s+inb)) - H)} e^{-x\ell_2}.
\end{align}
Combining the factors $\pm n$ together, we can play the same game, and combine the denominators into:
\begin{align}
x^2 - 2H \cos 2\pi n b^2 x + H^2 - \sin^2 2\pi n b^2 
= (x - \cosh 2\pi b (s\pm i n b)).
\end{align}
If $2j+1$ is even, then these are all of the factors. If $2j+1$ is odd, then we have one additional factor $(x-H)$ in the denominator. What is left is just a sum of $2j+1$ residues, where the denominator is a polynomial in $H$ of order $2j$. 
The previous procedure can be done for any $j \in \mathbb{N}/2$ and we get the complicated general expression:\footnote{We have conventionally divided by the partition function $Z$ in this equation.}
\begin{align}
\label{gendeg}
&\lb \mathcal{B}_{2j+1,1}\mathcal{B}_{2j+1,1} \rb  \nonumber \\
&= \frac{1}{Z}\int_0^{+\infty} ds \rho(s) \, e^{-\ell_1 \cosh 2\pi b s}\sum_{n=-j}^{+j} \frac{(2j)!e^{-\ell_2 \cosh 2\pi b (s+i nb)}}{\prod_{\stackrel{m=-j}{m\neq n}}^{j} (\cosh 2\pi b (s+i nb) - \cosh 2\pi b (s+imb))}.
\end{align}
One can check that in the UV limit $\ell_2 \to 0$, the entire sum becomes $\ell_2^{2j} + \mathcal{O}(\ell_2^{2j+1})$, and the expression reduces to
\begin{equation}
\label{UVdeg}
\lb \mathcal{B}_{2j+1,1}\mathcal{B}_{2j+1,1} \rb \to \ell_2^{2j},
\end{equation}
matching the general analysis in section \ref{s:bostwo}. \\
In the JT limit, the pole contributions and exponentials are expanded as:
\begin{align}
\label{reslim}
\cosh 2\pi b (s+inb) - \cosh 2\pi b (s+imb)\quad &\to \quad b^4 2\pi^2 (m-n)\left(n+m -2 ik \right) + \mathcal{O}(b^6), \\
e^{- \ell \cosh 2\pi b (s+inb)} \quad &\to \quad e^{- \ell_{\rm JT} k^2} e^{\ell_{\rm JT} n^2 }e^{- 2 i \ell_{\rm JT} n k}.
\end{align}
This is precisely the structure expected for a degenerate Schwarzian insertion \cite{Mertens:2020pfe}: the denominators \eqref{reslim} produce a polynomial in $k$, while the $m-n$ factors conspire to give a binomial coefficient. In the end, we can identify this matrix insertion with the degenerate Schwarzian bilocal as:
\begin{equation}
\mathcal{B}_{2j+1,1}\mathcal{B}_{2j+1,1} \quad \to \frac{1}{b^{8j}(2\pi^2)^{2j}} \, \mathcal{I}^{j}(0)\mathcal{I}^{j}(\tau), 
\end{equation}
where the prefactor is also readily determined from \eqref{UVdeg} combined with the relation between $\ell_{\rm JT}$ and $\ell$. Here $\mathcal{I}^j$ indicates an operator in the Schwarzian theory of dimension $\Delta=-j/2$. 
\\~\\
This structure of the minimal string correlators \eqref{gendeg} matches with the continuum approach by using the fusion property \eqref{qfus}. 
As an example, for the first minimal string $j=1/2$ insertion, a single application of \eqref{qfus} leads to the identity:
\begin{align}
\int_{-\infty}^{+\infty} dx \psi^{\epsilon}_{s_1}(x) \psi^{\epsilon * }_{s_2}(x) e^{- \pi b x} = \frac{\pi b}{i S_0{}^{s_1}}\left[\frac{\delta(s_1-s_2-ib/2)}{\sinh 2 \pi  b s_2 }- \frac{\delta(s_1-s_2 + ib/2)}{\sinh 2 \pi  b s_2 }.
\right]
\end{align}
The delta-function enforces the correct dependence in the exponential factor in \eqref{tocmp}. We also see the $1/\sinh 2 \pi b s$ factor in the denominator of \eqref{tocmp} appearing. \\
It is clear that for generic $j\in \mathbb{N}/2$, we will find a similar result. As an example, in Appendix \ref{app:one} we work out the formulas for $j=1$, and check indeed that the methods match.

\section{Other topologies}\label{sec:othertopo}
In this section we will extend previous calculations to situations with more general topologies and multiple boundaries. We will focus here on the minimal string theory since it has a direct interpretation as a one-matrix integral.

\subsection{Cylinder} 
We will first study minimal string theory on a cylinder between fixed length boundaries. This was computed from a continuum approach by Martinec \cite{Martinec:2003ka} and from a discrete approach by Moore, Seiberg and Staudacher \cite{Moore:1991ir}. We will present a technically simplified derivation from the continuum limit and make a connection with JT gravity for the $(2,p)$ string with large $p$. As another example, we will apply our method to the crosscap spacetime in Appendix \ref{app:crosscap}, also reproducing the matrix model result. 

Using the boundary state formalism \cite{Ishibashi:1988kg, Cardy:1989ir} we can described a boundary labeled by an FZZT parameter $s$ and matter labels $(n,m)$ by the following combination of Ishibashi states 
\beq
|s; n,m \rb = \sum_{n',m'} \int_0^{\infty} dP \hspace{0.1cm}\Psi_s(P) \frac{S_{n,m}^{n',m'}}{(S_{1,1}^{n',m'})^{1/2}}  |P\rb\hspace{-0.1cm}\rb_L |n',m' \rb\hspace{-0.1cm}\rb_M.
\eeq
As pointed out by Seiberg and Shih this state can be simplified as a sum of matter identity branes over shifted FZZT parameters, modulo BRST exact terms that cancel when computing physical observables, see equation (3.8) in \cite{Seiberg:2003nm}. Therefore in the end of the calculation we will focus on the matter sector identity brane.
\begin{figure}[t!]
\centering
\begin{tikzpicture}[scale=0.9]
\node at (-3.5,0) {\small $s_1{}_{(n_1,m_1)}$};
\node at (3.5,0) {\small $s_2{}_{(n_2,m_2)}$};
\draw[thick] (-2,0) ellipse (0.4 and 1.5);
\draw[thick] (2,0) ellipse (0.4 and 1.5);
\draw[thick] (-1.95,1.49) to [bend right=50] (1.95,1.49);
\draw[thick] (-1.95,-1.49) to [bend left=50] (1.95,-1.49);
\node at (0,-1.985) {};
\end{tikzpicture}
\caption{The figure shows the cylinder amplitude we are computing between two FZZT boundaries with boundary cosmological constants $\mu_B(s_1)$ and $\mu_B(s_2)$ and matter boundary conditions labeled by $(n_1,m_1)$ and $(n_2,m_2)$.}
\label{fig:cylinder}
\end{figure}
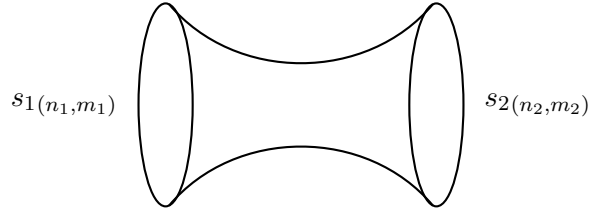

As explained in \cite{Martinec:2003ka} the annulus partition function between unmarked $s_1; n_1,m_1$ and $s_2;n_2,m_2$ branes is computed as the overlap of the boundary states, integrated over the moduli. For the annulus, there is a single real modulus $\tau$ parametrizing the length along the cylinder. Notice that this is a coordinate on the worldsheet and it is integrated over. In the end we will find dependence on \emph{physical} lengths instead as emphasized in the Introduction. 
Before integration the answer factorizes into the Liouville (L), matter (M) and ghost (G) contributions
\beq\nonumber
\lb Z(s_1;n_1,m_1)^{\U} Z(s_2;n_2,m_2)^{\U}\rb = \int d\tau Z_{L} Z_{M}  Z_{G},~~\begin{cases} Z_{L} = \int_0^\infty \frac{dP}{\pi} \frac{\cos 4 \pi s_1 P \cos 4 \pi s_2 P}{\sqrt{2}\sinh 2 \pi P b \sinh 2 \pi \frac{P}{b}}\chi_P(q), \\ 
~~\vspace{-0.5cm}\\
Z_{M} = \sum_{n,m}\mathcal{N}_{n,m}^{(n_1,m_1)(n_2,m_2)} \chi_{n,m}(q'),\\
~~\vspace{-0.5cm}\\
Z_{G} =\eta(q)^2, \end{cases}
\eeq 
where $q'=e^{-2 \pi i/\tau}$ and $\mathcal{N}_{n,m}^{(n_1,m_1)(n_2,m_2)} $ denote the fusion coefficient of the matter theory. In the matter sector, we used the Verlinde formula \cite{Verlinde:1988sn} to simplify the boundary state inner product as a sum over the dual channel characters weighted by the fusion numbers. This simplifies the calculation compared to \cite{Martinec:2003ka}. 
 We will write $\tau = i t$ where $t$ is integrated over the positive real line. Then using the modular property of the Dedekind eta function the contribution from descendants cancel up to a factor of $t^{-1/2}$ and we can write 
\bea
\lb Z(s_1;n_1,m_1)^{\U} Z(s_2;n_2,m_2)^{\U}\rb&=& \int_0^\infty \frac{dP}{\pi} \frac{\cos 4 \pi s_1 P \cos 4 \pi s_2 P}{\sqrt{2}\sinh 2 \pi P b \sinh 2 \pi \frac{P}{b}} \nonumber\\
&&\hspace{-3.5cm}\times \sum_{n,m}\mathcal{N}_{n,m}^{(n_1,m_1)(n_2,m_2)} \sum_k \int_0^\infty \frac{dt}{\sqrt{t}}   e^{-2\pi t P^2} \big(e^{-\frac{2\pi}{t}a_{n,m}(k)}-e^{-\frac{2\pi}{t} a_{n,-m}(k)}\big),
\eea
where $a_{n,m}(k)$ was defined in equation \eqref{degcharacters}. We first integrate over $t$. The answer depends on whether $k>0$, $k<0$ or $k=0$ so each case has to be considered separately. We then sum over $k$ taking this into account. The final answer is very simple
\beq\nonumber
\sum_k \int \frac{dt}{\sqrt{2t}} e^{-2\pi t P^2} (e^{-\frac{2\pi}{t}a_{n,m}(k)}-e^{-\frac{2\pi}{t} a_{n,-m}(k)}) = \begin{cases} \frac{\sinh 2 \pi b P (p'-n) \sinh 2 \pi \frac{P}{b} m }{P \sinh 2 \pi p \frac{P}{b}} ~~~{\rm if}~~np>mp'\\ 
~~\vspace{-0.2cm}\\
\frac{\sinh 2 \pi b P n \sinh 2 \pi \frac{P}{b} (p-m) }{P \sinh 2 \pi p \frac{P}{b}} ~~~{\rm if}~~np<mp'\end{cases}
\eeq
When we sum over the primaries we can use the fundamental domain $E_{p'p}$ which corresponds precisely to the first case in the result above. Then the final answer for the annulus partition function becomes 
\bea
\lb Z(s_1;n_1,m_1)^{\U} Z(s_2;n_2,m_2)^{\U}\rb &=& \sum_{(n,m)\in E_{p'p}}\mathcal{N}_{n,m}^{(n_1,m_1)(n_2,m_2)} \int_0^\infty \frac{dP}{\pi} \nonumber\\
&&\hspace{-3cm}\times  \frac{\cos(4 \pi s_1 P) \cos(4 \pi s_2 P) \sinh(2 \pi b P (p'-n)) \sinh(2 \pi \frac{P}{b} m) }{P \sinh(2 \pi p \frac{P}{b}) \sinh(2 \pi b P) \sinh(2 \pi \frac{P}{b})},
\eea
where we make explicit that this expression is valid when $(n,m)$ are in the fundamental domain $E_{p'p}$. This generalizes the formula derived by Martinec, which only includes boundary states of the form $(n,1)$ to an arbitrary boundary state and also matches in this case with the expression derived in reference \cite{Kutasov:2004fg}.  We can use the Seiberg-Shih relation between boundary states to justify focusing on the matter identity branes, and the partition function simplifies to
 \beq
\lb Z(s_1)^{\U} Z(s_2)^{\U}\rb_{(p,p')} = \int_0^\infty \frac{dP}{\pi} \frac{\cos(4 \pi s_1 P) \cos(4 \pi s_2 P)  \sinh(2 \pi \frac{P}{b} (p-1)) }{P \sinh(2 \pi p \frac{P}{b}) \sinh(2 \pi \frac{P}{b})}. 
 \eeq
 This is valid for the $(p,p')$ minimal string. Since we will be interested mostly in theories dual to single matrix models we can further take the $(2,p)$ minimal model and get 
\beq
\lb Z(s_1)^{\U} Z(s_2)^{\U}\rb_{(2,p)} = \int_0^\infty dP\frac{\cos(4 \pi s_1 P) \cos(4 \pi s_2 P)}{2\pi P \sinh 2 \pi  \frac{P}{b} \cosh 2 \pi  \frac{P}{b}}.
\eeq
In the rest of this section we will analyze this expression. 

As indicated, these are unmarked FZZT boundaries. Using the methods described above we can first compute the marked boundary amplitude which is more directly related to the matrix integral. Taking derivatives with respect to the boundary cosmological constant using \eqref{fion} we get 
\bea\label{annulusunmarked}
\lb Z(s_1)^{\M} Z(s_2)^{\M}\rb_{(2,p)} &=&\frac{2}{b^2} \int_0^\infty \frac{dP}{\pi} \frac{\sin(4 \pi s_1 P) \sin(4 \pi s_2 P)}{\kappa \sinh \pi b s_1 \kappa \sinh \pi b s_2} \frac{2P }{\sinh 4 \pi  \frac{P}{b}} ,\nonumber\\
&=& \frac{1}{8\pi} \frac{1}{\sqrt{-\mu_1 + \kappa}\sqrt{-\mu_2 + \kappa}}\frac{1}{(\sqrt{-\mu_1+\kappa} + \sqrt{-\mu_2 + \kappa})^2},
\eea
where $\mu_i=\mu_B(s_i)$. The expression in the second line is precisely the connected component to the resolvent two point function (see for example equation (47) of \cite{Saad:2019lba}). When written in the appropriate variables this result is completely independent of $p$ and therefore independent of the precise density of states. This is evident  in the matrix integral approach but unexpected from the continuum approach.
\begin{figure}[t!]
\centering
\begin{tikzpicture}[scale=0.9]
\node at (-2.6,0) {\small $\ell_1$};
\node at (2.6,0) {\small $\ell_2$};
\draw[thick] (-2,0) ellipse (0.3 and 1.5);
\draw[thick] (2,0) ellipse (0.3 and 1.5);
\draw[thick] (-1.937,1.47) to [bend right=50] (1.937,1.47);
\draw[thick] (-1.937,-1.47) to [bend left=50] (1.937,-1.47);
\node at (0,-1.985) {};
\node at (4.5,0) {\large $=\int $ {\small $d\mu(\lambda)$}};
\draw[thick] (6.5,0) ellipse (0.3 and 1.5);
\draw[thick] (8.5,0) ellipse (0.1 and 0.7);
\draw[thick] (6.54,1.49) to [bend right=20] (8.5,0.7);
\draw[thick] (6.54,-1.49) to [bend left=20] (8.5,-0.7);
\draw[thick] (11.5,0) ellipse (0.3 and 1.5);
\draw[thick] (9.5,0) ellipse (0.1 and 0.7);
\draw[thick] (11.46,1.49) to [bend left=20] (9.5,0.7);
\draw[thick] (11.46,-1.49) to [bend right=20] (9.5,-0.7);
\node at (5.9,0) {\small $\ell_1$};
\node at (12.1,0) {\small $\ell_2$};
\node at (8.5,-1) {\small $\lambda$};
\node at (9.5,-1) {\small $\lambda$};
\end{tikzpicture}
\caption{We depict the cylinder amplitude in physical space between fixed length boundaries. The final answer can be interpreted as gluing minimal string trumpets generalizing the procedure of JT gravity}
\label{fig:trumpetgluing}
\end{figure}
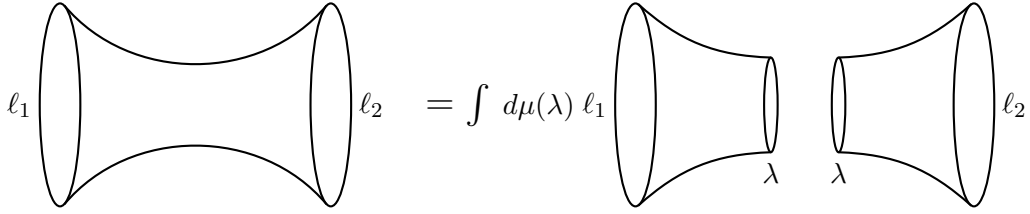

We can compute the fixed length amplitude in two ways. Firstly, we can apply the method above to compute the inverse Laplace transform through the discontinuity before integrating over $P$. In order to do this we can use the expression \eqref{eq:disccos} for the discontinuity. Secondly, we can apply this directly to the second line of \eqref{annulusunmarked}. Either way the result is the same, given after relabeling $\lambda = 2P/b$ by the formula 
 \bea\label{eq:2loopcorr}
 \lb Z(\ell_1) Z(\ell_2) \rb &=&\frac{2}{\pi} \int_0^\infty \lambda d\lambda  \tanh \pi \lambda ~K_{i\lambda}(\kappa \ell_1) K_{i \lambda}(\kappa \ell_2),\\
 &=&  \frac{\sqrt{\ell_1 \ell_2}}{\ell_1+\ell_2}e^{-\kappa (\ell_1+ \ell_2)}.
 \eea
The first line of the previous equation has a very familiar form when comparing with JT gravity. As we explained before, inserting a bulk operator in the disk can be interpreted as creating a hole in the physical space which in the JT gravity limit becomes a geodesic boundary of length $\sim \lambda$. Therefore we can compare the integral above after replacing $ \lambda = b_{\rm JT}/(2\pi b^2)$ and $ \ell = \ell_{\rm JT}/(2\kappa \pi^2 b^4)$  as gluing two minimal string trumpets with a deformed measure\footnote{The prefactors of this equation can be tracked by using the integral representation \eqref{idthree}.}
\bea
e^{\kappa \ell} K_{i\lambda}(\kappa \ell) &\to& \pi b^2 \sqrt{\frac{\pi}{\ell_{\rm JT}}}e^{- \frac{b_{\rm JT}^2}{4\ell_{\rm JT}}},\\
\lambda d\lambda \tanh \pi \lambda &\to& \frac{1}{4\pi^2b^4}b_{\rm JT}db_{\rm JT} ,~~~{\rm for}~\lambda \to \infty.
\eea
Liouville CFT is deeply intertwined with Teichm\"uller theory (the universal cover of the moduli space of Riemann surfaces), see e.g. \cite{Verlinde:1989ua,Teschner:2002vx}.\footnote{A related observation is the following. The partition function of group $G$ Chern-Simons theory on an annulus times $\mathbb{R}$ is known to be describable through the diagonal modular invariant of the $\hat{G}$ (non-chiral) WZW model, where the chiral sectors of the WZW model are each associated to one of the boundary cylindrical walls \cite{Elitzur:1989nr}. Something similar was observed in \cite{Blommaert:2018iqz} for Liouville CFT: the Liouville diagonal torus partition function yields the two-boundary sector of 3d gravity, but glued within Teichm\"uller space.} Here we see that when Liouville is combined with the minimal model into a full gravitational theory the integral becomes the WP measure over the moduli space instead, in accordance with the matrix model expectation. This is clear in the JT gravity limit, and it would be interesting to understand the origin of this tanh measure for finite $p$ minimal string, and to confirm this is its correct normalization.

\subsection{Multiple boundaries}  

It will be useful to rephrase the minimal string as a matrix integral in the double scaling limit using the formalism of \cite{Brezin:1990rb, Banks:1989df}. A central object from this approach is the heat capacity $u(x)$ appearing in the string equation. This is related to the density of states as  
\beq\label{eqdeff}
\rho_0(E) = \frac{1}{2\pi} \int_{E_0}^E \frac{du}{\sqrt{E-u}} f(u),
\eeq
where $\partial_x u = - f(u)^{-1}$ (see \cite{Johnson:2019eik, *Johnson:2020heh,Okuyama:2019xbv,*Okuyama:2020ncd} for recent discussions). It will be convenient for us to define shifted and rescaled quantities that will have a finite large $p$ limit, as:
\begin{equation}
\label{ujt}
E = \kappa \Big(1  + \frac{8\pi^2}{p^2} E_{\rm JT} \Big), \qquad u = \kappa\Big(1  + \frac{8\pi^2}{p^2} u_{\rm JT} \Big).
\end{equation}
For ease of notation, we set $\kappa=1$ in the following. With these conventions the undeformed minimal string will correspond to $x\to0$. The minimal string density of states according to the Liouville calculation is  
\bea
\rho_0(E_{\rm JT}) &=&\frac{1}{4\pi^2} \sinh \Big( \frac{p}{2} {\rm arccosh} \Big(1  + \frac{8\pi^2}{p^2} E_{\rm JT} \Big) \Big),\\
&=& \sum_{j=0}^\mathfrak{m}  \frac{(2\pi)^{2j-3}}{(2j-1)!}  \frac{4^{j-1}(\mathfrak{m}+j-2)!}{(2\mathfrak{m}-1)^{2j-2}(\mathfrak{m}-j)!} (\sqrt{E_{\rm JT}})^{2j-1} ~~~~{\rm with}~~p=2\mathfrak{m}-1,
\eea
where $\mathfrak{m}\in \mathbb{Z}$. For large $\mathfrak{m}$ we get the JT gravity density of states. 
We can find the function $f(u)$ by solving \eqref{eqdeff} and get 
\begin{equation}
f(u_{\rm JT}) = \frac{1}{2} {}_2F_1\left(\frac{1-p}{2},\frac{1+p}{2},1, - \frac{4\pi^2}{p^2}u_{\rm JT}\right),
\end{equation} where ${}_2F_1(a,b,c,x)$ is the hypergeometric function. Integrating this relation we can get an implicit formula for the minimal string heat capacity
\beq
\frac{u_{\rm JT}}{2}  \hspace{0.0cm} {}_2F_1 \Big(\frac{1-p}{2},\frac{1+p}{2},2, - \frac{4\pi^2}{p^2}u_{\rm JT}\Big) = - x. 
\eeq
This can be written in a more familiar form recognizing that for these values of parameters the hypergeometric function becomes a Legendre polynomial, e.g.:
\begin{equation}
f(u_{\rm JT}) = \frac{1}{2} P_{\mathfrak{m}-1}\Big(1+\frac{8\pi^2}{p^2}u_{\rm JT} \Big)
\end{equation}
The relation above becomes the string equation \cite{Brezin:1990rb} (to leading order in genus expansion) written in the usual form, given by
\beq
\sum_{j=0}^\mathfrak{m} t_j u_{\rm JT}^j = 0 ,~~~~~~t_j\equiv \frac{1}{2} \frac{\pi^{2j-2}}{ j! (j-1)!}  \frac{4^{j-1}(\mathfrak{m}+j-2)!}{(\mathfrak{m}-j)! (2\mathfrak{m}-1)^{2j-2}}
\eeq 
where $p=2\mathfrak{m}-1$, we introduced the couplings $t_j$ and defined $t_0=x$. As explained in \cite{Moore:1991ir}\cite{Belavin:2008kv} this is an analytic redefinition of coupling constants of the $\mathfrak{m}$'th multicritical point and for large $x$ behaves as $u \sim x^{1/\mathfrak{m}}$, as expected. Knowing the couplings $t_j$, it is possible to also compute higher genus corrections by replacing the power law in the equation above by the KdV hierarchy operators $u^j \to R_j[u]$ derived in \cite{Gelfand1975,*Gelfand2}.

Knowing the heat capacity for the minimal string $u(x)$ derived from the density of states a surprising formula can be written for the $n$th loop correlator first proposed by \cite{Ambjorn:1990ji, Moore:1991ir}. The relation can be written in different ways but we found a useful version to be 
  \beq \label{nbdyformula}
  \Big\lb \prod_i Z(\ell_i)^{\M} \Big\rb =- \frac{\sqrt{\ell_1\ldots \ell_n}}{2 \pi^{n/2}}  \Big( \frac{\partial}{\partial x} \Big)^{n-3} u'(x) e^{-u(x)(\ell_1+\ldots+\ell_n)} \Big|_{u\to 1}.
  \eeq
From now one we will only work with marked boundaries and omit the $\M$ suffix. \eqref{nbdyformula} is based on the discrete approach and its surprising such a simple answer exists from the continuum approach. Shifting to our variable $u_{\rm JT}$ as $u = 1  + \frac{8\pi^2}{p^2} u_{\rm JT}$, the final answer is evaluated at $u_{\rm JT}(x\to0)=0$.
To apply this formula we need the derivatives $\partial_x u_{\rm JT}$ but the relation $u_{\rm JT}(x)$ is given only implicitly. To find the necessary derivatives, we can apply the Lagrange inversion theorem to write 
\beq\label{eq:inversederums}
\left.\partial_x^n u_{\rm JT}\right|_{u_{\rm JT} =0} = \lim_{u_{\rm JT}\to 0} \frac{d^{n-1}}{du_{\rm JT}^{n-1}} \left(- \frac{2}{{}_2F_1(\frac{1-p}{2},\frac{1+p}{2},2, - \frac{4\pi^2}{p^2}u_{\rm JT})} \right)^n.
\eeq
This can be used order by order to find all terms appearing in the loop correlators. 

We will now use this to generate some $n$ loop correlators for fixed boundary length. The case $n=1$ is special and actually is used to fix $u(x)$. The case $n=2$ is also special and gives $\lb Z(\ell_1) Z(\ell_2) \rb = \frac{1}{2\pi} \frac{\sqrt{\ell_1 \ell_2}}{\ell_1+\ell_2}$, which coincides with \eqref{eq:2loopcorr} after appropriate shifts and redefinitions mentioned above. The cases $n=3,4,5$ give\footnote{We have defined the length parameter $\ell^{\rm JT} = \kappa \frac{8\pi^2}{p^2} \ell = 2\pi^2b^4\kappa \ell$ as in \eqref{JTparam}, and have redefined the overall normalization by dropping a factor of $\big(\frac{8\pi^2}{p^2}\big)^{1-\frac{n}{2}} e^{-\kappa \sum_i \ell_i}$.}
\begin{align}
\Big\lb \prod_{i=1}^3 \frac{Z(\ell^{\rm JT}_{i})}{\sqrt{\ell^{\rm JT}_{i}} } \Big\rb &=-\frac{1}{2\pi^{3/2}} \left.\frac{\partial u_{\rm JT} }{ \partial x}\right|_{u_{\rm JT}=0} =\frac{1}{2\pi^{3/2}}  2 , \\
\Big\lb \prod_{i=1}^4 \frac{Z(\ell^{\rm JT}_{i})}{\sqrt{\ell^{\rm JT}_{i}} } \Big\rb &= \frac{1}{2\pi^{2}}  \Big( 4 \Big(\sum_{i=1}^4\ell^{\rm JT}_{i}\Big) +4\pi^2\Big(1-\frac{1}{p^2} \Big)  \Big), \\
\Big\lb \prod_{i=1}^5 \frac{Z(\ell^{\rm JT}_{i})}{\sqrt{\ell^{\rm JT}_{i}} } \Big\rb &= \frac{1}{2\pi^{5/2}}  \Big(8 \Big(\sum_{i=1}^5\ell^{\rm JT}_{i}\Big)^2 +24 \pi^2 (1-\frac{1}{p^{2}})\Big(\sum_{i=1}^5\ell^{\rm JT}_{i}\Big) + 4\pi^4 (5-\frac{2}{p^{2}} -\frac{3}{p^{4}})  \Big).
\end{align}
At this point it should be clear how to generalize it to arbitrary boundaries. 

As a further check of these expressions we will take the JT gravity limit $p \to \infty$. First we will take the JT limit of the string equation. Using the following identity (Abramowitz and Stegun eq (9.1.71)):
\begin{equation}
\label{AS}
\lim_{\nu \to+\infty} P_\nu \left(\cos \frac{x}{\nu}\right) = J_0(x), \qquad \text{with }\cos \frac{x}{\mathfrak{m}-1} = 1 + \frac{8\pi^2}{p^2}u_{\rm JT},
\end{equation}
one shows that $f(u) \to \frac{1}{2} I_0(2\pi \sqrt{u})$. For large $p$ the couplings become $t_j \to \frac{1}{2} \frac{\pi^{2j-2}}{j!(j-1)!}+\mathcal{O}(1/p)$. The sum can be done explicitly and the JT gravity string equation becomes 
\beq
\sum_{j=1}^\infty \frac{1}{2} \frac{\pi^{2j-2}}{ j! (j-1)!} u_{\rm JT}^j = \frac{\sqrt{ u_{\rm JT} }}{2\pi} I_1(2\pi \sqrt{ u_{\rm JT}}) = -x. 
\eeq   
The $n$-boundary JT gravity partition function to leading order in the genus expansion is then 
 \beq \label{nbdyformulaJT}
  \Big\lb \prod_i Z_{\rm JT}(\ell^{\rm JT}_{i}) \Big\rb =- \frac{\sqrt{\ell^{\rm JT}_{1}\ldots \ell^{\rm JT}_{n}}}{2\pi^{n/2}} \Big( \frac{\partial}{\partial x} \Big)^{n-3} u'_{\rm JT}(x) e^{-u_{\rm JT}(x)(\ell^{\rm JT}_{1}+\ldots+\ell^{\rm JT}_{n})} \Big|_{u_{\rm JT}\to 0}.
  \eeq
This can be seen as a generating function for the genus $0$ WP volumes with $n$ geodesic boundaries $V_{g=0,n}(\mathbf{b})$ with length $\mathbf{b}=(b_1,\ldots, b_n)$, after an appropriate Laplace transform we will do explicitly in the next section.

We can check this formula computing some simple cases with $n=3,4,5,\ldots$. This can be obtained either from the $p\to\infty$ limit of the minimal string or directly using the JT string equation $u_{\rm JT}(x)$. The result is 
\bea
\Big\lb \prod_{i=1}^3 \frac{Z_{\rm JT}(\ell^{\rm JT}_{i})}{\sqrt{\ell^{\rm JT}_{i}} } \Big\rb &=&\frac{1}{2\pi^{3/2}} 2 ,\\
\Big\lb \prod_{i=1}^4 \frac{Z_{\rm JT}(\ell^{\rm JT}_{i})}{\sqrt{\ell^{\rm JT}_{i}} } \Big\rb &=& \frac{1}{2\pi^{2}}  \Big( 4 \Big(\sum_{i=1}^4\ell^{\rm JT}_{i}\Big) +4\pi^2  \Big), \\ 
\Big\lb \prod_{i=1}^5 \frac{Z_{\rm JT}(\ell^{\rm JT}_{i})}{\sqrt{\ell^{\rm JT}_{i}} } \Big\rb &=& \frac{1}{2\pi^{5/2}} \Big(8 \Big(\sum_{i=1}^5\ell^{\rm JT}_{i}\Big)^2 +24 \pi^2\Big(\sum_{i=1}^5\ell^{\rm JT}_{i}\Big) + 20\pi^4  \Big).
\eea
It is surprising that these correlators match with the direct JT gravity calculation \cite{Saad:2019lba} where 
\beq
\Big\lb \prod_{i=1}^n Z_{\rm JT}(\ell^{\rm JT}_{i}) \Big\rb = \int \prod_{i=1}^n b_i db_i Z_{\rm trumpet}(\ell^{\rm JT}_{i}, b_i) V_{0,n}(\mathbf{b}),
\eeq
where the trumpet partition function is given by $Z_{\rm trumpet}(\ell^{\rm JT}_{i}, b_i) =e^{-b_i^2/4\ell^{\rm JT}_{i}}/2\sqrt{\pi \ell^{\rm JT}_{i}} $. For the WP volumes we used the expressions in \cite{do2011moduli} and we also check this works for $n=6$ and $7$, although we did not write it here. Therefore we see that \eqref{nbdyformulaJT} gives a simple generating function for (a simple integral transform) of WP volumes. 

\subsection{$p$-deformed Weil-Petersson volumes}
In this section we will point out some interesting structure in the minimal string multi-loop correlator. One can write the amplitude in two ways:
\begin{align}
\label{Wppdef}
\left\langle \prod_{i=1}^n Z(\ell_i)\right\rangle &= -\left. \frac{\sqrt{\ell_1 \hdots \ell_n}}{2\pi^{n/2}}\left(\frac{\partial}{\partial x}\right)^{n-3} u ' (x) e^{-u(x) (\ell_1 + \hdots \ell_n)}\right|_{u=1} \\
\label{defwp}
&= \frac{2^{n-1}}{\pi^{n/2}} \prod_i \int_{0}^{+\infty} d\lambda_i \lambda_i \sinh \pi \lambda_i K_{i\lambda_i}(\ell_i) \frac{V_{0,n}(\bm{\lambda})}{\cosh \pi \lambda_i}
\end{align}
where in the second line we have written the integral in terms of multiple gluing cycles $\lambda_i$, with the gluing measure $d\mu(\lambda) = d\lambda \lambda \tanh \pi \lambda$ which we have written suggestively. The quantity $V_{0,n}(\bm{\lambda}) \equiv V_{0,n}(\lambda_1,...\lambda_n)$ will turn out to be a polynomial in $\lambda_i^2$ and can be viewed as a generalization of the WP volumes to the $p$-deformed setup. The numerical prefactors were chosen such that the $p\to+\infty$ limit directly yields back the WP volumes. \\
We can find explicit expressions for the $V_{0,n}(\bm{\lambda})$ by applying the Kontorovich-Lebedev (KL) transform:
\begin{align}
\label{KL}
g(y) = \int_0^{+\infty} \frac{dx}{x} f(x) K_{iy}(x), \qquad f(x) = \frac{2}{\pi^2} \int_{0}^{+\infty} dy g(y) K_{iy}(x) y \sinh \pi y
\end{align}
leading to
\begin{align}
\label{wpfi}
V_{0,n}(\bm{\lambda}) =  \left.\left(\frac{\pi}{2}\right)^{n/2} \left(\frac{\partial}{\partial x}\right)^{n-3} u ' \prod_i \mathcal{L}_u\left(\frac{K_{i\lambda_i}(\ell_i)}{\sqrt{\ell_i}}\right) \cosh \pi \lambda_i\right|_{u=1}
\end{align}
where we need the following Laplace transform:\footnote{The integral is convergent at $x=0$ since $\left|K_{i\lambda}\right|$ is bounded close to zero.}
\begin{align}
\mathcal{L}_u\left(\frac{K_{i\lambda_i}(\ell_i)}{\sqrt{\ell_i}}\right) = \int_0^{+\infty} \frac{dx}{\sqrt{x}}K_{i\lambda}(x) e^{-u x} &= \frac{(2\pi)^{3/2}}{4\cosh \pi \lambda} {}_2F_{1}\left(\frac{1}{4} + \frac{i \lambda}{2}, \frac{1}{4} - \frac{i \lambda}{2}, 1, 1-u^2\right),\\
\label{legp}
&= \frac{(2\pi)^{3/2}}{4\cosh \pi \lambda} P_{-\frac{1}{2}-i\lambda}(u)
\end{align}
where $ u\geq 1$. Notice that the KL transform is invertible, and hence the WP volumes are unambigously defined by the above relation \eqref{Wppdef}. It is convenient as before to work with the shifted and rescaled variable $u_{\rm JT}$ defined by the relation $u=1+\frac{8\pi^2}{p^2}u_{\rm JT}$. We hence write:
   \beq
	\label{WPp}
	\boxed{
	  V_{0,n}(\bm{\lambda}) =\lim_{u_{\rm JT}\to 0}- \frac{1}{2} \Big( \frac{\partial}{\partial x} \Big)^{n-3} u'_{\rm JT}(x) \prod_{i=1}^n P_{-\frac{1}{2}-i\lambda}\left(1+\frac{8\pi^2}{p^2}u_{\rm JT}(x)\right)},
  \eeq
where an explicit formula for $\partial_x^n u_{ \rm JT}(x=0) $ was given above in equation \eqref{eq:inversederums}.
Zograf proved a theorem about a generating function for WP volumes in the sphere with n punctures $V_{0,n}(\mathbf{0})$ \cite{zograf1998weilpetersson}. This formula \eqref{WPp} gives a minimal string version of it and extends it to finite size boundary lengths. \\

To find explicit formulas from \eqref{WPp}, we can have to differentiate and evaluate at $u_{\rm JT} = 0$ in the end. To that effect, we can use the result:
\begin{align}
\partial_{u_{\rm JT}}^m \left. P_{-\frac{1}{2}-i\lambda}(u)\right|_{u_{\rm JT} = 0}  &= (-)^m \left(\frac{8\pi^2}{p^2}\right)^m\frac{4\cosh \pi \lambda}{(2\pi)^{3/2}}\int_0^{+\infty} dx x^{m-1/2}K_{i\lambda}(x) e^{-x} \\
\label{toins}
&= \left(\frac{8\pi^2}{p^2}\right)^m (-)^m \frac{1}{2^m m!}\prod_{j=1}^{m}(\lambda^2+(2j-1)^2/4)
\end{align}
The equality in the last line is the KL transform of equations written in Appendix C of \cite{Moore:1991ag}. Importantly, this produces a polynomial in $\lambda_i^2$, mirroring the analogous situation for the WP volumes. \\

Finally, in order to make contact with the Weil-Petersson volumes at $p\to + \infty$, we define
\begin{equation}
\label{geolim}
\lambda_i = \frac{p}{4\pi}b_i
\end{equation}
in terms of the geodesic length $b_i$ that stays finite as we take the limit.
As explicit examples, for $n=4,5,6$ we obtain by inserting \eqref{eq:inversederums} and \eqref{toins} into \eqref{WPp} :
\begin{align}
V_{0,4}(\bm{\lambda}) &= \Big( 2\pi^2  + \frac{6 \pi^2}{p^2}\Big)+\frac{1}{2} \sum_i b_i^2 \\
V_{0,5}(\bm{\lambda}) &=\Big( 10\pi^4 + \frac{56\pi^4}{p^2} + \frac{104 \pi^4}{p^4} \Big) +\left(3\pi^2+ \frac{10\pi^2}{p^2}\right) \sum_i b_i^2 + \frac{1}{2}\sum_{i<j}b_i^2b_j^2 + \frac{1}{8} \sum_i b_i^4\\
V_{0,6}(\bm{\lambda}) &= \Big(\frac{244}{3}\pi^6 +  \frac{1972\pi^6}{3p^2} + \frac{6604\pi^6}{3p^4} + \frac{3060 \pi^6}{p^6}\Big) + \left(26\pi^4+ \frac{160\pi^4}{p^2} + \frac{916\pi^4}{3p^4}\right) \sum_i b_i^2  \\
&\hspace{-1cm} + \left(6\pi^2+ \frac{21\pi^2}{p^2}\right) \sum_{i<j} b_i^2 b_j^2 + \left(\frac{3\pi^2}{2}+ \frac{31\pi^2}{6p^2}\right) \sum_i b_i^4 + \frac{3}{4}\sum_{i<j<k}b_i^2b_j^2b_k^2 + \frac{3}{16}\sum_{i,j, i \neq j}b_i^4b_j^2  + \frac{1}{48}\sum_i b_i^6 \nonumber
\end{align}
All of these satisfy the correct $p\to+\infty$ WP limit, as can be seen by comparing to Appendix B of \cite{do2011moduli}, see also \cite{Mirzakhani:2006fta, *Mirzakhani:2006eta}.

\begin{center}
\textbf{Adding handles} 
\end{center} 

We will show some more evidence of the structure identified here. We will derive the simplest correction for a single boundary and higher genus $g=1$, and then discuss some properties of the generic higher genus result. This is very hard to do from the continuous approach but we can assume the duality is true and obtain the leading handle correction to the partition function using the matrix model.

To find higher genus amplitudes, we can use Eynard's topological recursion relations as follows \cite{Eynard:2004mh,Eynard:2007kz}. Provided the two quantities:
\begin{equation}
W_{0,1}(z) = 2 zy(z), \qquad W_{0,2}(z_1,z_2) = \frac{1}{(z_1-z_2)^2},
\end{equation}
the generic amplitude for a double-scaled matrix integral can be found recursively by computing the residue
\begin{align}
\label{eynard}
&W_{g,n}(z_1,J) = \\
&\text{Res}_{z\to 0}\left\{\frac{1}{(z_1^2-z^2)} \frac{1}{4y(z)}\left[W_{g-1,n+1}(z,-z,J) + \sum_{h,I,h',I'} W_{h,1+I}(z,I)W_{h',1+I'}(-z,I')\right]\right\}, \nonumber
\end{align}
where $h+h'=g$ and $I \cup I' =J$ denoting a subset of the labels $z_2, \hdots z_n$, and the sum excludes the cases $(h=g,I=J)$ and $(h'=g,I'=J)$. 

Using the minimal string spectral curve as seed, and applying it to genus one with one boundary, we get the following correction to the partition function  
 \beq\label{eq:Z11}
 Z(\ell_{\rm JT})_{g=1,n=1} = \frac{\sqrt{\ell_{\rm JT}}}{12\sqrt{\pi}} (\ell_{\rm JT} + \pi^2(1-p^{-2})),
 \eeq
which we wrote in term of the normalized length $\kappa\ell=\ell_{\rm JT} \frac{p^2}{ 8\pi^2}$. Using the Kontorovich-Lebedev transform this correction can be written
 \beq
Z(\ell)_{g=1} \sim   \int \lambda d\lambda \tanh \pi \lambda  K_{i \lambda} (\ell) V_{1,1}(\lambda),
 \eeq
 where we will not worry about the overall normalization. The $p$-deformed WP volume appearing from \eqref{eq:Z11} is given by 
 \beq
\label{pgen1}
 V_{1,1}(\lambda)= \frac{\pi^2}{12} + \frac{\pi^2}{3p^2}\lambda^2 .
 \eeq
After calling $\lambda = \frac{p}{4\pi} b_{\rm JT}$ this expression becomes exactly the WP volume for a torus with one geodesic boundary of length $b_{\rm JT}$, namely $V_{1,1}(\lambda)= (b_{\rm JT}^2 + 4\pi^2)/48$. It is interesting to note that this particular volume does not get deformed at finite $p$, when written in terms of $b_{\rm JT}$.\footnote{We thank A. Artemev for pointing out an error in an earlier version of this paper. This correction is important for the conclusion that the volume is not modified at finite $p$.}
\\~\\
This $p$-deformed WP volume \eqref{pgen1} is again a polynomial in $\lambda_i^2$ as before. Using the recursion relation \eqref{eynard}, we can give an argument why this is so for arbitrary genus $g$ and boundaries $n$.
The resolvents $W_{g,n}(z_1, \hdots z_n)$ for a one-cut matrix model with edges at $z=a,b$ are symmetric rational functions of the $z_i$ with poles only at $z_i = a,b$, see e.g. section 4.2.3 in \cite{Eynard:2015aea}.\footnote{Except of course $W_{0,2}$.} In addition, they decay to zero as $z_i \to \infty$.
For a double-scaled matrix integral, for which we shift the edge to $z_i=0$, these properties fix the $W_{g,n}(z_1, \hdots z_n)$ to be multivariate polynomials of $1/z_i$. 

If the spectral curve $y(z)$ is in addition an odd function of $z$, then the $W_{g,n}(z_1, \hdots z_n)$ are polynomials with only even powers of $1/z_i$, making it a polynomial in $1/z_i^2$.\footnote{The reason for this constraint is that $W_{0,2}$ is not an even function of the $z_i$, but for $y(z)$ odd, when computing the residue in \eqref{eynard}, the Taylor series of $W_{0,2}(z,z_1)$ around $z=0$ needs to select an even power of $z$ (and hence of $z_1$) in order to contribute to the residue.}
For the minimal string case at hand, the spectral curve is odd and hence this is true.

The resolvent $W_{g,n}(z_1, \hdots z_n)$ is related to the multi-loop amplitude $Z_{g,n}(\ell_1 \hdots \ell_n)$ through
\begin{equation}
W_{g,n}(z_1, \hdots z_n) = 2^n z_1\hdots z_n \int_{0}^{+\infty} \prod_i d\ell_i e^{- \ell_i z_i^2} Z_{g,n}(\ell_1 \hdots \ell_n),
\end{equation}
which is in turn related to the WP volume $V_{g,n}(\bm{\lambda})$ by \eqref{defwp}. Each such $1/z_i^{2(m+1)}$ term in $W_{g,n}(z_1, \hdots z_n)$, where $m=0,1,\hdots$, then gets inverse Laplace transformed and Kontorovich-Lebedev transformed to the WP volumes using consecutively:\footnote{The $e^{-\ell}$ factor is explained by our choice to shift the spectral edge to $z=0$.}
\begin{align}
\frac{2 (m+1)!}{z^{2(m+1)}} &= 2z \int_{0}^{+\infty} d\ell e^{- \ell z^2} \ell^{m+1/2}, \qquad m=0,1,\hdots, \\
\ell^{m+1/2} e^{-\ell} &= \sqrt{\frac{2}{\pi}} \frac{1}{2^m m!}\int_{0}^{+\infty} d\lambda \lambda \tanh \pi \lambda \prod_{j=1}^{m} (\lambda^2 + (2j-1)^2/4) K_{iE}(\ell).
\end{align}
Hence if $W_{g,n}$ is a multivariate polynomial in the $1/z_i^{2}$, as happens for the minimal string, then the $p$-deformed WP volumes are polynomials in the $\lambda_i^2$:
\begin{equation}
W_{g,n}(z_1,\hdots z_n) = \sum_{i_1\hdots i_n} \frac{c_{i_1\hdots i_n}}{z_1^{2i_1}z_2^{2i_2} \hdots z_n^{2i_n}} \qquad \to \qquad V_{g,n} = \sum_{i_1\hdots i_n \hdots =0}^{n+3g-3}\tilde{c}_{i_1\hdots i_n} \lambda_1^{2i_1}\lambda_2^{2i_2} \hdots \lambda_n^{2i_n},
\end{equation}
as was to be shown.
 
\begin{center}
\textbf{Classical WP volumes} 
\end{center} 

As a final application of these results we will write an explicit formula for WP volumes in the sphere. One can take the JT limit directly at the level of the generating functions. Considering the description in terms of a Legendre function \eqref{legp}, inserting \eqref{geolim} and \eqref{ujt}, and using \eqref{AS}, we get:
\begin{equation}
P_{-\frac{1}{2}-\frac{ip}{4\pi}b_i}\Big(1+\frac{8\pi^2}{p^2}u_{\rm JT}\Big) \,\, \to \,\, J_0(b_i\sqrt{u_{\rm JT}}),
\end{equation}
leading to the closed formula for the (undeformed) WP volumes:
   \beq
	\boxed{
	  V_{0,n}(\mathbf{b}) =\lim_{x\to0} -\frac{1}{2} \Big( \frac{\partial}{\partial x} \Big)^{n-3} u'_{\rm JT}(x) \prod_{i=1}^n J_0(b_i \sqrt{u_{\rm JT}(x)})},
  \eeq
  where the derivatives of $u_{\rm JT}(x)$ are equal to 
  \beq
\partial_x^{n} u_{\rm JT}(x=0) =   \lim_{u\to0} \frac{d^{n-1}}{du^{n-1}} \Big( -\frac{2\pi \sqrt{u}}{I_1(2\pi\sqrt{u})}\Big)^{n}.
  \eeq
For each value of $n$ it is easy to take the appropriate derivatives and obtain a formula for WP volumes with $n$ holes. We computed these explicitly for $n=1,\ldots,7$ matching previous results that use the loop equations presented, for example, in Appendix B of \cite{do2011moduli}. This is surprising since even though we derived this formula from the matrix model we did not use the loop equations explicitly. 

As a special case, we can take the WP volume on the sphere with $n$ punctures which is equivalent to taking the limit $\mathbf{b}\to0$. It is easy to see that this gives $V_{0,n}(\mathbf{0}) =-\frac{1}{2}\partial_x^{n-2} u_{\rm JT}(0)$. Using the expression above for these derivatives using the Lagrange inversion theorem gives a somewhat more explicit formula 
\beq
V_{0,n}(\mathbf{0}) =\lim_{u\to0} \frac{1}{2} \frac{d^{n-3}}{du^{n-3}} \left( \frac{2\pi \sqrt{u}}{J_1(2\pi\sqrt{u})}\right)^{n-2},
\eeq
where we used that the minus signs can be absorbed in a shift of Bessel functions $I_1\to J_1$. This result is equivalent to the WP volume extracted from the generating function derived by Zograf \cite{zograf1998weilpetersson}, which is precisely the string equation of JT gravity.   

\begin{center}
\textbf{Summary}
\end{center}	
With these polynomials, we can now explicitly decompose the $n$-loop amplitude as:
\begin{align}
\left\langle \prod_{i=1}^n Z(\ell_i)\right\rangle_g = 2^n (2\pi)^{n-3}(\pi b^2)^{n} \prod_{i=1}^{n} \int_0^\infty \lambda_i d\lambda_i  \tanh \pi \lambda_i \,V_{g,n}(\bm{\lambda})  \left\langle \mathcal{T}_{\alpha_{Mi}}\right\rangle_{\ell_i},
\end{align}
in terms of the $p$-deformed gluing measure $d\mu(\lambda) \sim d\lambda_i \lambda_i \tanh \pi \lambda_i$, the $p$-deformed WP-volume polynomial $V_{g,n}(\bm{\lambda})$, and the bulk one-point functions \eqref{bulkone} with $\lambda = 2P/b$. Graphically, we have the situation:
\begin{align}
\left\langle \prod_{i=1}^n Z(\ell_i)\right\rangle \quad = \quad \prod_{i=1}^{n}\int d\mu(\lambda_i) \qquad \raisebox{-25mm}{\includegraphics[width=0.3\textwidth]{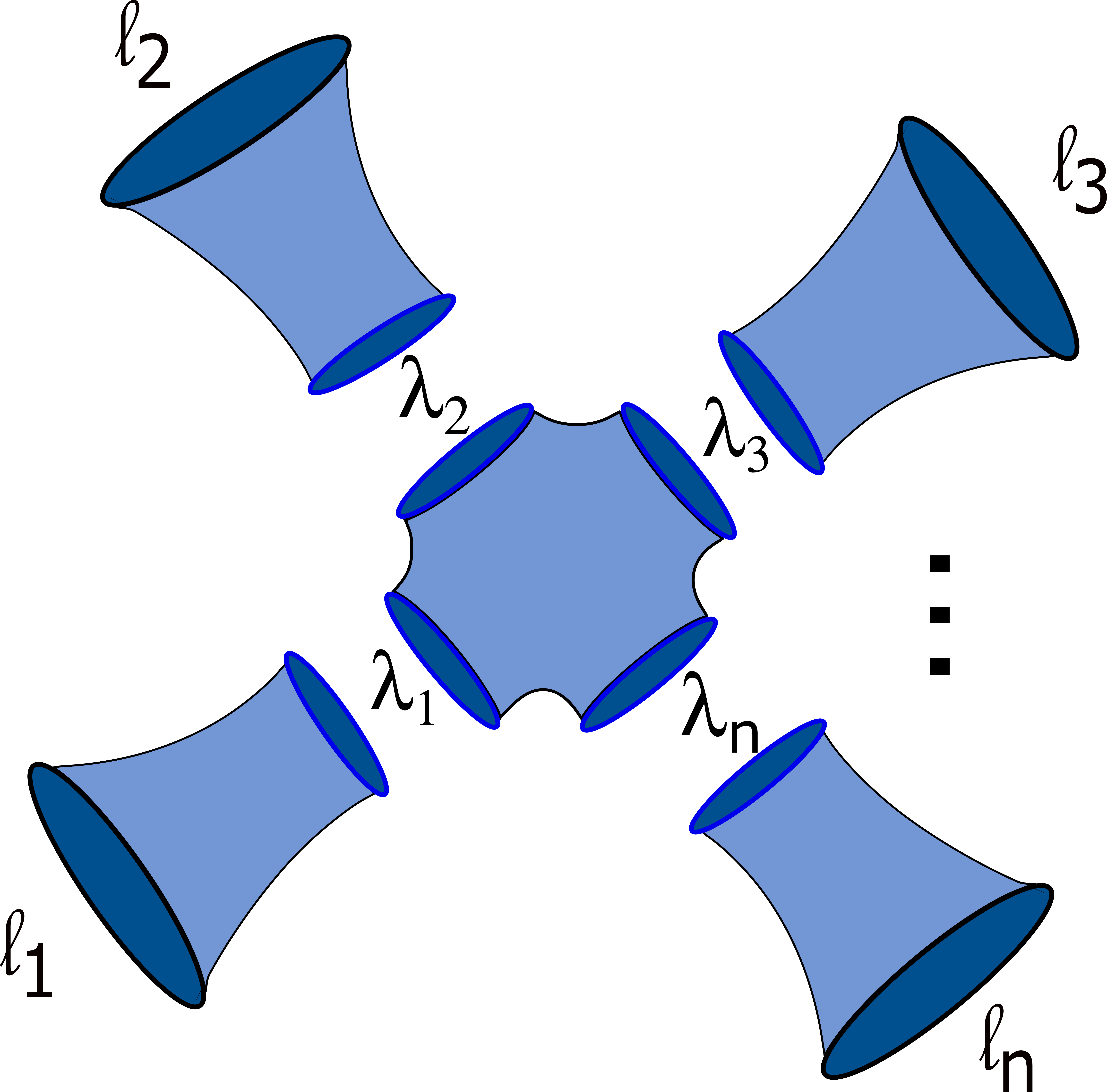}}
\end{align}

Notice that one only integrates over the macroscopic labels where $\alpha_M = -q/2 + iP$ with $P \in \mathbb{R}$, in analogy with the JT limit. For finite $p$, one can deform the contour of integration and replace the integral by a discrete sum over minimal string physical operators \cite{Moore:1991ir}. 

We studied this mainly for $g=0$, where we found explicit expressions \eqref{WPp}, but proposed a very similar structure for higher genus contributions, which we checked explicitly by computing $V_{1,1}(\lambda)$ and utilizing general arguments based on the topological recursion relations of the matrix model.
  
\section{Conclusions} \label{sec:conclusions}

Throughout this work, we have presented fixed length amplitudes of Liouville gravities, and in particular of the minimal string. We have developed both the continuum approach and the discrete matrix model approach. A particular emphasis was placed on the interpretation in terms of Euclidean gravity amplitudes at fixed temperature $\beta^{-1}$, and in their JT parametric limit. \\
We here present some open problems and preliminary results that will be left to future work.

\begin{center}
\textbf{Heavy boundary operators and cusps}
\end{center}	
We have seen that taking $\beta_M = bh$ in \eqref{eq:2pt} and letting $b\to 0$, one finds the JT boundary two-point function. However, the expression \eqref{eq:2pt} is more general. In particular, if we set $\beta_M = Q-bh$, we would find a finite $b\to 0$ limit as well:
\begin{equation}
\mathcal{A}_{\beta_M}(\ell_1,\ell_2) \sim \int dk_1 dk_2 \rho_{\rm JT}(k_1)\rho_{\rm JT}(k_2)e^{-k_1^2 \ell_{{\rm JT}1}}e^{-k_2^2 \ell_{{\rm JT}2}}\frac{\Gamma(2h)}{\Gamma(h \pm i k_1 \pm i k_2)},
\end{equation}
with \emph{inverted} vertex functions. This corresponds to taking a heavy boundary insertion. Since we know heavy bulk insertions correspond geometrically to conical singularities in the Euclidean JT geometry, it is natural to suspect that the situation here corresponds geometrically to having cusps in the boundary at the location of the operators. Such expressions are ill-defined when $h \in -\mathbb{N}/2$. 

\begin{center}
\textbf{Quantum groups}
\end{center}	
In section \ref{s:qg} we have developed the quantum group perspective on these amplitudes, mirroring the structure of JT gravity based on SL$(2,\mathbb{R})$. An interesting question is to understand precisely how this structure persists for four- and higher-point functions. This is dependent on understanding how the moduli summation for multiple ($>3$) boundary insertions works when combining the Liouville and the matter sectors. \\
The group theoretic structure  $\mathcal{U}_{q}(\mathfrak{sl}(2,\mathbb{R}))$ is present in 3d gravity as well \cite{Jackson:2014nla}.\footnote{Another connection with 3d (and higher dimensional) gravity was developed for example in \cite{Ghosh:2019rcj, Iliesiu:2020qvm}, but only works in the Schwarzian limit.} In that case however, one has angular dependence on all correlators, requiring a more complicated combination of these group theoretic building blocks. Our setup is based on the same (quantum) group structure, but does not require additional features. As such, it is one of the simplest quantum extensions of the SL$(2,\mathbb{R})$ case. \\
Another setting that generalizes JT gravity through $q$-deformation is the double-scaled SYK model, explicitly solved in \cite{Berkooz:2018jqr}. In that case the vertex functions were found to be of the form:
\begin{equation}
\frac{\Gamma_b(h\pm is_1 \pm is_2)}{\Gamma_b(2h)},
\end{equation}
which is not quite the same as the structure we have. This can be explained since that work argues that double-scaled SYK is governed by the $q$-deformation into SU$_q(1,1)$, which is a different quantum group theoretical structure than ours. In the classical regime $q\to 1$, both groups coincide since we have the classical isomorphism SL$(2,\mathbb{R}) \simeq$ SU$(1,1)$.

\begin{center}
\textbf{Multi-boundary and higher genus amplitudes}
\end{center}	
In the last section \ref{sec:othertopo}, we have investigated the structure of multi-loop amplitudes, both in the continuum approach and through matrix model techniques. This leads to several unanswered questions. \\
We found the gluing measure for the minimal string for genus zero multi-loop amplitudes to be $d\mu(\lambda) = \lambda\, d\lambda\,  \tanh\pi \lambda$, limiting to the Weil-Petersson measure $d\mu_{\rm WP}(b)=b\, db$ in the semi-classical limit where $\lambda \to \infty$. The quantity $b$ has a geometric interpretation as circumference of the gluing tube, and the factor of $b$ in $b\, db$ represents the sum over all possible twists, ranging from 0 to $b$, happening before gluing two tubes together. It would be interesting to find a similar geometric interpretation for the measure $d\lambda \lambda \tanh\pi \lambda$, perhaps as a gluing formula on quantum Riemann surfaces. \\
In the same vein, we can observe that for generic $c_M<1$ matter, the two-loop amplitude for fixed matter momentum $p$, can be written suggestively as \cite{Moore:1991ag,Martinec:2003ka}
\begin{equation}
\left\langle Z(\ell,p)Z(\ell',-p)\right\rangle \sim \int_{0}^{+\infty}dE \rho_{\widetilde{SL(2)}}\left( \frac{E}{2}, \frac{p}{2} \right) K_{iE}(\ell) K_{iE}(\ell'),
\end{equation}
with gluing measure the Plancherel measure of the universal cover of SL$(2,\mathbb{R})$:
\begin{equation}
\rho_{\widetilde{SL(2)}}(s,\mu) = \frac{s\sinh 2\pi s}{\cosh 2\pi s - \cos 2\pi \mu}, \quad 0\leq \mu \leq 1.
\end{equation}
For the $(2,2\mathfrak{m}-1)$ minimal string, the matter momentum takes on values $p=\pm 1/2$, and hence $\cos 2\pi \mu=0$. We do not understand the significance of this. \\

When summing over higher genus, it remains to be seen whether a simplification occurs. For the case of $c=1$ ($b=1$) several expressions for the all-genus result are known in a very concise form, see e.g. \cite{Moore:1991sf} for early work and \cite{Betzios:2020nry} for a recent account. \\

Finally, the expression \eqref{Wppdef} has some interesting implications. A different way to write it is the following 
\beq
\Big\langle \prod_{i=1}^n Z(\ell_i)\Big\rangle =\lim_{x\to 0}\sqrt{\frac{\ell_1\ldots \ell_n}{\ell_1+\ldots+\ell_n}}  \Big(\frac{\partial}{\partial x}\Big)^{n-1} \langle Z(x;\ell_1+ \ldots+ \ell_n)\rangle. 
\eeq
Each derivative can be interpreted as an insertion for each boundary of the KdV operator associated to the parameter $x$ (corresponding to $t_0$ in the usual nomenclature). The undeformed ($x=0$) version of $Z(\ell_1+ \ldots+ \ell_n)$ is, in the JT limit, the answer one would obtain from a multi-loop amplitude in BF theory associated to (the universal cover of) $SL(2,\mathbb{R})$, as derived in \cite{Verlinde:2020upt}. It would be interesting to understand the BF nature of this KdV operator, since it allows to go from the moduli space of flat connections to the WP one, up to the simple length dependent prefactor in the equation above.

This formula also predicts a very simple behavior for the higher order spectral form factor correlator $ \lb |Z(\beta+ i T)|^{2n} \rb_{\rm conn} \sim (\beta^2+T^2)^{n/2}$, which (to leading order in genus expansion) is valid for all times.

A possible application of the multi-loop amplitudes computed here is to study the structure of the baby universe Hilbert space introduced in \cite{Coleman:1988cy,*Giddings:1988cx,*Giddings:1988wv} (and recently further developed in \cite{Saad:2019pqd} and \cite{Marolf:2020xie}), which we leave for future work. These euclidean wormholes were recently found to be relevant towards understanding unitarity of black hole evaporation \cite{Saad:2019pqd, Almheiri:2019qdq, *Penington:2019kki, Marolf:2020xie}.\footnote{Although their Lorenzian interpretation is not clear \cite{Giddings:2020yes}.} Also, adding brane boundaries can be interpreted as fixing eigenvalues of the random matrix integral \cite{Maldacena:2004sn}, which allows one to simulate an underlying discrete system \cite{Blommaert:2019wfy,*Blommaert:2020seb}. 

\begin{center}
\textbf{Supersymmetric versions}
\end{center}	
Our construction of fixed length amplitudes can be generalized to the $\mathcal{N}=1$ minimal superstring, composed of $\mathcal{N}=1$ super-Liouville with a superminimal model, mimicking most of the steps in this work. The comparison to JT gravity can be made since both the disk partition functions, the bulk one-point function and the boundary two-point functions are all known \cite{Stanford:2017thb,Mertens:2017mtv,Stanford:2019vob}. The resulting structure of the amplitudes is quite analogous and is presented in \cite{Mertens:2020pfe}.

\begin{center}
\textbf{Dilaton gravity interpretation}
\end{center}	
It would be of high interest to get a better understanding of the bulk gravitational interpretation of the Liouville gravities, with the holographic interpretations made in this work. We point out a connection of Liouville gravity to dilaton gravity in Appendix \ref{app:connliouville}, derived in \cite{StanfordSeiberg}, where we combine the Liouville $\phi$ and matter field $\chi$ into the conformal factor of the metric $\rho$ and the dilaton field $\Phi$. In particular, the dilaton potential is $V(\Phi) \sim \sinh 2b^2 \Phi$.

Assuming such a connection to dilaton gravity exists, we can substantiate the precise form of the potential purely from bulk gravity considerations as follows. It is known that for a generic model with dilaton potential $V(\Phi)$
\begin{equation}
\label{act}
S = - \frac{1}{2} \int d^2 \sqrt{g}(\Phi R + V(\Phi)),
\end{equation}
every classical solution to this system can be written in the form \cite{Gegenberg:1994pv,Witten:2020ert}:
\begin{equation}
\label{bhgen}
ds^2 = A(r) dt^2 + \frac{dr^2}{A(r)}, \qquad \Phi(r) = r,
\end{equation}
where the asymptotic region $r\to +\infty$, has a linearly diverging dilaton field, like in JT gravity. The classical solution is determined by the equations of motion in terms of the potential $V$ as:
\begin{equation}
A(r) = \int_{r_h}^{r}dr' V(r'),
\end{equation}
where $r_h$ is the horizon location. Moreover, the energy-temperature relation of the black hole is determined by
\begin{equation}
E = \frac{1}{2} \int^{V^{-1}(4\pi T)} V(\Phi) d\Phi,
\end{equation}
in terms of the dilaton potential $V(\Phi)$, where $V^{-1}$ denotes the inverse function. Given an $E(T)$ relation, one can solve this functional equation to find the dilaton potential $V(\Phi)$.\footnote{In fact, there is an explicit solution for the inverse function $\Phi(V)$. First computing the canonical entropy $S(T)$ as a function of temperature $T$, one finds:
\begin{equation}
\Phi = \frac{1}{2\pi} S\left(\frac{V}{4\pi}\right),
\end{equation}
which is uniquely invertible into $V(\Phi)$ given an assumption of monotonicity of $S(T)$ as a function of $T$.
} 
Taking
\begin{equation}
V(\Phi) = 4\pi b^2 \kappa \, \sinh 2 \pi b^2\Phi,
\end{equation}
we indeed find
\begin{equation}
E = \sqrt{ T^2/b^4+  \kappa^2} ,
\end{equation}
reproducing the first law \eqref{firstlaw} we found for the fixed-length disk partition function, but now coming from a (thermodynamically stable) bulk black hole solution. This provides substantial evidence to our claim that the bulk gravity is a 2d dilaton gravity model with a $\sinh$ dilaton potential.\footnote{The precise coefficients in the sinh potential can be changed by rescalings and are not important for our purposes here.}

The (real-time) classical black hole solution \eqref{bhgen} is then:
\begin{equation}
\label{geom}
ds^2 = - 2\kappa \left[\cosh 2\pi b^2 r - \cosh 2 \pi b^2 r_h \right] dt^2 + \frac{dr^2}{2\kappa \left[\cosh  2\pi b^2 r - \cosh 2\pi b^2 r_h \right]}, \qquad \Phi(r) = r,
\end{equation}
where the horizon radius $r_h$ is related to the temperature $T$ as
\begin{equation}
r_h =  \Phi_h = \frac{1}{2\pi b^2} \text{arcsinh} \frac{T}{\kappa b^2}.
\end{equation}
The thermal entropy of the system can be found as the Bekenstein-Hawking entropy, or directly by using the first law, and we get:
\begin{equation}
S = 2 \pi \Phi_h +S_0 = \frac{1}{b^2} \text{arcsinh} \frac{T}{\kappa b^2} +S_0.
\end{equation}
One checks that the Ricci scalar of this solution is indeed
\begin{equation}
R = - 8 \pi^2 b^4 \kappa\,  \cosh 2\pi b^2 r  = - V'(\Phi),
\end{equation}
as required by the $\Phi$ equation of motion of \eqref{act}. The geometry \eqref{geom} interpolates between the JT black hole for $r,r_h \ll 1/b^2$ with constant negative Ricci scalar, and an exponentially rising Ricci-scalar closer to the boundary.
This black hole solution has been written before in \cite{Kyono:2017jtc,*Kyono:2017pxs,*Okumura:2018xbh} in the context of a Yang-Baxter deformation of JT gravity.\footnote{The thermodynamical relations are not the same as there due to a coordinate transformation in the time coordinate.} It would be interesting to understand this connection  and the dilaton gravity description better, which we postpone to future work.

To further probe the bulk gravitational dynamics, we can mention the following.
Heavy operator insertions serve as interesting probes of backreaction effects, which are expected to have a gravitational interpretation in terms of classical energy injections. For JT gravity, this setup was analyzed in \cite{Lam:2018pvp,Goel:2018ubv}. 
In \cite{Blommaert:2019hjr,*Mertens:2019bvy,*Blommaert:2020yeo} JT bulk observables and their correlators were introduced, exploiting a radar definition to anchor bulk points to the holographic boundary. This relied strongly on the specifics of JT gravity as a theory of boundary frames (the Schwarzian description). While the bulk here would not be so easily treated, it would be very interesting to understand whether a similar construction in the bulk would be viable, and in particular whether bulk physics behaves similarly. Since the IR of the Liouville gravities studied here matches that of JT gravity, we do not expect strong deviations from conclusions made there.

Finally, it would be interesting to apply these methods to understanding closed universes. This can be done by considering fixed length boundaries with imaginary length \cite{Maldacena:2019cbz, Cotler:2019nbi}. In particular, the CFT perspective on Liouville gravity might help finding the correct inner product between no-boundary states.

\paragraph{Acknowledgements} 

We thank V. Gorbenko, K. Jensen, S. Okumura, S. Shenker, D. Stanford, M. Usatyuk, H. Verlinde and W. Weng for useful discussions. We also thank A. Blommaert for initial collaboration.  We thank A. Artemev for pointing out a mistake in a previous version. TM gratefully acknowledges financial support from Research Foundation Flanders (FWO Vlaanderen). GJT is supported by a Fundamental Physics Fellowship. 
 
\appendix 

\section{Degenerate branes}
\label{s:degbrane}
In the main text, we investigated fixed length brane segments found by applying the integral transform \eqref{eq:deffixlength} to the fixed $\mu_B$ FZZT brane segments. This leaves the question how precisely the degenerate ZZ $(m,n)$ branes relate to the fixed length boundaries. By contemplating the classical Liouville geometry, a disk with a ZZ-brane boundary corresponds to the full pseudosphere geometry \eqref{ZZZgeom}, and can be viewed as the $\ell \to + \infty$ limit of a fixed length geometry. Here we entertain this possibility and look at whether ZZ-brane segments can indeed be viewed as $\ell_i \to +\infty$ limits of fixed length amplitudes. \\
It is long known that the degenerate (ZZ) branes can be found from the FZZT branes as \cite{Martinec:2003ka}
\begin{equation}
\left|m,n\right\rangle = \left|\mu_B(m,n)\right\rangle - \left|\mu_B(m,-n)\right\rangle,
\end{equation}
where one takes the FZZT brane at the imaginary value $s = i \left(\frac{m}{2b}+\frac{nb}{2}\right)$ and hence $\mu_B(m,n) = (-)^m \kappa \cos \pi n b^2$.
Using this equality, one can readily write down the \emph{marked} ZZ-disk partition function by using \eqref{Zmarked} with the ZZ-values for the brane parameters:
\begin{equation}
Z_{m,n} \sim \cosh \frac{2\pi s(m,n)}{b} - \cosh \frac{2\pi s(m,-n)}{b} = -2 \sin \frac{\pi m}{b^2} \sin \pi n = 0.
\end{equation}
To further motivate this definition also for correlation functions, we make the following two remarks.
\begin{itemize}
\item
Subtracting the values of $s(m,n)$ and $s(m,-n)$ for the bulk one-point function $U_s(\alpha)$ \eqref{Lonep}, we write:
\begin{align}
&U_{m,n}(\alpha) = \left\langle ZZ_{m,n}\right|\left. V_\alpha \right\rangle \nonumber \\
&= \frac{\sin \pi b n (2\alpha-Q) \sin \frac{\pi m}{b}(2\alpha-Q)}{\sin \pi b (2\alpha-Q) \sin \frac{\pi}{b}(2\alpha-Q)}\frac{4\pi^2(\pi \mu \gamma(b^2))^{\frac{Q}{2b} - \frac{\alpha}{b}}}{\Gamma(bQ-2b\alpha)\Gamma(Q/b-2\alpha/b)(Q-2\alpha)}.
\end{align}
This matches the bulk one-point function on the pseudosphere \cite{Zamolodchikov:2001ah}, provided we normalize this object such that $U_{m,n}(0)= 1$.
\item
For the bulk-boundary correlator, Hosomichi proposed the following strategy of stripping off the FZZT brane wavefunction to go to the Ishibashi state $\left|p\right\rangle \hspace{-0.5mm} \rangle$ and then convolving this with the ZZ-brane wavefunction \cite{Hosomichi:2001xc,Ponsot:2003ss}:
\begin{align}
\tilde{R}(p|\alpha,\beta) &= \frac{1}{2}\int_{-\infty}^{+\infty}ds \, e^{4\pi s p} \, R(s|\alpha,\beta), \\
R(m,n|\alpha,\beta) &= \int_{-i\infty}^{+i\infty} dp \, \sin 2\pi \frac{m}{b}p \, \sin 2\pi n b p \, \tilde{R}(p|\alpha,\beta).
\end{align}
This combined strategy, upon shifting the $s$-integration contour in the imaginary direction (where no poles are crossed in the $s$-plane), corresponds precisely to taking the difference between the $s(\pm m,\pm n)$ analytically continued FZZT-branes, which is indeed Martinec's boundary state prescription.
\end{itemize}
Motivated by these results, we will consider an arbitrary boundary $n$-point function where any segment $i$ is replaced by a degenerate ZZ-brane by subtracting the $s_i(m,n)$ and $s_i(m,-n)$ FZZT-brane amplitudes. Generically, the resulting amplitudes vanish unless the boundary vertex operator is fine-tuned to satisfy the degenerate fusion rules. As an explicit example, the two-point function where any segment is a ZZ-brane vanishes. Indeed, inserting $s_1(m,\pm 1)= i \left( \frac{m}{b} \pm  b\right)$ in \eqref{lou2}, the computation reduces to a variant of \eqref{discmain}. The result boils down to:
\begin{align}
&\sin(\pi b (Q-\beta \pm i \frac{s_2}{2} - \frac{m}{2b} -\frac{b}{2} )) - \sin(\pi b (Q-\beta \pm i \frac{s_2}{2} + \frac{m}{2b} -\frac{b}{2} )) \nonumber \\
&= \sin(2\pi b (Q-\beta) - \pi b^2) \sin \pi b \frac{m}{b} \equiv 0.
\end{align}
All of this is consistent with the $\ell \to + \infty$ limit of any brane segment within a correlation function, bringing the entire correlator down to zero. And indeed, in the classical fixed-length geometry the $\ell \to +\infty$ limit is the pseudosphere geometry, see Appendix \ref{app:unif} for some formulas.

\section{Degenerate insertions and uniformization} \label{app:unif}
The bulk operator insertion becomes a degenerate Virasoro primary \eqref{eq:liouvdeg} when
\begin{equation}
P = \frac{i}{2} \left( \frac{m}{b} + nb\right), \qquad n,m \geq 1, n,m \in \mathbb{N}.
\end{equation}
In that case, we define the fixed length bulk tachyon vertex operator as the linear combination:
\begin{equation}
\mathcal{T}_{(m,n)}^{\text{deg}} \equiv \mathcal{T}_{\alpha_{(m,n)}} - \mathcal{T}_{\alpha_{(m,-n)}},
\end{equation}
where $\mathcal{T}_{P =  \frac{i}{2} \left( \frac{m}{b} \pm nb\right)}$ are defined in \eqref{defbulk}. Note that these are degenerate operators in the \emph{Liouville} sector. This is the same procedure as how degenerate ZZ-branes are found from the FZZT branes \cite{Martinec:2003ka}. Defining the bulk operator as the zero-length limit of a ZZ-brane, we are led to studying this combination of vertex operators. Using \eqref{bulkone}, one immediately evaluates this to\footnote{A fun way of writing this equivalently is as:
\begin{align}
\frac{4}{b}\int_{0}^{+\infty}ds e^{-\ell \kappa\cosh 2 \pi  b s} \sinh 2 \pi n b s \sinh \frac{2\pi m s}{b} = (-)^{n-1}U_{n-1}\left(\partial_{\kappa \ell} \right) \left[\frac{m}{2\pi b^3} \frac{1}{\kappa \ell} K_{\frac{m}{b^2}}(\kappa \ell)\right],
\end{align}
in terms of the Chebyshev polynomial of the second kind $U_{n-1}$ of the differential operator $\partial_{\kappa \ell}$ applied to the $n=1$ result.
}
\begin{align}
\label{degvir}
\left\langle \mathcal{T}_{(m,n)}^{\text{deg}}\right\rangle = \frac{4}{b}\int_{0}^{+\infty}ds e^{-\ell \kappa\cosh 2 \pi  b s} \sinh 2 \pi n b s \sinh \frac{2\pi m s}{b} 
&= \frac{1}{\pi b^2}\left[ K_{\frac{m}{b^2}+n}(\kappa \ell) - K_{\frac{m}{b^2}-n}(\kappa \ell)\right].
\end{align}

For the particular case of $n=1$, there is a second way of evaluating this amplitude. Using the identity
\begin{align}
\frac{2\alpha}{\ell} K_{\alpha}(\ell) = K_{\alpha+1}(\ell) - K_{\alpha-1}(\ell),
\end{align}
the case $n=1$ can be equivalently written as 
\begin{equation}
\label{n1alt}
\left\langle \mathcal{T}_{(m,1)}^{\text{deg}}\right\rangle = \frac{2m}{b^2 \kappa}\frac{1}{\ell}\frac{1}{\pi b^2} K_{\frac{m}{b^2}}(\kappa \ell),
\end{equation}
which can be read as the $P = i \frac{\theta}{2b}$ where $\theta=m$ amplitude with one less marking (due to the factor $1/\ell$). The case $\theta \in \mathbb{N}$ is a discrete subset of the microscopic Liouville punctures, but it is special in that it shouldn't be marked additionally, unlike the $\theta \notin \mathbb{N}$.
Let us indeed show that this is true.
Inspired by the above relation, we define a differently normalized bulk operator as:
\begin{equation}
\mathcal{T}_{\alpha_{(m,0)}}^{\U} \equiv \frac{2m}{b^2 \kappa}\, \mathcal{T}_{P =  \frac{im}{2b}},
\end{equation}
to be used in a fixed length computation without additional marking.
Deforming the contour as before, we have the relation:
\begin{equation}\label{eq:disccosexc}
\text{Disc } \left[\cosh \frac{2\pi m s}{b}\right] = 2i \sin \frac{\pi m}{b^2} \sinh \frac{m}{b^2} \text{arccosh} \frac{\left|\mu_B\right|}{\kappa},
\end{equation}
valid for $\mu_B<-\kappa$. As before, for $\mu_B \in (-\kappa,0)$, there is no discontinuity. Finally the bulk one-point function at fixed length is directly given by 
\begin{equation}
\label{bulkexc}
\left\langle \mathcal{T}_{\alpha_{(m,0)}}^{\U}\right\rangle_\ell =  \frac{4}{b} \int_{0}^{\infty} ds\hspace{0.1cm} e^{-\ell \kappa \cosh(2\pi b s)} \sinh 2 \pi b s \sinh\frac{2 m \pi s}{b}, 
\end{equation}
indeed reproducing \eqref{n1alt}.

\subsection{Uniformization and markings}
\label{app:mark}
To gain intuition for why the $\theta \in \mathbb{N}$ case is with one less marking operator than the generic case $\theta \notin \mathbb{N}$, we can think about the classical Liouville geometry and its uniformization map. Bulk operator insertions of the type $P = i \frac{\theta}{2b}$ correspond semi-classically to introducing conical defects in the bulk geometry. In particular, we will be interested in the fixed length geometry.
The pseudosphere geometry $\left|h\right| \leq 1$ is:
\begin{equation}
\label{ZZZgeom}
ds^2 = \frac{1}{(1-(h \bar{h}))^2} dh d\bar{h}
\end{equation}
The pseudosphere geometry $\left|z\right| \leq 1$ with conical deficit angle $2\pi (1-\theta)$ is given by \cite{Zamolodchikov:2001ah}:\footnote{Our parameter $\theta$ is related to their $\eta$ as $\theta=1-2\eta$.}
\begin{equation}
\label{ZZgeom}
ds^2 = \frac{\theta^2}{(z\bar{z})^{1-\theta}(1-(z \bar{z})^{\theta})^2} dz d\bar{z}
\end{equation}
The fixed-length geometry with boundary length $\ell$ and area $A$ is given by \cite{Fateev:2000ik}:
\begin{equation}
\label{S1}
ds^2 = \frac{\ell^2(a-1/a)^2}{4\pi^2((z\bar{z})^{\frac{1-\theta}{2}}/a - a(z \bar{z})^{\frac{1+\theta}{2}}))^2}dz d\bar{z}
\end{equation}
where $a^2 = 1- \frac{4\pi A}{\ell^2}\theta$. The boundary is at $\left|z\right|=1$ and it ranges from $0\leq \text{arg} z \leq 2\pi$. In the infinite length limit $\ell \to + \infty$ in \eqref{S1}, we obtain the full pseudosphere geometry \eqref{ZZgeom}:
\begin{equation}
ds^2 \to \frac{4A^2\theta^2}{\ell^2 (z\bar{z})^{1-\theta}(1-(z \bar{z})^{\theta})^2}dwd\bar{w} = \theta^2\frac{1}{(z\bar{z})^{1-\theta}(1-(z \bar{z})^{\theta})^2}dzd\bar{z}
\end{equation}
where we also have that area and length scale the same in AdS: $A = 2 \pi \int_{0}^{R}\frac{r}{(1-r^2)^2}dr \approx \frac{\pi}{1-R^2}$, and $\ell = \frac{2 \pi}{1-R^2} = 2A$. \\
This is intuition we also use in Appendix \ref{s:degbrane} where we analyze marked ZZ-brane boundary segments and illustrate that the result vanishes, in accordance with taking indeed $\ell \to + \infty$ in the amplitude explicitly.
\\~\\
Performing the uniformization map
\begin{equation}
\label{map}
w = z^{\theta},
\end{equation}
the geometry becomes:
\begin{equation}
ds^2 = \frac{\ell^2(a-1/a)^2}{4\pi^2\theta^2(1/a - a(w \bar{w}))^2}dwd\bar{w}
\end{equation}
without conical defect, but with new angular periodicity $0 \leq \text{arg} w \leq 2\pi \theta$.\footnote{In case $\theta=0$, one has parabolic monodromy and a cusp in the geometry \eqref{S1}:
\begin{equation}
ds^2 = \frac{4 \ell^2 A^2 dzd \bar{z}}{(z\bar{z}(4\pi A - \ell^2\log(z\bar{z})))^2}
\end{equation}
which can be uniformized by setting $w=\log(z)$ and $\bar{w} = -1/\log(\bar{z})$, into:
\begin{equation}
ds^2 = \frac{4 \ell^2 A^2dwd \bar{w}}{(4\pi A \bar{w}^2 - \ell^2(w \bar{w}-1))^2}.
\end{equation}
 } \\

\begin{figure}[h]
\centering
\includegraphics[width=0.4\textwidth]{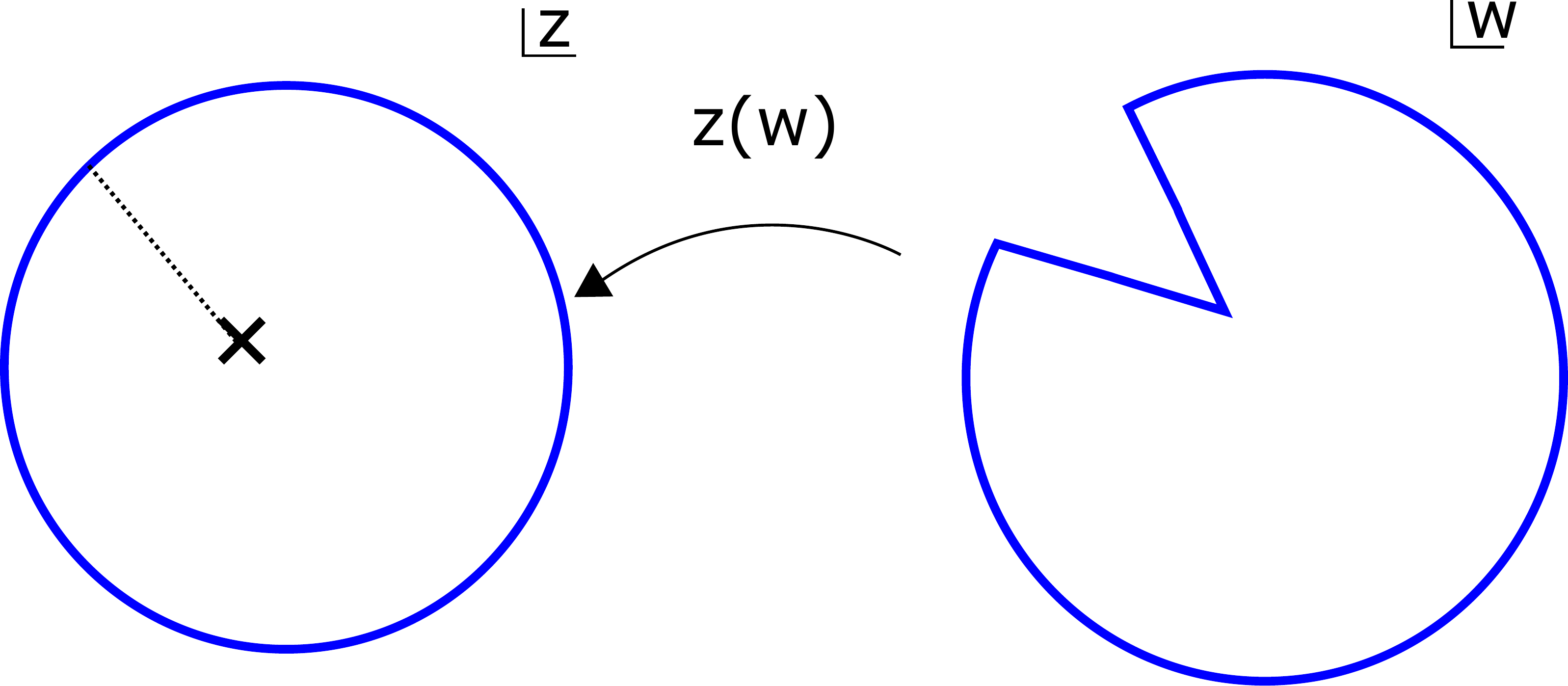}
\caption{The transformation $w(z) = z^{\theta}$ replaces the twisted angular periodicity in the $w$-coordinates with a conical defect in the $z$-coordinates. The $z$-plane has a branch cut which hits the boundary somewhere and marks it.}
\label{coneuniform}
\end{figure}

Generically, the coordinate transformation $w = z^{\theta}$, has a branch cut in the $z$-frame, reaching the boundary (Figure \ref{coneuniform} ).
This procedure marks a point on the boundary in the $w$-coordinates. In the special cases $\theta \in \mathbb{N}$, the uniformization map is regular and no branch cut needs to be defined. Note that the geometries themselves never contain branch cuts since they cancel between holomorphic and anti-holomorphic contributions. 
This provides intuition into why for $\theta \in \mathbb{N}$ we need to consider the unmarked transformation, whereas for other values of $\theta$, we consider the marked fixed length one-point function instead.

\section{Pole contribution from the two-point function} \label{app:poles}
\begin{figure}[h]
\centering
\includegraphics[width=0.7\textwidth]{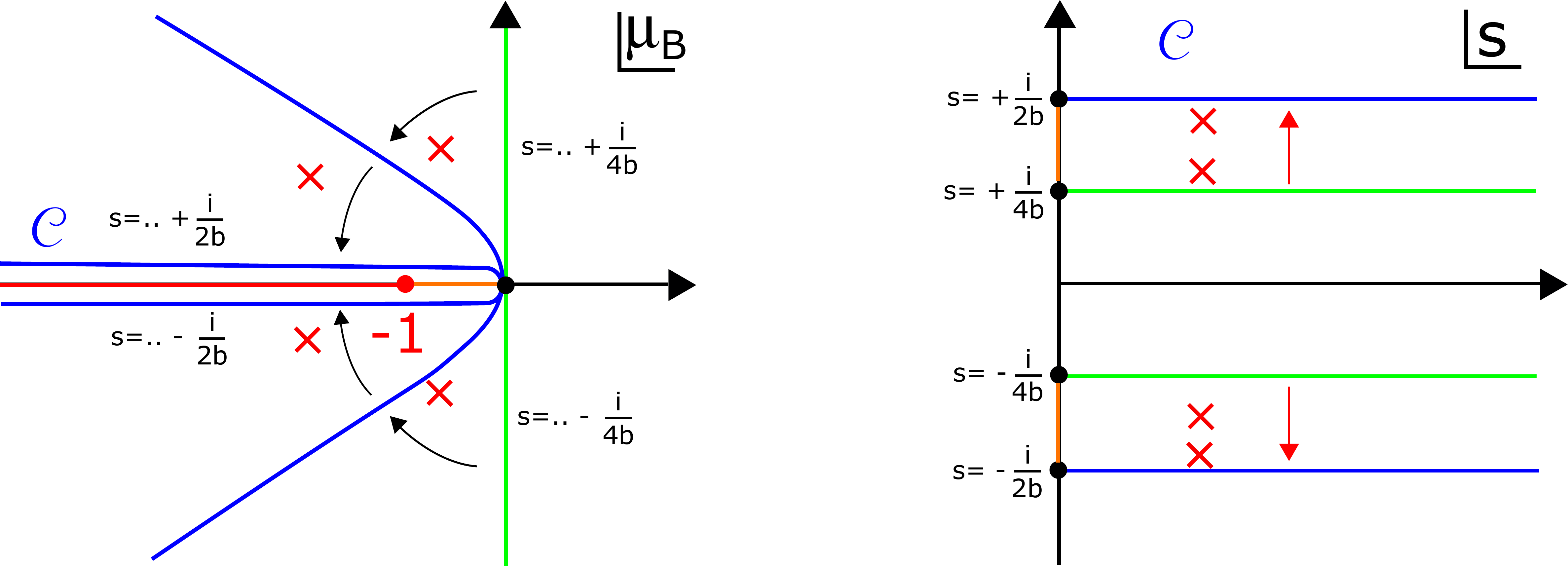}
\caption{Left: Contour deformation ($\mathcal{C}$) in the $\mu_B$-plane, and range of the variable $s$ in the process. Right: Contour deformation in the $s$-plane and possible poles. The small (orange) segments cancel out in the computation and can be dismissed.}
\label{contourDeform}
\end{figure}

In the main text, we chose to define the fixed-length contour to exclude all of the poles of the integrand. Here we show that in the JT limit, these pole contributions disappear in any case. \\
We focus in particular on the boundary two-point function \eqref{lou2}, for which the incriminating factor is
\begin{equation}
\label{incri}
S_b(Q-\beta \pm i s_1 \pm i s_2), \qquad \beta = b- \beta_M.
\end{equation}
The $S_b(x)$-function has poles at $x=-nb - m/b$ and zeroes at $x=Q+nb+m/b$ for $n,m \in \left\{0,1,2\hdots \right\}$.
The region of contour deformation is $\Re(\mu_B) <0$ where $\mu_B=\kappa \cosh 2 \pi b s$ (we set $\kappa=1$ here for convenience), for which the parameter $s$ has the following properties $\Re(s) > 0$ and:
\begin{align}
\label{polespan}
\frac{i}{4b} \leq \Im(s) &\leq \frac{i}{2b}, \qquad \Im(\mu_B) > 0, \nonumber\\
-\frac{i}{2b} \leq \Im(s) &\leq -\frac{i}{4b}, \qquad \Im(\mu_B) < 0.
\end{align}
This is illustrated in Figure \ref{contourDeform}.
We need to perform two consecutive contour deforms for the two variables $\mu_{Bi}$. 
Fix any contour for $s_2$ as in the figure, with hence $\Re(s_2) >0$, and focus on deforming the $s_1$ contour.
Since $\Re(s_i)>0$, poles can only occur for two of the four double-sine functions in \eqref{incri}:
\begin{align}
S_b\left(Q-\beta + i s_1 - i s_2\right)S_b\left(Q-\beta - i s_1 + i s_2\right),
\end{align}
at the values of $s_1$:
\begin{equation}
s_1 \equiv s_{2(m,n)} = s_2 \pm i\left[ (nb + \frac{m}{b}) +  (Q-\beta)\right], \quad n,m \in \left\{0,1,2,\hdots\right\}.
\end{equation}
Since $\beta \sim b$, one can absorb all $1/b$-dependence in $m=1,2,\hdots$. Then since $\frac{i}{4b} < \left|\Im(s_2)\right| < \frac{i}{2b}$, and since $m\geq1$, it is impossible for any of the poles in $s_1$ to be contained in the region of interest, in the $b\to 0$ limit.\footnote{When arguing for this, we assumed the $n$-label is parametrically less than $1/b^2$ and hence we do not allow that label to be so large such that it can counteract the effect of the $m$-label.}
\\~\\
For finite $b$, we will get a contribution from crossing poles. Let us write an explicit expression for this contribution. 
Denote by $I$, the subset of poles that are included in the region crossed during the $s_1$-contour deformation \eqref{polespan}. The sum of the two pole series (coming from the two $S_b$-functions with poles in the crossed region), then gives the pole contributions:
\begin{align}
\label{poleterm}
R_{s_2} \equiv &\sum_{n,m \in I} \text{Res}\left.S_b\right|_{x=-nb - \frac{m}{b}} \\
&\times S_b(2(Q-\beta)+nb + m/b) S_b(-nb - m/b + is_2) S_b(2(Q-\beta)+nb + m/b-is_2) - (cc), \nonumber
\end{align}
where the residue has the explicit expression
\begin{equation}
\text{Res}\left.S_b\right|_{x=-nb - \frac{m}{b}} = \frac{1}{2\pi}\frac{(-)^{nm+n+m}}{2^{n+m}} \prod_{r=1}^{n}\frac{1}{\sin r \pi b^2} \prod_{s=1}^{m}\frac{1}{\sin s \frac{\pi}{b^2}} .
\end{equation}
We get
\begin{align}
\label{halfway}
\int_{i\mathbb{R}}d\mu_{B2} e^{\ell_2 \mu_{B2}(s_2)} \int_{\mathcal{C}} d\mu_{B1} e^{\ell_1 \mu_{B1}(s_1)} S_b\left(Q-\beta\pm i s_1 \pm i s_2\right) + \int_{i\mathbb{R}} d\mu_{B2} e^{\ell_2 \mu_{B2}(s_2)+\ell_1\mu_{B_1}(s_{2(m,n)})} R_{s_2},
\end{align}
where the contour $\mathcal{C}$ is now wrapping the negative real axis (figure \ref{contourDeform}).

After this, we deform the $s_2$-contour in a similar way to wrap the negative real axis. There is a term similar to \eqref{poleterm} coming from picking up the residues of the first term in \eqref{halfway}. Due to $s_1 \leftrightarrow s_2$ symmetry, this term is identical to \eqref{poleterm}. Since there are no crossed poles in $s_2$ from the second term in \eqref{halfway},\footnote{Because $\Re(s_2)$ only vanishes on the real axis on $(-1,1)$, which is not traversed during the contour deformation (see Figure \ref{contourDeform}).} the net result is merely accounting for the pole term \eqref{poleterm} twice:
\begin{align}
\label{allway}
\int_{\mathcal{C}} d\mu_{B2} \int_{\mathcal{C}} d\mu_{B1} &e^{\ell_1 \mu_{B2}(s_1)} e^{\ell_1 \mu_{B2}(s_2)} S_b\left(Q-\beta\pm i s_1 \pm i s_2\right) \nonumber \\
&+ \int_{i\mathbb{R}} d\mu_{B} \left(e^{\ell_2 \mu_{B}(s)+\ell_1\mu_{B}(s_{(m,n)})} + e^{\ell_2 \mu_{B}(s_{(m,n)})+\ell_1\mu_{B}(s)} \right)R_{s},
\end{align}
where the first term can be done as in the main text by evaluating the discontinuity across the cut, leading to \eqref{eq:2pt}. The second line represents a discrete addition. It would be interesting to have a more intuitive understanding of it. Regardless, in the main text and motivated by the match to the matrix integral, we define our amplitudes and in particular the integration contour $\mathcal{C}$ to exclude this contribution.

\section{Degenerate fusion versus matrix model: a $j=1$ case study}
\label{app:one}
In this appendix, we work out the formula for the second minimal string operator insertion at $j=1$ using the fusion algebra \eqref{qfus} and match to the discrete matrix model result \eqref{gendeg}.

The general boundary two-point function has the schematic structure:
\begin{align}
\label{stram}
\int ds_1 ds_2 \rho(s_1) \rho(s_2) e^{-\ell_1 \mu_B(s_1)}e^{-\ell_2\mu_B(s_2)} \, \mathcal{A}_{\beta_M}(s_1,s_2)
\end{align}

For $j=1$, we apply the fusion algebra \eqref{qfus} twice to derive the identity:
\begin{align}
\mathcal{A}_{\beta_M}(s_1,s_2) &= \int_{-\infty}^{+\infty} dx \psi^{\epsilon}_{s_1}(x) \psi^{\epsilon * }_{s_2}(x) e^{- 2\pi b x} \\
&= \frac{-\pi b^2}{\rho(s_1)}\left[\frac{\delta(s_1-s_2-ib)}{\sinh 2 \pi  b s_2 \sinh 2 \pi b (s_2+ib/2) }- \frac{\delta(s_1-s_2)}{\sinh 2 \pi  b s_2 \sinh 2 \pi b (s_2+ib/2)} \right. \nonumber \\
&\left. - \frac{\delta(s_1-s_2)}{\sinh 2 \pi  b s_2 \sinh 2 \pi b (s_2-ib/2) } + \frac{\delta(s_1-s_2+ib)}{\sinh 2 \pi  b s_2 \sinh 2 \pi b (s_2-ib/2) } \nonumber
\right]
\end{align}
Inserting this in the amplitude \eqref{stram}, we get:
\begin{align}
\label{fusone}
\int ds_2  \rho(s_2) e^{-\ell_2\mu_B(s_2)}&\left[\frac{e^{-\ell_1 \mu_B(s_2+ib)}}{\sinh 2 \pi  b s_2 \sinh 2 \pi b (s_2+ib/2) } + \frac{e^{-\ell_1 \mu_B(s_2-ib)}}{\sinh 2 \pi  b s_2 \sinh 2 \pi b (s_2-ib/2)} \right. \nonumber \\
&\left.- \frac{e^{-\ell_1 \mu_B(s_2)}}{\sinh 2 \pi  b s_2} \left( \frac{1}{\sinh 2 \pi b (s_2+ib/2)}+\frac{1}{\sinh 2 \pi b (s_2-ib/2)}\right)\right]
\end{align}
This matches with the matrix model result \eqref{gendeg} with $j=1$. To see this, one uses the following hyperbolic identities:
\begin{align}
&\frac{-C}{\displaystyle{\prod_{\stackrel{m\in \left\{-1,0,1\right\}}{m\neq 1}}}(\cosh 2\pi b (s+ib) - \cosh 2\pi b (s+imb)))} = \frac{1}{\sinh 2\pi b s \sinh 2\pi  b(s+i\frac{b}{2})}\\
&\frac{-C}{\displaystyle{\prod_{\stackrel{m\in \left\{-1,0,1\right\}}{m\neq -1}}}(\cosh 2\pi b (s-ib) - \cosh 2\pi b (s+imb)))} = \frac{1}{\sinh 2\pi b s \sinh 2\pi  b(s-i\frac{b}{2})} \nonumber \\
&\frac{C}{\displaystyle{\prod_{\stackrel{m\in \left\{-1,0,1\right\}}{m\neq 0}}}(\cosh 2\pi b s - \cosh 2\pi b (s+imb)))} = \frac{1}{\sinh 2\pi b s}\left(\frac{1} {\sinh 2\pi  b(s+i\frac{b}{2})}-\frac{1}{ \sinh 2\pi  b(s-i\frac{b}{2})}\right) \nonumber
\end{align}
with a proportionality factor $C= 4 \sin \pi b^2 \sin 2\pi b^2$ that only gives an overall normalization. The lhs is the form of the correlator obtained using the matrix model description \eqref{gendeg}, whereas the rhs is the form obtained using successive applications of the degenerate fusion algebra \eqref{fusone}. The rhs also directly limits to the JT structure by setting $s = bk$:
\begin{equation}
\sinh \pi  b^2(2k - in) \to -i\pi b^2 (2ik + n)
\end{equation}
generating a polynomial of the type $\prod_n (2ik+n)$ indeed obtained in the degenerate JT bilocal correlators \cite{Mertens:2020pfe}.

\section{Crosscap spacetime} \label{app:crosscap}
In this section we will compute the partition function in the crosscap spacetime from the continuous approach. This will give a contribution to the resolvent when considering the unoriented minimal string. For the $(2,p)$ series, the theory is dual to a GOE or GSE random matrix integral (depending on how unoriented contributions are weighted) \cite{Harris:1990kc, *Brezin:1990xr, *Brezin:1990dk}. We will find a precise agreement with the continuous calculation and the discrete one. Interestingly, this contribution is not universal (as opposed to the cylinder amplitude) and a precise match depends on using the correct density of states.  

 To simplify we will consider from the beginning an FZZT brane with identity brane matter boundary conditions. The crosscap boundary state is given by 
 \beq
|{\rm xcap} \rb = \sum_{n,m} \int_0^{\infty} dP \hspace{0.1cm}\Psi_{\rm xcap}(P) \frac{P_{1,1}^{n,m}}{(S_{1,1}^{n,m})^{1/2}}  |P\rb\hspace{-0.1cm}\rb_L |n,m \rb\hspace{-0.1cm}\rb_M,
\eeq
where we used the modular $P$-matrix defined by $P=\sqrt{T} S T^2 S \sqrt{T}$ (not to be confused with the Liouville momentum). The matter crosscap state for minimal models was derived in \cite{Bianchi:1991rd} and the Liouville crosscap state $\Psi_{\rm xcap}(P)$ can be found in equation (4.16) of \cite{Hikida:2002bt}. The overlap, shown in figure \ref{fig:xcap}, between the crosscap and FZZT boundary state gives a factorized answer before moduli integration 
\beq
\lb Z(s)^{\U} \rb_{\rm xcap} = \int d\tau Z_{L} Z_{M}  Z_{G},~~~~\begin{cases} Z_{L} =\int_0^\infty \frac{dP \cos 4 \pi s P}{4\pi \sinh \pi b P \sinh \pi \frac{P}{b}}\hat{\chi}_P(q), \\ 
~~\vspace{-0.5cm}\\
Z_{M} = \hat{\chi}_{1,1} (\tau'=-1/\tau),\\
~~\vspace{-0.5cm}\\
Z_{G} =\eta^2(-\sqrt{q}). \end{cases}
\eeq
where following convention we defined the modified character $\hat{\chi}_h(q) = e^{-i \pi (h-\frac{c}{24})} \chi_h(-\sqrt{q})$. We stress this calculation gives the unmarked partition function. Up to a simple factor the descendant contributions cancel and the final amplitude is given by  
\beq\nonumber
Z(s)^{\U}_{\rm xcap} = \int_0^\infty dP \frac{\cos 4 \pi s P }{4\pi \sinh \pi b P \sinh \pi \frac{P}{b}} \sum_{k\in \mathbb{Z}} (-1)^k \int_0^\infty \frac{dt}{\sqrt{t}}   e^{-\pi t P^2} \big(e^{-\frac{\pi}{t}a_{1,1}(k)}+e^{-\frac{\pi}{t} a_{1,-1}(k)}\big),
\eeq
where $a_{m,n}(k)$ was defined in equation \eqref{degcharacters}.  We can do the moduli integral and the final answer for the $(p,p')$ minimal string is
\beq
 Z(s)^{\U}_{\rm xcap} = \pm \int_0^\infty dP\cos 4 \pi s P \frac{\sinh \pi (p-1)\frac{P}{b} \coth \pi b P }{2\pi P \sinh \pi \frac{P}{b} \cosh \pi p \frac{P}{b}}
\eeq
 and for the $(2,p)$ series we get
 \bea
Z(s)^{\U}_{\rm xcap} &=&\pm \int_0^\infty \frac{dP}{\pi}  \frac{\cos 4 \pi s P \coth \pi b P }{2 P \cosh 2 \pi \frac{P}{b}},\\
Z(s)^{\M}_{\rm xcap} &=&\pm \frac{1}{b} \int_0^\infty \frac{dP}{\pi} \frac{\sin 4 \pi s P}{\kappa \sinh \pi b s} \frac{ \coth \pi b P }{ \cosh 2 \pi \frac{P}{b}},
\eea
where in the second line we also wrote the marked partition function by taking a derivative with respect to $\mu_B(s)$. 
\begin{figure}[t!]
\centering
\begin{tikzpicture}[scale=0.9]
\node at (-3.4,0) {{\small $\lb Z(s)^{\M} \rb_{\rm xcap}$} $=$};
\draw[thick] (0,0) ellipse (0.3 and 1.5);
\draw[thick] (2,0) ellipse (0.2 and 0.7);
\draw[thick] (0.04,1.49) to [bend right=20] (2,0.7);
\draw[thick] (0.04,-1.49) to [bend left=20] (2,-0.7);
\draw[thick] (1.84,0.4) -- (2.16,-0.4);
\draw[thick] (1.84,-0.4) -- (2.16,0.4);
\node at (-1,0) {\small $s_{(1,1)}$};
\end{tikzpicture}
\caption{We depict the crosscap spacetime amplitude with an FZZT boundary and identity brane matter boundary condition. The boundary cosmological constant is $\mu_B(s)$ and the fixed length version is obtained by integrating over $s$. }
\label{fig:xcap}
\end{figure}
This latter quantity is equal to the crosscap contribution to the resolvent of the matrix integral. The prediction from the matrix integral is 
 \beq
 R_{\frac{1}{2}}(x) = - \frac{1}{2\pi \sqrt{-x}} \int_0^\infty \frac{\sqrt{x'}dx'}{x'-x} \frac{\partial_{x'}y(x')}{y(x')},
 \eeq
which can be found in \cite{Stanford:2019vob} and $y(x)$ is the density of states and $x$ the matrix eigenvalue. For the minimal string one should take $y(x) =\frac{1}{4\pi^2}\sinh(\frac{p}{2} {\rm arccosh}(1+\frac{8\pi^2}{p^2} x))$. Even though we were not able to perform the integrals explicitly we checked numerically that both quantities match when parameters are appropriately identified  
\beq
Z(s)^{\M}_{\rm xcap} = R_{\frac{1}{2}} (x= \mu_B(s) - \kappa).
\eeq
This check depends crucially on the details of the minimal string density of states since this contribution is not universal in the double scaling limit.

Finally we can also compute this contribution with a fixed length boundary. Doing the inverse Laplace transform and using the discontinuity of the integrand we get 
 \beq\label{xcapfixedl}
Z(\ell)_{\rm xcap} = \int_0^\infty \frac{d\lambda}{\pi} K_{i\lambda}(\kappa \ell) \coth \frac{\pi b^2 \lambda}{2}    \tanh \pi\lambda, 
\eeq
where we redefined $\lambda = 2P/b$. As we noticed for the cylinder this again has the form of an integral over a trumpet contribution with parameter $\lambda$ and a minimal string generalization of the WP measure for a crosscap. Following the notation used in the main text we define this as 
\beq
V_{\frac{1}{2}}(\lambda) =\coth \frac{\pi b^2 \lambda}{2}    \tanh \pi\lambda.
\eeq
We can look at the JT gravity limit where the Bessel function becomes the trumpet partition function and this volume becomes 
\beq
V_{\frac{1}{2}}(\lambda=b_{\rm JT}/(2\pi b^2)) \to \coth \frac{b_{\rm JT}}{4} ,
\eeq
 in the $b\to0$ limit, for fixed $b_{\rm JT}$. This matches with the answer found directly by Stanford and Witten \cite{Stanford:2019vob} up to an appropriate order one rescaling of $b_{\rm JT}$. As pointed out in Appendix F of \cite{Stanford:2019vob} from a matrix integral perspective, we see here directly that the answer is finite for finite $b$ since the $\tanh \pi \lambda$ factor in \eqref{xcapfixedl} makes the $\lambda\to0$ limit smooth. 

\section{JT vs Liouville gravity} \label{app:connliouville}
The connection between Liouville gravity and JT gravity seems to be very robust. We have checked this for several observables finding a match in each case. Is there a derivation then of this correspondence? In this appendix we want to make some comments in this direction \footnote{A different connection between JT and Liouville gravity was pointed out in \cite{Mandal:2017thl}.}. 

A possible derivation was done by Seiberg and Stanford \cite{StanfordSeiberg}. The idea is to start with the action for the gravitational Liouville field $\phi$ and the matter field $\chi$ written as a time-like Liouville field as in \eqref{eq:timeLioaction}. For simplicity we can pick the fiducial metric to be a flat disk $\hat{g}_{\mu\nu}=\delta_{\mu\nu}$. If we parametrize the fields as $\phi = b^{-1} \rho - b\pi  \Phi$ and $\chi = b^{-1} \rho + b\pi  \Phi$, the sum of the matter and gravitational Liouville actions is 
\beq
S = - \int \partial \Phi \cdot \partial \rho  +  \int e^{2 \rho} ( \mu_L e^{- 2\pi b^2 \Phi} + \mu_M e^{2\pi b^2 \Phi}) ,
\eeq  
which represents a more general dilaton gravity with dilaton potential 
\begin{equation}
V(\Phi) \sim (\mu_M-\mu_L)\sinh 2 \pi b^2 \Phi + (\mu_L+\mu_M)\cosh 2\pi  b^2 \Phi.
\end{equation}
To connect with JT gravity we want to interpret $\rho$ as a scale factor of a 2D metric $g_{\rm JT} = e^{2 \rho} \hat{g}$ and $\Phi$ as the dilaton. We can first look at the kinetic term in the action. It is easy to check that the first term above is precisely equal to $-\frac{1}{2}\int  \Phi R$ (including the appropriate GHY boundary terms) where the integral is done over a disk with metric $g_{\rm JT}$ and $R$ is the Ricci scalar corresponding to that metric. In the small $b$ limit the interaction term is approximately 
\beq
\int e^{2 \rho} ( \mu_L e^{-2\pi b^2 \Phi} + \mu_M e^{2\pi b^2 \Phi}) \sim (\mu_L+\mu_M) \int \sqrt{g_{\rm JT}} + 2\pi b^2(\mu_M-\mu_L) \int \sqrt{g_{\rm JT}} \hspace{0.1cm} \Phi  + \ldots 
\eeq
The second term in the right hand side is precisely the JT gravity linear dilaton potential with a cosmological constant $\Lambda_{\rm JT} =2 \pi b^2 (\mu_L- \mu_M)$. The first term in the right hand side is proportional to the area of the surface of metric $g_{\rm JT}$ and can be easily accounted for.
For obtaining the actual JT gravity action, we remove this term by picking $\mu_M=-\mu_L$. This leads to the prediction that the Liouville gravity models studied in this work can be written as a 2d dilaton gravity model with dilaton potential:
\begin{equation}
V(\Phi) = \frac{\Lambda_{\rm JT}}{2\pi b^2}  \sinh \left(2 \pi b^2 \Phi\right).
\end{equation}

Regarding boundary conditions, the fixed-length boundary is equivalent to Dirichlet boundary conditions fixing roughly $\ell \to (2\pi) e^{\rho|_{\partial}} e^{b^2 \Phi|_{\partial}}$. Meanwhile, the fact that we pick the matter identity brane means we are taking the combination $\chi |_{\partial} \to \infty$, since this is the ZZ brane boundary condition.\footnote{It would be interesting to study more general matter states, that might reproduce finite cutoff JT gravity in the semiclassical limit \cite{Iliesiu:2020zld, Stanford:2020qhm}} 

Even though this derivation seems reasonable at the level of the path integral, we would like to point out a subtlety that appears when we consider boundary correlators. In terms of the time-like and gravitational Liouville fields, a boundary insertion has the form 
\beq
\mathcal{B} \sim \int e^{\beta_M \chi} e^{\beta \phi}\sim \int \sqrt{h_{\rm JT}}\hspace{0.1cm} e^{-  2\pi b^2 (\frac{1}{2}-\frac{\beta_M}{b}) \Phi } ,~~~\beta=b-\beta_M,
\eeq
where $h_{\rm JT}$ is the boundary JT metric. From the perspective presented above, we are picking Dirichlet boundary conditions for both the boundary JT metric and dilaton. It is an open question (at least for us) to explain why this observable would match the JT gravity boundary correlator at all in the $b\to0$ limit. The observable in JT gravity usually appears when integrating out matter. On the other side the argument in this section would imply that the minimal string is dual to \textit{pure} JT gravity, with no matter. We leave a better understanding of this issue for future work.

\mciteSetMidEndSepPunct{}{\ifmciteBstWouldAddEndPunct.\else\fi}{\relax}
\bibliographystyle{utphys}
{\small \bibliography{references}{}}

\end{document}